%% file: main.tex
\documentclass[twocolumn,twocolappendix]{openjournal}

\usepackage{amsmath,amssymb,bm,natbib,latexsym,times}

\usepackage[frozencache]{minted}
\usepackage{listings}
\lstdefinestyle{yaml}{
     basicstyle=\ttfamily\color{black}\scriptsize,
     rulecolor=\color{black},
     keywordstyle=\color{blue}\bfseries,
     comment=[l]{\#},
     commentstyle=\itshape\color{black},
     showstringspaces=false
}

\usepackage[T1]{fontenc}
\usepackage{aecompl}

\usepackage{verbatim}
\usepackage{hhline}
\usepackage[dvipsnames]{xcolor}
\usepackage{todonotes}
\usepackage{xspace}
\usepackage{hyperref}
\usepackage{cuted}
\usepackage{eso-pic}% http://ctan.org/pkg/eso-pic

\hypersetup{colorlinks=true,linkcolor=blue,citecolor=blue,filecolor=blue,urlcolor=blue}

\newcommand\edit[1]{#1}

\numberwithin{equation}{section}

\DeclareMathSymbol{:}{\mathord}{operators}{"3A}

\newcommand{\dTt}{\ensuremath{\delta T^{(2)}/T^{(2)}}}
\newcommand{\dTf}{\ensuremath{\delta T^{(4)}/T^{(4)}}}
\newcommand{\deto}{\ensuremath{\delta e^{(2)}_1}}
\newcommand{\dett}{\ensuremath{\delta e^{(2)}_2}}
\newcommand{\defo}{\ensuremath{\delta e^{(4)}_1}}

\newcommand{\piff}{\textsc{Piff}}

\newcommand{\pixmappy}{\textsc{Pixmappy}}
\newcommand{\sex}{\textsc{SExtractor}}

\newcommand{\extmash}{\texttt{EXT\_MASH}}
\newcommand{\sizemag}{\texttt{SizeMag}}
\newcommand{\mdet}{\textsc{Metadetection}}
\newcommand{\optatmo}{\texttt{OptAtmo}}

\newcommand{\gaia}{\textit{Gaia}}

\newcommand*\justify{%
  \fontdimen2\font=0.4em% interword space
  \fontdimen3\font=0.2em% interword stretch
  \fontdimen4\font=0.1em% interword shrink
  \fontdimen7\font=0.1em% extra space
  \hyphenchar\font=`\-% allowing hyphenation
}

\newcommand\code[1]{\texttt{\small\justify #1}}

\newcommand\eqn[1]{equation~\ref{#1}}
\newcommand\eqnb[2]{equations~\ref{#1}~\& \ref{#2}}

\newcommand\Eqn[1]{Equation~\ref{#1}}   % If you need to start a sentence with this...

\newcommand\fig[1]{Figure~\ref{#1}}
\newcommand\figb[2]{Figures~\ref{#1}~\& \ref{#2}}
\newcommand\tab[1]{Table~\ref{#1}}

\newcommand\app[1]{Appendix~\ref{#1}}
\newcommand\sect[1]{\S \ref{#1}}

\allowdisplaybreaks

\defcitealias{y3-piff}{J21}

\shorttitle{DES Year 6 Results: PSF Modeling}
\shortauthors{Schutt et al. (DES Collaboration)}

\begin{document}

%%% add Fermilab and DES pre-print number
\AddToShipoutPictureBG*{%
  \AtPageUpperLeft{%
    \hspace*{18.25cm}%
    \raisebox{-9.4\baselineskip}{%
      \makebox[0pt][l]{\textnormal{DES 2024-0874}}
 
}}}%

\AddToShipoutPictureBG*{%
  \AtPageUpperLeft{%
    \hspace*{20.45cm} % Adjust this to position it correctly
    \raisebox{-10.5\baselineskip}{%
      \makebox[0pt][r]{\textnormal{FERMILAB-PUB-25-0009-PPD}}
}}}%

\label{firstpage}
\title{Dark Energy Survey Year 6 Results: Point-spread Function Modeling}
%% authors
\include{authors}

\begin{abstract}

We present the point-spread function (PSF) modeling for weak lensing shear measurement using the full six years of the Dark Energy Survey (DES Y6) data.
We review the PSF estimation procedure using the \piff\ (PSFs In the Full FOV) software package and describe the key improvements made to \piff\ and modeling diagnostics since the DES year three (Y3) analysis: (i) use of external \gaia\ and infrared photometry catalogs to ensure higher purity of the stellar sample used for model fitting, (ii) addition of color-dependent PSF modeling, the first for any weak lensing analysis, and (iii) inclusion of model diagnostics inspecting fourth-order moments, which can bias weak lensing measurements to a similar degree as second-order modeling errors.
Through a comprehensive set of diagnostic tests, we demonstrate the improved accuracy of the Y6 models evident in significantly smaller systematic errors than those of the Y3 analysis, in which all $g$ band data were excluded due to insufficiently accurate PSF models.
For the Y6 weak lensing analysis, we include $g$ band photometry data in addition to the $riz$ bands, providing a fourth band for photometric redshift estimation.
Looking forward to the next generation of wide-field surveys,
we describe several ongoing improvements to \piff, which will be the default PSF modeling software for weak lensing analyses for the Vera C. Rubin Observatory's Legacy Survey of Space and Time.

\end{abstract}

% \begin{keywords}
% surveys -- catalogs --
% methods: data analysis -- techniques: image processing --
% gravitational lensing: weak -- cosmology: observations
% \end{keywords}

\section{Introduction}
\label{sec:intro}

Weak gravitational lensing (WL) is a powerful probe of cosmology, allowing us to study the distribution of matter in the Universe by measuring the correlated distortions in the shapes of background galaxies caused by the gravitational influence of foreground matter.
One method for measuring WL caused by the large-scale structure of the matter distribution is by measuring the two-point correlation function of galaxy ellipticities, also called ``cosmic shear.''
Several successful cosmic shear measurements have been conducted in recent years \citep[e.g.,][]{Asgari21, Dalal23, Li23}.
In particular, the Dark Energy Survey (DES) collaboration's cosmic shear analysis using the first three years of data (Y3) constrained the key quantity describing the amplitude of matter fluctuations, $S_8 \equiv \sigma_8\sqrt{\Omega_m/0.3}$, to $0.776\pm0.017$ \citep*{y3-cosmicshear1, y3-cosmicshear2}, where $\sigma_8$ is the amplitude of the linear matter fluctuations smoothed at the scale 8~$h^{-1}$Mpc, and $\Omega_m$ is the present day matter density.

Achieving a 2\% error on $S_8$ required strict control of multiple sources of systematic bias and uncertainty.
In order to fully benefit from the greater data volume available with the full six years of DES data (Y6), requirements on measurement and modeling systematic uncertainties must be even more exacting.
As the cosmic shear analysis relies on measurement of the ellipticities of individual galaxies, a critical component of this analysis is modeling the point-spread function (PSF), which describes the response of the full optical system (atmosphere, optics and sensors) to a point source.
It acts as a convolution kernel for each object in an image and is a function of a source's (sky and pixel) coordinates.
In addition, the PSF profile varies with wavelength, so the effective PSF for any particular source in the sky is a function of its spectral energy distribution (SED).
Finally, since the timescale for changes in the atmosphere is much shorter than typical exposure times, the PSF also varies temporally between exposures.
For more details on the role of the PSF in weak lensing analyses, see recent reviews, e.g., \citet{Mandelbaum18, Liaudat23}.

Generally, the PSF for a given image is estimated by fitting a model of the surface brightness profile to a high-purity sample of stars in the image.
The stars are point sources and thus directly measure the PSF at their positions for their particular SED.
To model the PSF for objects at other positions, the fitted surface brightness profiles are interpolated across the image and potentially adjusted in some way to account for the differences in SEDs, e.g., by interpolating with respect to broadband photometric colors.

As the observed image of each galaxy is a convolution of its true image with the PSF, we need to understand the PSF to high precision to infer the shearing of galaxies accurately.
It is critical to minimize modeling errors and to characterize those that remain as they directly contribute to shape measurement bias.
Errors in PSF size cause a multiplicative shear bias by making the object more or less round than the truth, and errors in PSF shape cause additive shear bias by imparting spurious ellipticity to shape measurements.
Further, the two-point spatial correlation of the PSF and PSF modeling residuals, often called ``$\rho$ statistics'' in the literature \citep{Rowe10, sv-shearcat}, can quantify and potentially correct for the additive bias such errors impart to the cosmic shear signal.
Thus, careful modeling and validation of the PSF is a key component of our and other recent experiments' WL shape measurement analyses \citep{Giblin21, Zhang23b, Li22}.

Although the PSF estimation process for DES Y3 (\citet{y3-piff}, \citetalias{y3-piff} hereafter) was sufficiently accurate to result in negligible biases \citep*{y3-cosmicshear1}, known outstanding issues from the Y3 analysis have motivated further development of our PSF estimation software, \piff\footnote{\url{https://github.com/rmjarvis/Piff}} (PSFs in the Full Field-of-View).
\citetalias{y3-piff} found the PSF size and ellipticity model residuals were color dependent, as expected from atmospheric phenomena like differential chromatic refraction (DCR) and chromatic aberrations in the optics, among other effects.
Further, galaxy contamination of the stellar population also slightly biased the modeling results.
These two effects were negligible in the $riz$ bands but were particularly strong in $g$ band images.
They were likely the chief cause of unacceptably high two-point correlations in the $g$ band PSF shape residuals, leading to the exclusion of $g$ band data from any Y3 WL analysis.

In this work, we address these shortcomings by improving the PSF modeling primarily in two ways:
(i) addition of color-dependent terms in the PSF interpolation, and (ii) improved star-galaxy separation using the \gaia\ Early Data Release 3 (EDR3) catalog \citep{gaia-edr3} and infrared (IR) $K$ band magnitude data.
Further, following \citet{Zhang22, Zhang23a, Zhang23b}, who found that residuals in the fourth-order moments of the PSF can contribute to PSF contamination of shear at a comparable level to second-order moments, we have added studies of fourth-order moments to our model diagnostics.

The DES Y6 PSF modeling uses \piff\ version 1.2.4. The fully specified configuration settings are provided in \app{app:config}.

The paper is organized as follows: \sect{sec:scheme} provides an overview of the PSF modeling algorithms used in the Y6 analysis, including new features added to \piff.
\sect{sec:data} outlines the DES dataset and data-specific analysis choices made for modeling.
\sect{sec:tests} presents a comprehensive set of diagnostics of the PSF models, including comparisons with the quality of the Y3 PSF modeling.
\sect{sec:future} discusses ongoing and future improvements to \piff, particularly geared toward the next generation of wide-field imaging surveys.
We conclude in \sect{sec:conclusion}.

%%%%%%%%%%%%%%%%%%%%%%%%%%%%%%%%%%%%%%%%%%%%%%%%%%%%%%%%%%%%%%%%%%%

\section{PSF Estimation Procedure}
\label{sec:scheme}

\edit{To accommodate diverse analysis needs, \piff\ has many methods available to specify the PSF modeling.}
%\piff\ is designed to be flexible, so there are many methods available for PSF modeling.
Here we give an outline of the \edit{specific} methods used for the DES Y6 analysis.
We note when a step is explained in greater detail in the following subsections.
\edit{Examples of how to specify each option are shown in the configuration file provided in \app{app:config}.}

For a given exposure in the survey, the PSF is modeled for each charge-coupled device (CCD) image separately.\footnote{As the name suggests, \piff\ has algorithms that fit and interpolate the PSF model across the full focal plane. However, we do not use these methods for this work because they would require significant additional testing and, as we will show, PSF modeling on an individual CCD basis is sufficient for the DES Y6 WL analysis.}
The general sequence of operations for a \piff\ run for each CCD is:
\begin{enumerate}
    \item Read in the image pixel data, a weight map for the pixels, and a catalog of detected objects in the image from which stars will be selected. The catalog must contain (at minimum) the information upon which the PSF depends -- in our case, the sky coordinates and color of each object. 
    \item Select the stars to be used for PSF modeling. This may be a trivial subset (or all) of the input catalog based on flags or other properties of each object, or \piff\ can perform a more sophisticated algorithm to select good PSF exemplars from among the input objects. There are also options to reject some objects according to various measurements that can be performed on each potential PSF star (\sect{sec:piff_star_select}).
    \item Reserve some fraction of the PSF stars to exclude from the fitting process. This step is optional, but it is generally recommended to use these reserve stars for diagnostic tests of the final PSF model.
    \item Fit a model to the pixel data at each star to yield a maximum-likelihood estimate of the PSF parameters at the location of the star. There are a variety of options for the form of the PSF model, which can be specified by the user (\sect{sec:psf_formalism} \& \sect{sec:pixelgrid}). \label{step:singlestar}
    \item Calculate the maximum-likelihood interpolation of the PSF parameters across the sample of stars using their positions and possibly other properties, such as color.  Again, the user has a choice of what kind of interpolation to use (\sect{sec:piff_color} \& \sect{sec:interp}). \label{step:interp}
    \item Identify and excise outlier stars if desired. Outliers are those that are deemed to be poor exemplars of the current best-fit PSF according to some metric (\sect{sec:outlier_reject}). \label{step:outlier}
    \item Iterate over steps~\ref{step:singlestar}-\ref{step:outlier} to refit and reject outliers until convergence is reached.
\end{enumerate}

\piff\ has several models and interpolation schemes available.
For the DES Y6 WL analysis, we use the \texttt{PixelGrid} parametric model and \texttt{BasisPolynomial} interpolation.
The \texttt{BasisPolynomial} algorithm actually performs the modeling and interpolation, steps~\ref{step:singlestar} and \ref{step:interp}, simultaneously.
We give a relevant overview of each below, but refer the reader to \citetalias{y3-piff} for further details on these and other modeling functions implemented as of \piff\ version 1.0.

Since \citetalias{y3-piff}, we have added new functionality to \piff: (i) model interpolation based on any arbitrary stellar property provided as a column in the input catalog and (ii) the ability to perform star selection within \piff\ rather than requiring star-galaxy separation to be completed in advance.
We give details on the implementation of these features in \sect{sec:piff_color} and \sect{sec:piff_star_select}.

%%%%%%%%%%%%%%%%%%%%%%%%%%%%%%%%%

\subsection{Color-dependent PSF modeling}
\label{sec:piff_color}
The main improvement from Y3 to Y6 PSF modeling is the added capability to interpolate the PSF model over arbitrary stellar properties provided in the input catalog.
This is done by specifying the appropriate column name(s) using \texttt{property\_cols} in the \texttt{input} configuration section and \texttt{keys} in the \texttt{psf}, \texttt{interp} section.

As PSFs are known to be wavelength dependent, we use this functionality to interpolate the PSF in color, the first-order approximation of wavelength dependence.
There are many factors that cause the PSF to be wavelength dependent, including chromatic seeing of the atmosphere \citep{Hardy98, Xin18, Carlsten18}, differential chromatic refraction (DCR) \citep{Plazas12, Meyers15b, Lee23}, chromatic aberrations in the Dark Energy Camera (DECam) optical elements \citep{DECam} and wavelength-dependent sensor effects \citep{DECam-ccds, Meyers15a}.
We model color dependence by interpolating the surface brightness profile model using a first-order polynomial in the stars' color.
As such, our model is empirical, not physically based, describing the net effect of all the wavelength-dependent phenomena noted above.
The application of this function in the Y6 analysis is described in further detail in \sect{sec:data:color}.

%%%%%%%%%%%%%%%%%%%%%%%%%%%%%%%%%

\subsection{Star selection}
\label{sec:piff_star_select}
We implemented the ability to perform star-galaxy separation on the input catalog to select stars to use for PSF fitting.  
Previously, this step had to be performed prior to using \piff, and \piff\ would use either all input objects for PSF fitting, or
possibly remove some according to a flag column in the input catalog.
The new \texttt{select} section of the configuration file provides several ways to select a subset of the input objects as probable stars.
Currently, \piff\ has four selection algorithms implemented.

\begin{enumerate}
\item \texttt{Flag}:
The default selection is essentially equivalent to the selection in previous versions of \piff.
The PSF stars to be used are specified using a flag column in the input file.  This column can be either a positive flag indicating
which objects should be considered stars, or a negative flag marking objects to exclude.
If there is no flag column specified, all input objects are taken to be stars.

\item \texttt{Properties}:
This selection allows the user to specify any calculation that can be done on one or more input columns for each object \edit{(e.g., a cut in color-color space)}.
The user specifies which columns should be read in for each object using \texttt{property\_cols} in the \texttt{input} section.
The user can specify any calculation that Python can evaluate
(using \texttt{eval}) using these property values in a \texttt{where} field, which should evaluate to true for any object that should
be used as a PSF star, and false otherwise.

\item \texttt{SizeMag}:
This selection algorithm uses the sizes and fluxes of the input objects to try to identify stars falling on a stellar locus of
(approximately) uniform size over a range of magnitudes. It is very closely based on the star selection algorithm used for DES Y3.
It requires an initial selection using one of the other selection types.
Then it iteratively improves upon the selection by finding additional stars with similar size, going as faint as possible before the
stellar locus merges with the galaxies, and removing stars that do not have a similar enough size.
The selection can be tuned with an \code{impurity} parameter, which controls how cleanly the stellar locus separates from the galaxies.
It should not be taken to be a quantitative estimate of the actual impurity in the sample, but larger values of impurity will correspond 
to larger samples with more non-stellar interlopers.
Smaller impurity will likely result in a purer selection of stars, but also one with fewer total stars.

\item \texttt{SmallBright}: 
This crude selection algorithm measures the flux and size of each object in the input catalog
and selects a given fraction of the brightest and smallest objects.
This algorithm is not intended to be the primary selection of PSF stars, but it is usually good enough to provide a decent initial selection
for the \texttt{SizeMag} algorithm.  It is the default method for that initial selection, and it is based on the initial selection algorithm 
that we used for DES Y3.

\end{enumerate}

There are also a number of rejection options that one can perform on the initial selection of stars to try to exclude ones that may not be
optimal for PSF modeling, e.g., because they are too noisy or might be blended with another object.  Some notable fields are
\begin{itemize}
\item \texttt{min\_snr} specifies the minimum allowed signal-to-noise ratio.
\item \texttt{max\_mask\_pixels} specifies the maximum number of pixels in the postage stamp that are allowed to be masked.
\item \texttt{max\_edge\_frac} specifies the maximum fraction of the flux allowed in pixels around the edge of the stamp.
The ``edge'' is defined as the set of pixels outside of the central circle of pixels whose radius is controlled by the ancillary setting \texttt{stamp\_center\_size}.
\item \texttt{hsm\_size\_reject} rejects stars that have measured sizes (using the HSM adaptive moments algorithm from \citealp{hsm}) that are very different from the rest of the stars.
\item \texttt{reject\_where} allows the user to specify any Python \texttt{eval}-able string to reject stars according to input properties.
This option is essentially the converse of the \texttt{Properties} selection type, and it can use any
property names specified with \texttt{property\_cols} in the \texttt{input} section.
\end{itemize}

For the Y6 analysis, we used the \texttt{SizeMag} algorithm with \texttt{impurity=0.05}.  The initial selection was the set of Gaia stars 
with appropriate colors (see \sect{sec:data:star_select} below for details) using the \texttt{Properties} method.  After the \texttt{SizeMag} selection, 
we reapplied the color selection using \texttt{reject\_where}.  We also used \texttt{hsm\_size\_reject} to remove blends and other problematic images and applied
\texttt{min\_snr=20}.

%%%%%%%%%%%%%%%%%%%%%%%%%%%%%%%%%

\subsection{General PSF estimation formalism}
\label{sec:psf_formalism}

The PSF describes the combined atmosphere, telescope, and sensor response to an ideal point source (i.e., a delta function), mapping this to a 2D image measured with respect to any point of the detector plane $(x_0, y_0)$ as $I(x-x_0, y-y_0)$ (in units of photons/pixel).
This mapping may be modeled in detector coordinates $(x,y)$ directly or, given a world coordinate system (WCS) solution, in sky coordinates $(u,v)$ via the transformation
\begin{equation}
    \label{eq:wcs_trans}
    I(x,y) = I(u,v)
    \left| \begin{matrix}
    %\partial u/ \partial x & \partial v/\partial x \\
    %\partial v / \partial y & \partial v/\partial y
    \frac{du}{dx} & \frac{dv}{dx} \\
    \frac{du}{dy} & \frac{dv}{dy}
    %du/dx & dv/dx \\
    %du/dy & dv/dy
    \end{matrix} \right|,
\end{equation}
where $I(u, v)$ is the surface brightness profile (in units of photons/arcsec$^2$), and the last factor is the determinant of the Jacobian of the coordinate transformation, which we identify as the pixel area, $A_{\rm pix}$ (in units of arcsec$^2$/pixel).

The data for star $i$ measured over a set of pixels, each with area $A_{\rm pix}$ and indexed by $\alpha$, consist of the (sky-subtracted) counts in each pixel $d_{i\alpha}$ around the star.
We normalize the PSF model to have unit flux
\begin{equation}
    \int du\,dv\,I(u,v) = \int dx\,dy\,I(x,y) = 1,
    \label{eq:psfnorm}
\end{equation}
such that the model estimator for the counts $\hat{d}_{i\alpha}$ of a star with flux $f_i$ at the sky coordinates $(u_i, v_i)$, is
\begin{equation}
    \hat{d}_{i\alpha} = f_i A_\mathrm{pix} I(u_{i\alpha}{-}u_i,v_{i\alpha}{-}v_i).
    \label{eq:dmodel}
\end{equation}
In \piff, $I(u,v)$ is a parametric model with a vector of model parameters $\bm{p}$, which are fitted using maximum likelihood estimation.
The likelihood, $\mathcal{L}(\bm{d}_i)$, of obtaining the data given the model is given by
\begin{equation}
    -2 \log \mathcal{L}(\bm{d}_i) =
    \sum_{\alpha \in i}
    \left[ \frac{\left(d_{i\alpha} {-}
    \hat{d}_{i\alpha}\right)^2}{
    \sigma_{i\alpha}^2 {+}
    \hat{d}_{i\alpha}} + \log\left( \sigma_{i\alpha}^2 {+}
    \hat{d}_{i\alpha}\right) \right],
\label{likelihood}
\end{equation}
where we have assumed that the model and data are in units of photoelectrons, such that the total variance of a pixel is the sum of the read/background variance $\sigma^2_{i\alpha}$ and the Poisson variance from the expected counts $\hat{d}_{i\alpha}$.

The full PSF model involves interpolating the model parameters $\bm{p}$ across the focal plane.
The parameters of the interpolation are constrained at the locations of the stars given the stars' properties (such as color) using \eqn{likelihood}.  Then the model parameters $\bm{p}$ may be evaluated at any arbitrary location and/or property value in the domain, e.g., at a galaxy's location and color.

\piff\ solves for the PSF model iteratively.  In each iteration,
the model and interpolation coefficients are updated keeping the
list of stars, their centroids, and their fluxes all fixed.  
Then the centroids and fluxes are updated given the current best-fit
model at the location of each star.  Finally, some stars may be
removed from consideration if they are deemed to be outliers (see \sect{sec:outlier_reject}).  The process continues until no outliers are found and the centroid and flux estimates have converged.
\edit{For the Y6 modeling, convergence is usually reached in $\sim$3-6 iterations.}

%%%%%%%%%%%%%%%%%%%%%%%%%%%%%%%%%

\subsection{PSF model: \texttt{PixelGrid}}
\label{sec:pixelgrid}

We use \texttt{PixelGrid} as the parametric model for individual star profiles.
It is a flexible model, using a 2D grid of points over the postage stamp image for each star, aligned with the $(u,v)$ coordinate frame.
We call the pixels of this model grid ``model pixels'' to differentiate them from the CCD pixels, or ``data pixels.''
A 1D smooth interpolation kernel, $L(x)$, is used to interpolate between model pixel centers (not to be confused with the function interpolating between stars described in the next section).
The functional form of the PSF model is
\begin{equation}
    I(u,v)=\sum_{k=1}^{N_{\rm{pix}}}p_kL(u-u_k)L(v-v_k),
\label{eq:pixelgrid}
\end{equation}
where $N_{\rm{pix}}$ is the total number of model pixels in the 2D grid, $p_k$ is the free parameter coefficient for the $k$th model pixel, and $(u_k,v_k)$ is the center of the $k$th model pixel in sky coordinates.
The grid interpolation kernel used in the Y6 analysis is the Lanczos kernel, $L(x) = L_n(x)$, with the form
\begin{linenomath*}
\begin{align}
L_n(x) &\equiv
\begin{cases}
1 & \mathrm{if} ~|x| = 0\\
\frac{n}{\pi^2 x^2} \sin \left(\pi x\right) \sin \left(\frac{\pi x}{n}\right) & \mathrm{if} ~0<|x|<n \\
0 & \mathrm{if} ~|x| \ge n
\end{cases}
\label{eq:lanczos}
\end{align}
\end{linenomath*}
where $n$ is a free integer parameter. We use the default\footnote{The default value was $n=3$ for the version of \piff\ used for the Y6 PSF modeling; however, the default has since been changed to $n=7$ in more recent versions of \piff.} value in \piff\ of $n=3$.

The model pixels do not need to match the CCD pixels in size or alignment.
Indeed, we find empirically using a grid scaled slightly larger than the CCD pixels (0.30\arcsec\ vs. 0.263\arcsec) results in more stable performance during fitting.
We use a 17x17 model grid for a 32x32 pixel image postage stamp. The extra area of the postage stamp allows room for recentering of the model grid during fitting.

As mentioned earlier, the interpolation method we use, \texttt{BasisPolynomial}, delays the process of solving for $\bm{p}_i$ and instead completes model fitting and interpolation simultaneously.
However, as this method effectively concatenates the optimization of the grid coefficients for all stars together, we review the fitting process for a single star here.
To constrain the pixel grid coefficients $\bm{p}_i$ for star $i$, we minimize

\begin{linenomath*}
\begin{align}
\chi^2 &= \sum_{\alpha}
    \frac{\left(d_{i\alpha} - \hat{d}_{i\alpha}(u_\alpha{-}u_c,v_\alpha{-}v_c) \right)^2}
{\sigma_{i\alpha}^2 + \hat{d}_{i\alpha}}
\label{eq:chisq_pixelgrid}
\end{align}
\end{linenomath*}
where $\hat{d}_{i\alpha}$ is the model, defined by \eqnb{eq:dmodel}{eq:pixelgrid}, evaluated with respect to the star's centroid $(u_c, v_c)$.
This expression is minimized by perturbing the coefficient parameters by $\delta \bm{\hat{p}}_i$ and solving the resulting design equation
\begin{equation}
    \label{eq:design}
    \bm{A}_i \delta \bm{\hat p}_i  = \bm{b}_i,
\end{equation}
where
\begin{linenomath*}
\begin{align}
\label{eq:design_ab}
A_{i\alpha k} & \equiv \frac{\partial \hat{d}_{i\alpha}}{\partial p_{ik}}
                  \left(\sigma^2_{i\alpha}{+}\hat{d}_{i\alpha}\right)^{-1/2}\\
b_{i\alpha} & \equiv \left(d_{i\alpha}{-}\hat{d}_{i\alpha}\right)\left(\sigma^2_{i\alpha}{+}\hat{d}_{i\alpha}\right)^{-1/2}.
\end{align}
\end{linenomath*}

$\bm{A}_i$ and $\bm{b}_i$ are computed using the values of $\bm{\hat{p}}_i$ and $\bm{\hat{d}}_{i}$ derived in the previous iteration.
The differential change $\delta \bm{\hat{p}}_i$ is then added to the solution to obtain the parameters for the next iteration.

%%%%%%%%%%%%%%%%%%%%%%%%%%%%%%%%%

\subsection{Interpolation: \texttt{BasisPolynomial}}
\label{sec:interp}

The procedure described above solves for the surface brightness profile of a single star.
However, the PSF varies across the FOV and as a function of stellar properties like color.
To model the PSF across the full CCD and across the full domain of a given stellar property, we must interpolate between the PSF star positions $(u_i,v_i)$ and property values $c_i$.
In \piff, the \texttt{Polynomial} interpolation method interpolates each coefficient in the pixel grid vector $\bm{p}$, $p_k$, according to a polynomial of specified order in $u$, $v$ and $c$.
This is expressed as
\begin{equation}
    p_{ik} = \sum_{m}Q_{km}K_{im}(u_i, v_i, c_i),
\label{eq:basiscoeffs}
\end{equation}
where  $\bm{K}_i$ is the polynomial basis vector for the $i$th star,
$m$ indexes the elements in $\bm{K}_i$, and $Q_{km}$ is the coefficient for the $m$th term in the polynomial interpolation function for the $k$th model pixel across all the stars.
Here the $p_{ik}$ grid coefficients are already fitted using the process described in \sect{sec:pixelgrid}, and a separate maximum likelihood fit for $\bm{Q}$ is completed.

In contrast, the interpolation method used for the Y6 PSF modeling, \texttt{BasisPolynomial}, delays solving the $\bm{p}$ coefficients and instead directly models $\bm{p}$ in terms of the interpolation coefficients $\bm{Q}$ via \eqn{eq:basiscoeffs}.
With this substitution, the iterative design equation for a single star, \eqn{eq:design}, is rewritten as
\begin{equation}
\bm{A}_i \delta \bm{Q} \bm{K}_i = \bm{b}_i,
\label{eq:starbasisdesign}
\end{equation}
where $\delta \bm{Q}$ is a differential change in the interpolation coefficients rather than the grid coefficients in \eqn{eq:design}.

Intuitively, \texttt{BasisPolynomial} concatenates the design equations of all the stars together and solves them simultaneously.
Doing so constrains the surface brightness profile model such that all the stars contribute information to the model globally, not just at their specific locations (and colors, for example).
While \texttt{BasisPolynomial} is more computationally intensive than the simpler \texttt{Polynomial} interpolation method, its main benefit is that the fitting process for each star also uses information from the other stars in its vicinity.
Thus, this method is more robust to errors stemming from masked or missing data that afflict modeling of stars individually.

For the DES Y6 PSF modeling, we interpolate using a polynomial that is second-order in $u$ and $v$ and linear in color.
\sect{sec:data:color} provides further details on how the color interpolation is implemented.

%%%%%%%%%%%%%%%%%%%%%%%%%%%%%%%%%

\subsection{Outlier rejection: \texttt{Chisq}}
\label{sec:outlier_reject}

At the end of each iteration, there is an option to remove stars that are determined
to be outliers and thus are probably not good exemplars of the PSF.
We use the same method as in the DES Y3 analysis, \texttt{Chisq}, which is described in greater detail in \citetalias{y3-piff}.

In short, this method evaluates the $\chi^2$ for each star (\eqn{eq:chisq_pixelgrid}) and removes the star from future iterations if its $\chi^2$ value exceeds a specified threshold given its number of degrees of freedom.
For DES Y6, we use the threshold specification of \code{nsigma=4}, corresponding to the value for which the probability that the star's measured $\chi^2$ would exceed it purely from statistical noise is $p=6.3\times10^{-5}$.

Since a small number of outliers can potentially skew the solution such that almost all of the stars have a bad $\chi^2$ value, we also set a maximum number of stars that may be rejected in each iteration.
For DES Y6, we limit the rejection to at most 1\% of the stars used for fiting (rounded up), typically 1 or 2 stars per iteration.

%%%%%%%%%%%%%%%%%%%%%%%%%%%%%%%%%%%%%%%%%%%%%%%%%%%%%%%%%%%%%%%%%%%

\section{Application to Y6 Data}
\label{sec:data}

The full six years of DES data cover nearly 5000 square degrees of the sky, primarily in the southern hemisphere.
The sky coverage was not significantly expanded between Y3 and Y6, but Y6 doubles the number of exposures in this area: 96,263 exposures using the DECam $grizY$ broadband filters.
The survey reaches an $i$ band 10$\sigma$ depth for point sources, measured in a 2\arcsec\ aperture, of 23.8 \citep{y6-gold}.

The images are processed using a standardized instrument signature removal (ISR) pipeline described in \citet{imageproc} and updated for the full six years of data in the Data Release 2 paper \citep{des-dr2}.
We refer the reader to these sources for details but describe a few elements of the processing that are particularly relevant to analysis choices and diagnostic tests described later in this work.

Part of the standard ISR pipeline is a correction for the ``brighter-fatter'' (BF) effect \citep{Antilogus14, Guyonnet15}, where drifting photoelectrons are deflected into neighboring pixels by the charge already accumulated at the pixel gate over a single exposure, thereby increasing the measured size of the source object.
This effect is stronger for high surface brightness objects like bright stars, and thus can significantly bias the measurement of the PSF if left unaccounted for.
Previous analyses of DECam images have shown that the brightest three magnitudes of unsaturated stars in a given image are strongly affected by the BF effect \citep{Melchior15, sv-shearcat, y1-shearcat}.
For the Y6 data, we apply the same correction algorithm as the Y3 analysis \citep{Antilogus14, Gruen15, imageproc}, which corrects about 90\% of the effect.
As such, we use the same procedure as in the Y3 PSF modeling to further mitigate the effect; we remove stars that are up to 1.2 mag fainter than the brightest unsaturated star in the image from the PSF stellar sample (as long as the brightest star has signal-to-noise (S/N) $>1000$).

We use the same world coordinate system (WCS) estimation software, \pixmappy\footnote{\url{ https://github.com/gbernstein/pixmappy}}, as was used in the Y3 analysis \citep{pixmappy}.
The solutions were updated for all science-quality images in the full six years of the wide-field survey data.

%%%%%%%%%%%%%%%%%%%%%%%%%%%%%%%%%

\subsection{Supplementing the input catalogs with additional data}
\label{sec:data:ext_data}

\begin{figure}
\includegraphics[width=\columnwidth]{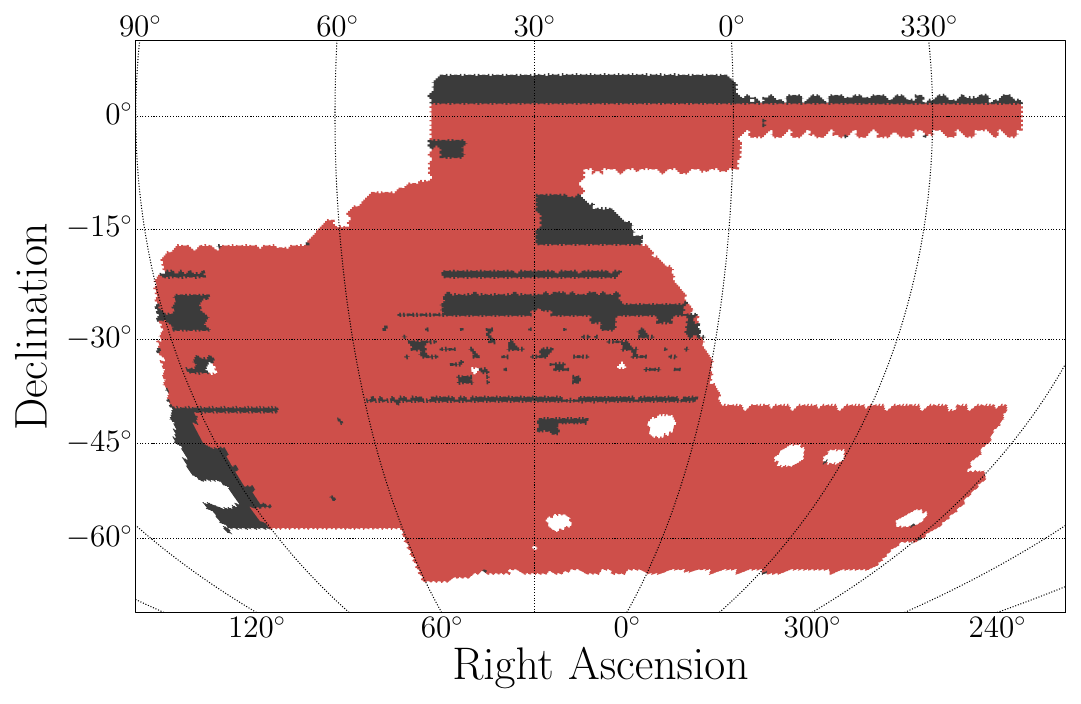}
\caption{
Footprint overlap of the DES Y6 survey (all shaded regions) and the VHS and \edit{VIKING} IR surveys (red). By area, there is an 87\% overlap of IR data with the DES footprint.
By match fraction, 74\% of the reserve stars have a match to a VHS or \edit{VIKING} source within 0.5\arcsec.
}
\label{fig:desvhs}
\end{figure}

The main difference in the PSF modeling between Y3 and Y6 is that now, to aid in stellar selection and enable color-based model interpolation, we use a significant amount of external or coadd-based data to supplement the information available in the single-epoch (SE) images.
This supplementary data consists of

\begin{itemize}
    \item high S/N DES photometry from the Y6 Gold catalog based on coadded images \citep{y6-gold},
    \item $Ks$ band magnitude data from the Visible and Infrared Survey Telescope for Astronomy (VISTA) Hemisphere Survey \citep[VHS DR5,][]{vhs} and VISTA Kilo-degree INfrared Galaxy \citep[VIKING DR1,][]{viking} survey catalogs,
    \item detected object coordinates from the \textit{Gaia} EDR3 catalog \citep{gaia-edr3}.
\end{itemize}

Using photometry from the Y6 Gold catalog allows us to leverage the statistical power of the full DES dataset when applying color selections and using colors for model interpolation.

As was explored in \citetalias{y3-piff}, IR data can be very informative in distinguishing quasars and active galactic nuclei (AGN) from true stellar sources.
In that work, IR information was only available for about half of the DES Y3 area and was only used as a diagnostic tool for estimating stellar purity.
Now, with 87\% sky overlap of the VHS and VIKING surveys with the DES footprint, we are able to use the IR photometry  to aid star-galaxy separation from the outset.
\fig{fig:desvhs} shows the overlap of the DES footprint with available VHS and VIKING catalog data.

Likewise, as a dedicated space-based stellar surveyor, \gaia\ \citep{gaia-instrument} can provide a very pure stellar sample.
\citetalias{y3-piff} matched the PSF stars to the \gaia\ DR2 catalog as a test of stellar purity.
For the Y6 PSF modeling we use match information with the \gaia\ EDR3 catalog to form the initial sample for the stellar selection algorithm \texttt{SizeMag}, described in \sect{sec:piff_star_select}.
The fraction of PSF stars that have a match within 0.5\arcsec\ to the \gaia\ EDR3 catalog or IR catalogs is listed in \tab{tab:cat_stats} for each band and the full reserve star catalog.
\begin{table}
    \centering
    \begin{tabular}{rrrrrr}
         Band & $N_*$ & $N_{\rm exp}$ & $N_*/\mathrm{CCD}$ & $f_{\rm IR}$ & $f_{\it Gaia}$ \\
         \hline
         $g$    & 22,503,685  & 18,547 & 139 & 0.73 & 0.92 \\
         $r$    & 28,158,845  & 19,158 & 176 & 0.72 & 0.81 \\
         $i$    & 30,368,812  & 19,381 & 191 & 0.72 & 0.73 \\
         $z$    & 28,758,584  & 20,956 & 176 & 0.79 & 0.79 \\
         $griz$ & 109,789,926 & 78,042 & 164 & 0.74 & 0.80 \\
    \end{tabular}
    \caption{Statistics for the catalog of reserve PSF stars (20\% of full sample) after masking and removal of problematic images. The number of stars $N_*$, number of exposures $N_{\rm exp}$, median number of (PSF and reserve) stars per sensor $N_*/\mathrm{CCD}$, and match fractions to the IR catalogs and Gaia catalog, $f_{\rm IR}$ and $f_{\it Gaia}$, are listed for each band and $griz$ combined. Note that stars are counted each time they appear in an exposure; thus, $N_{\rm star}$ should be interpreted as the total number of star \textit{instances} in the survey.
    }
    \label{tab:cat_stats}
\end{table}

A suggestion given by \citetalias{y3-piff} for potential improvement was to fit the PSF models using only the stars matched to \gaia\ sources (see Section 8.3 in \citetalias{y3-piff}).
However, initial test runs of \piff\ limiting the training sample to only \gaia\ matches resulted in PSF models with larger modeling errors than those measured in the Y3 analysis.
We found that 4-5\% of the \gaia\ EDR3 catalog sources were not good PSF examplars.
They may actually be binary stars or galaxies with bright bulge components, or they were blended with another nearby object in DES images.
This high level contamination (compared to $\sim1\%$ in Y3) was likely the leading cause of the large PSF model residuals.
Further, the \gaia\ survey is not as deep as DES, and the \gaia\ $G$ bandpass is significantly broader than that of any DECam filter.
Thus using only \gaia\ sources imparted an artificial magnitude and color selection.
For these reasons, we choose to only supplement our stellar selection using \gaia\ information rather than replace it.

To form an input catalog (which will include stars and galaxies) for each CCD image, we begin with the output \texttt{SExtractor} \citep{BertinSExtractor1996} catalog for the image.
Any objects that are flagged as saturated or blended or have flagged pixels in the object's footprint are removed.
We then find matches within 0.5\arcsec\ between objects detected in the SE image and the three supplemental catalogs.
From the DES Y6 Gold catalog, we record the object extendedness classification based on \texttt{EXT\_MASH} and the $griz$ magnitudes\footnote{\texttt{PSF\_APER8\_MAG} in the Y6 Gold catalog, the PSF magnitudes calibrated to 5.8437\arcsec-diameter aperture magnitudes. See \cite{y6-gold} for details on this and the \texttt{EXT\_MASH} morphological classifier.}.
From the VHS/VIKING catalogs, we record the $Ks$ band magnitude\footnote{\texttt{KSAPERMAG3} in the VHS and VIKING catalogs, the 2\arcsec-diameter aperture-corrected $Ks$ magnitude.}.
For the optical and IR photometry data, the corresponding information is added to the input catalog for all matches; those objects with no matching entry in a particular catalog or with missing magnitude information are assigned a sentinel value.
From the \gaia\ catalog, matches within 0.5\arcsec\ are recorded as a boolean flag in the catalog.

\begin{figure}
\includegraphics[width=\columnwidth]{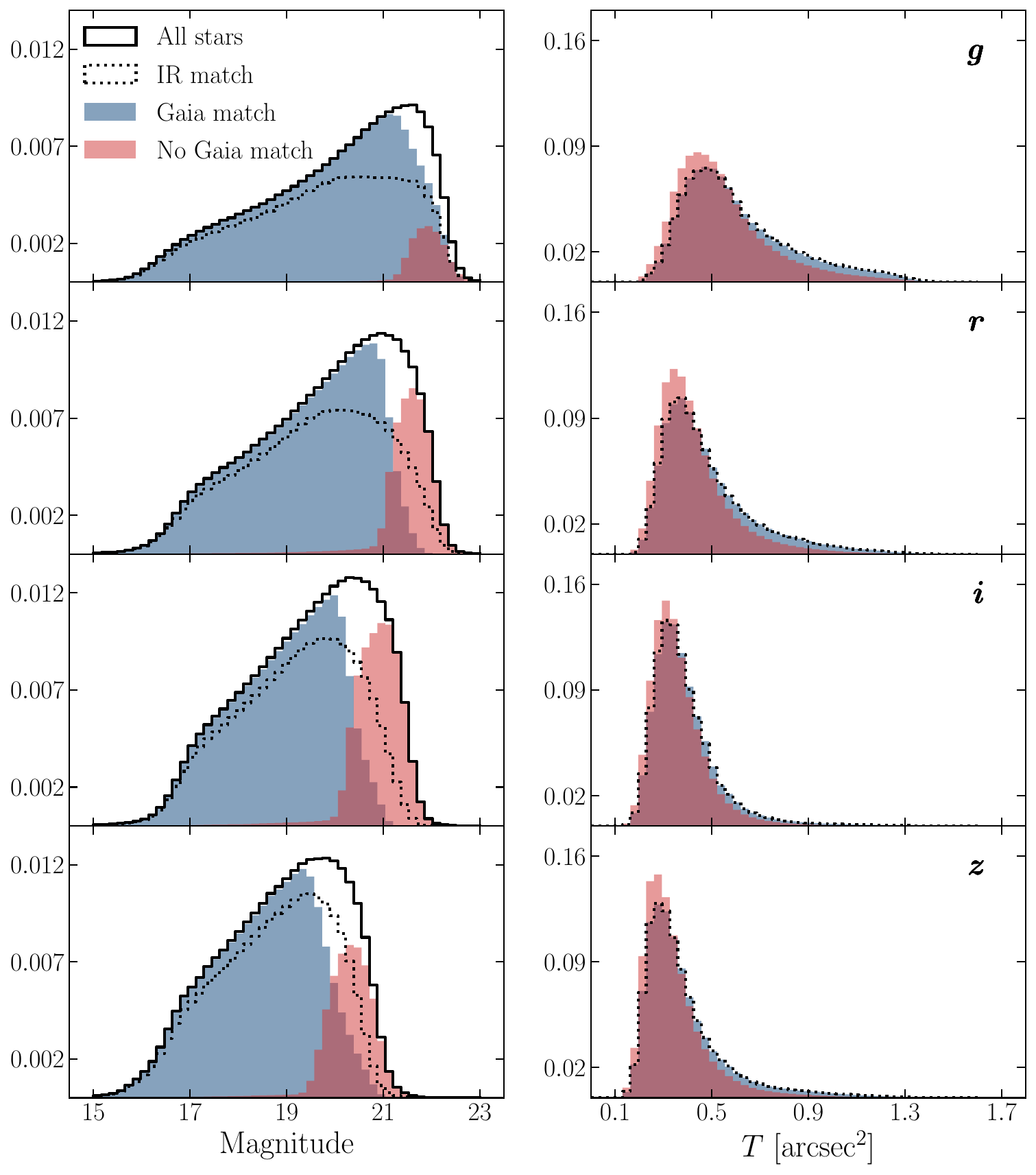}
\caption{
\textbf{Left:} Histograms of magnitudes in each band $griz$ for PSF stars with a match to \gaia\ within 0.5\arcsec\ (blue), no match to \gaia\ (red), a match to the IR catalogs within 0.5\arcsec\ (black dotted line) and all stars (black solid line).
A star can have both a \gaia\ match and IR match.\\
\textbf{Right:} Histograms of the measured size $T=2\sigma^2$ for PSF stars with (blue) and without (red) a match to \gaia.
}
\label{fig:maghist}
\end{figure}

As shown in \fig{fig:maghist}, the PSF star sample has a much larger match fraction to the \gaia\ catalog than in the Y3 analysis, which used the \gaia\ DR2 catalog (c.f. Figure 4 in \citetalias{y3-piff}).
Further, \fig{fig:maghist} shows the large match fraction to the IR catalogs spanning the majority of the magnitude range of the DES stellar sample.
The right-side panels show the histograms in the HSM-measured size $T$ (defined as $M_{11}$ in \eqn{eq:m11}) for objects with (blue) and without (red) a match in the \gaia\ catalog.
Both populations have similar size distributions, and the objects without a match to \gaia\ do not display the tail of large $T$ values that was evident in Y3, particularly for $g$ band.
Indeed, these objects' size distribution actually peaks at a lower value.
The unmatched objects are systematically fainter than the rest of the sample; they likely have a smaller measured size because they were only detectable during good seeing conditions.
Limiting the \gaia-matched objects to $g>21$ results in a nearly identical size distribution (not shown) to the unmatched objects.

%%%%%%%%%%%%%%%%%%%%%%%%%%%%%%%%%

\subsection{Data for color interpolation}
\label{sec:data:color}

\begin{figure}
\includegraphics[width=\columnwidth]{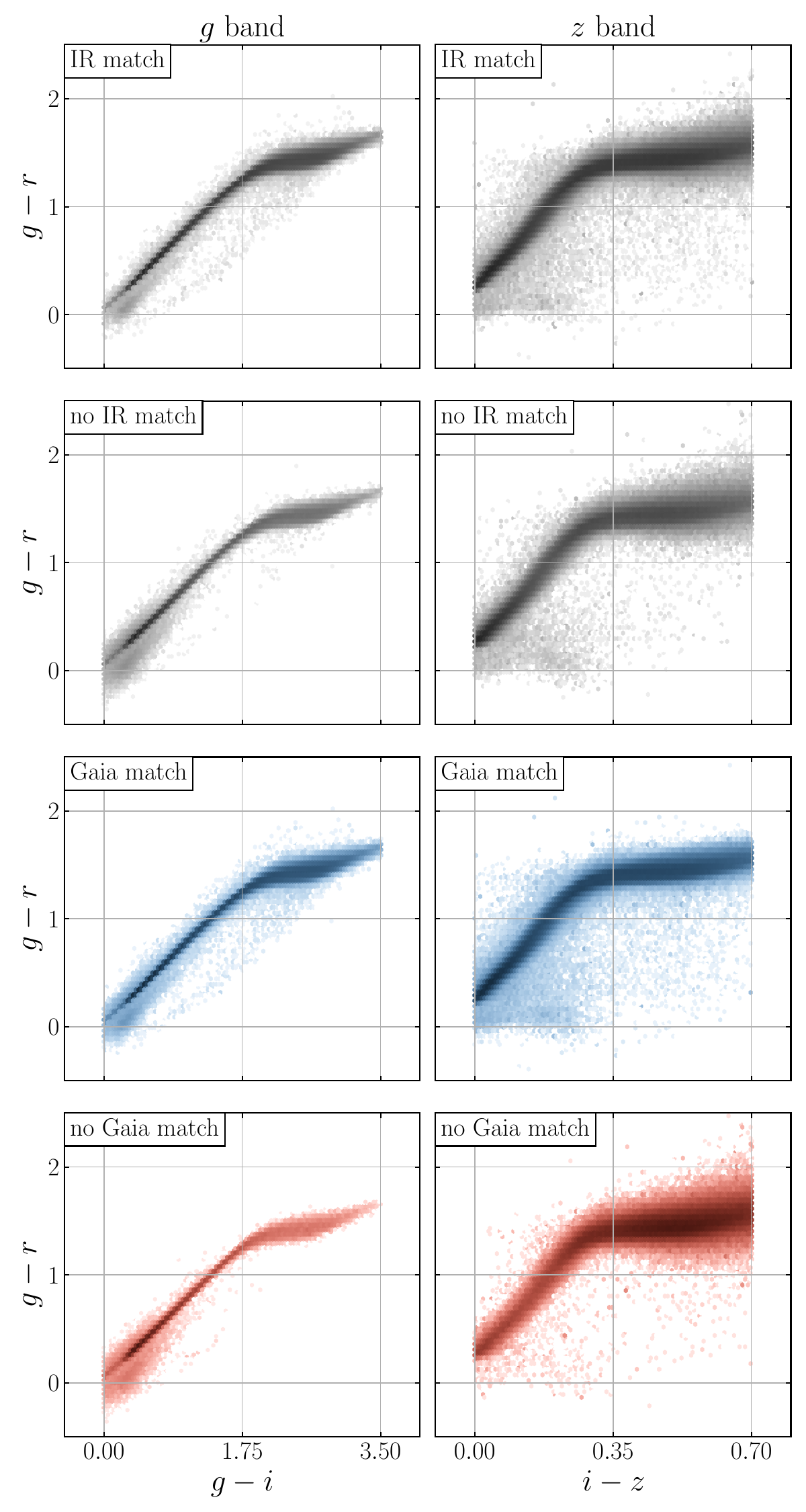}
\caption{
    Color-color plots for the same splits of the PSF stars as shown in Figure \ref{fig:maghist}.
    The left column shows for stars detected in $g$ band $g-r$ versus $g-i$, the color used in PSF interpolation for the $gri$ bands.
    The right column shows for stars detected in $z$ band $g-r$ versus $i-z$, the color used in PSF interpolation for the $z$ band.
    The sources with no IR or \gaia\ match do not exhibit any extra population not seen among the sources with matches.
}
\label{fig:gmr_vs_interp_color}
\end{figure}

As noted in \sect{sec:interp}, we interpolate our PSF models to first order in a given color.
We choose linear interpolation because the stellar sample for each CCD is generally not able to constrain the larger number of coefficients in a higher-order polynomial.
For the color interpolation, we use $g-i$ for $gri$ band exposures and $i-z$ for $z$ band exposures, using the DES Y6 Gold photometry.
For objects that do not have a match in the Y6 Gold catalog (usually because they are outside the edge of the nominal survey footprint) or those missing magnitude information in a necessary band (e.g., $g$ and $i$ to form a $g-i$ color), we assign these objects the median color value measured on the PSF stars that do have a Gold catalog match.
These median stellar color values are $g-i=1.3$ and $i-z=0.2$. 

The $g-i$ color is chosen for the $gri$ bands because it has a fairly linear mapping to stellar SED types \citep{Rykoff23}, as shown in the color-color plots on the left panels of \fig{fig:gmr_vs_interp_color}.
Other colors like $g-r$ can exhibit degeneracies in describing stellar types.

For $z$ band, we tested using $g-i$ as well but found requiring a detection in the $g$ band imposed an artificial color selection against very red stars in $z$ band.
In this test, following our scheme for stars with missing band magnitudes in the Y6 Gold catalog, these stars were assigned the default median stellar $g-i$ value of 1.3, much bluer than their true color, causing a bias in the PSF solutions.
To avoid this, we use $i-z$ since this color also has a relatively linear mapping to SED types, as shown in the right panels of \fig{fig:gmr_vs_interp_color}.

%%%%%%%%%%%%%%%%%%%%%%%%%%%%%%%%%

\subsection{Selection of PSF stars}
\label{sec:data:star_select}

PSF estimation requires a very pure sample of training stars.
Including even only a few percent of galaxies in the stellar sample will lead to biases in the PSF and ultimately shear measurements.
Indeed, the star-galaxy separation process for the Y3 PSF estimation led to increased PSF uncertainties.
While this effect was found to be negligible for the $riz$ bands, a separate population of larger objects was apparent in diagnostic plots for $g$ band (see Figure 4 of \citetalias{y3-piff}).
In order to use $g$ band for the Y6 analysis, we need to further mitigate galaxy contamination in our stellar sample.
To do so, we implement a multi-step procedure for selecting stars to use in the PSF estimation.

Before supplying the input catalog to \piff, we apply three selection criteria on the catalog that are difficult to apply through the \piff\ \code{select} module.

The first is a cut on the brightest objects.
As mentioned earlier, the standard image processing applies a brighter-fatter correction.
This correction mitigates about 90\% of the effect.
To ensure the residual effect does not bias our PSF modeling, we remove objects that are 1.2 mag fainter than the brightest unsaturated star in the image (as long as that brightest star has S/N $>1000$).
The median magnitude for this cut is $\sim17$ across all bands.

The second and third selections are performed only on those objects that have a match with a \textit{Gaia} source.
We wish to provide \piff\ a very pure sample of isolated stars to initialize the \texttt{SizeMag} algorithm; we do this by limiting the initial sample to only \textit{Gaia} matches.
To better ensure the sample is made of isolated stars, we remove \gaia\ sources that are within 2\arcsec\ of each other. To remove likely galaxies or other non-stellar objects, we also flag those objects with sizes more than 3$\sigma$ away from the 4$\sigma$-clipped average size of the \textit{Gaia} matches.
As these latter objects are only flagged, they may be readded to the star sample by the \texttt{SizeMag} selection described below.

At this point, the input catalog is prepared to be read in by \piff.
All further selections happen through the \piff\ \code{select} module:

\begin{enumerate}
    \item Apply two extra cuts to the \textit{Gaia} initial selection: \label{step:gaia_cuts}
    \begin{enumerate}
        \item remove any object with an \extmash\ classification $>0$, indicating it is not a high-confidence star through morphological classification, and
        \item remove any object with $g-i<0$ or $g-i>3.5$ for $gri$ band exposures or $i-z<0$ or $i-z>0.7$ for $z$ band exposures.
    \end{enumerate}
    \item Run the \sizemag\ algorithm, which adds objects from the full input catalog (and potentially removes objects from the initial sample) until the stellar sample's locus starts merging with the galaxy locus.
    We use an impurity parameter of 0.05.
    The algorithm is described in greater detail in \sect{sec:piff_star_select}.
    \item Apply the previous color and morphology cuts from step \ref{step:gaia_cuts} to the objects added to the stellar sample by \sizemag.
    \item Also reject objects that are 3$\sigma$ outliers based on the HSM adaptive moments measurement for the object.
    \item For those objects with a match to the VHS/VIKING catalogs and valid $r$ and $z$ magnitude measurements, apply a second, IR-based color cut where objects that satisfy the following inequality are removed:
        \begin{equation}
            \label{eq:ir_cut}
            z-K > 0.5\times(r-z).
        \end{equation} \label{step:ir_cut}
    \item The S/N is measured for each object based on its postage stamp image.
    \begin{enumerate}
        \item Remove objects with S/N $<20$.
        \item For objects with S/N $>100$, alter their pixel weight map such that their nominal S/N is pinned to 100. This prevents these sources from dominating the model fitting.
    \end{enumerate}
    \item Reserve 20\% of the remaining stars for testing and remove them from the training set.
    \item During iterative model fitting, reject outliers according to the procedure in \sect{sec:outlier_reject}. Note that this step does not apply to the reserve stars. \label{step:outlier_reject}
    \item When measuring moments on the reserve stars for diagnostic purposes, reject any object whose centroid moves more than 1\arcsec, indicating the algorithm migrated to fit a close neighbor instead.
\end{enumerate}

\begin{figure}
\includegraphics[width=\columnwidth]{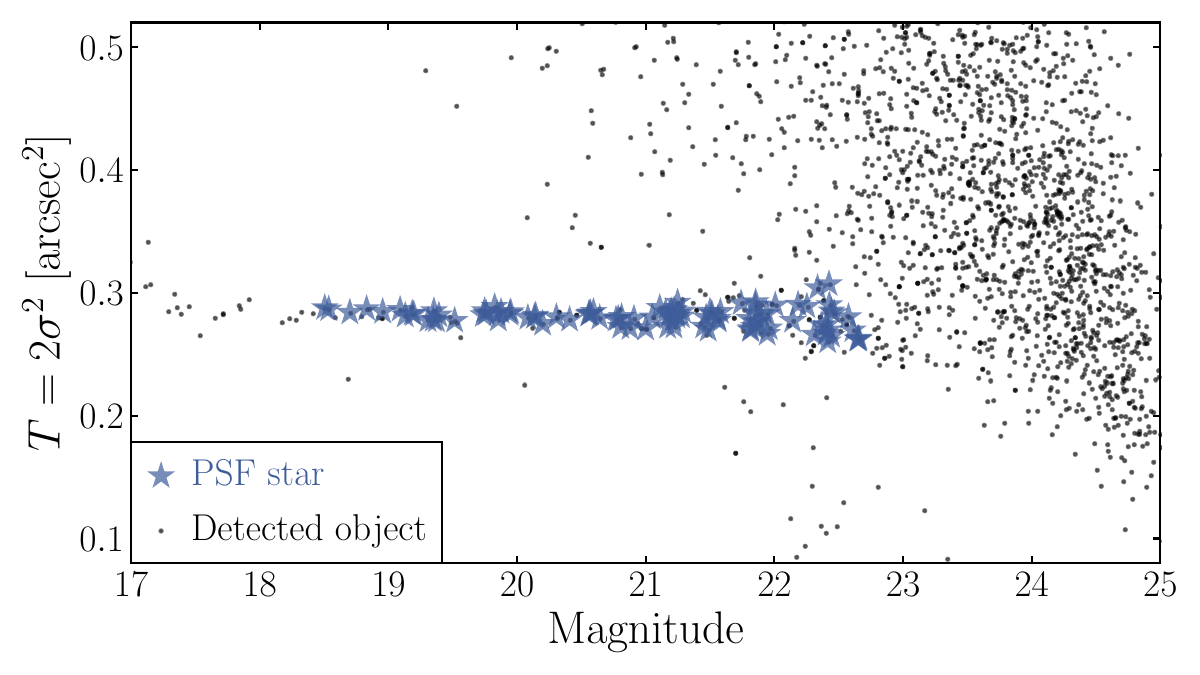}
\caption{
Example size-magnitude diagram for a single typical $r$ band CCD image,
showing all objects detected on the CCD (black) and the subset of objects selected as stars suitable for PSF fitting (blue).
For this image, there are 98 PSF stars total, 20 of which are reserved for model testing.
}
\label{fig:sizemag}
\end{figure}

The DES Y3 PSF modeling employed similar elements of this selection, specifically the bright magnitude cut and a very similar \sizemag\ selection (though the latter was not implemented in \piff\ at the time).
The similarity in \edit{the Y3 and Y6} selection criteria can be seen in \fig{fig:sizemag} (\edit{see} Figure 3 in \citetalias{y3-piff}).
\fig{fig:sizemag} shows the size-magnitude diagram for a typical $r$ band CCD image.
The black points show all objects detected in the image using \sex. 
The blue stars mark the PSF star sample resulting from this selection process.
The cut on bright but unsaturated objects and the identification of the stellar locus are very similar between the two methods.

\begin{figure}
\includegraphics[width=\columnwidth]{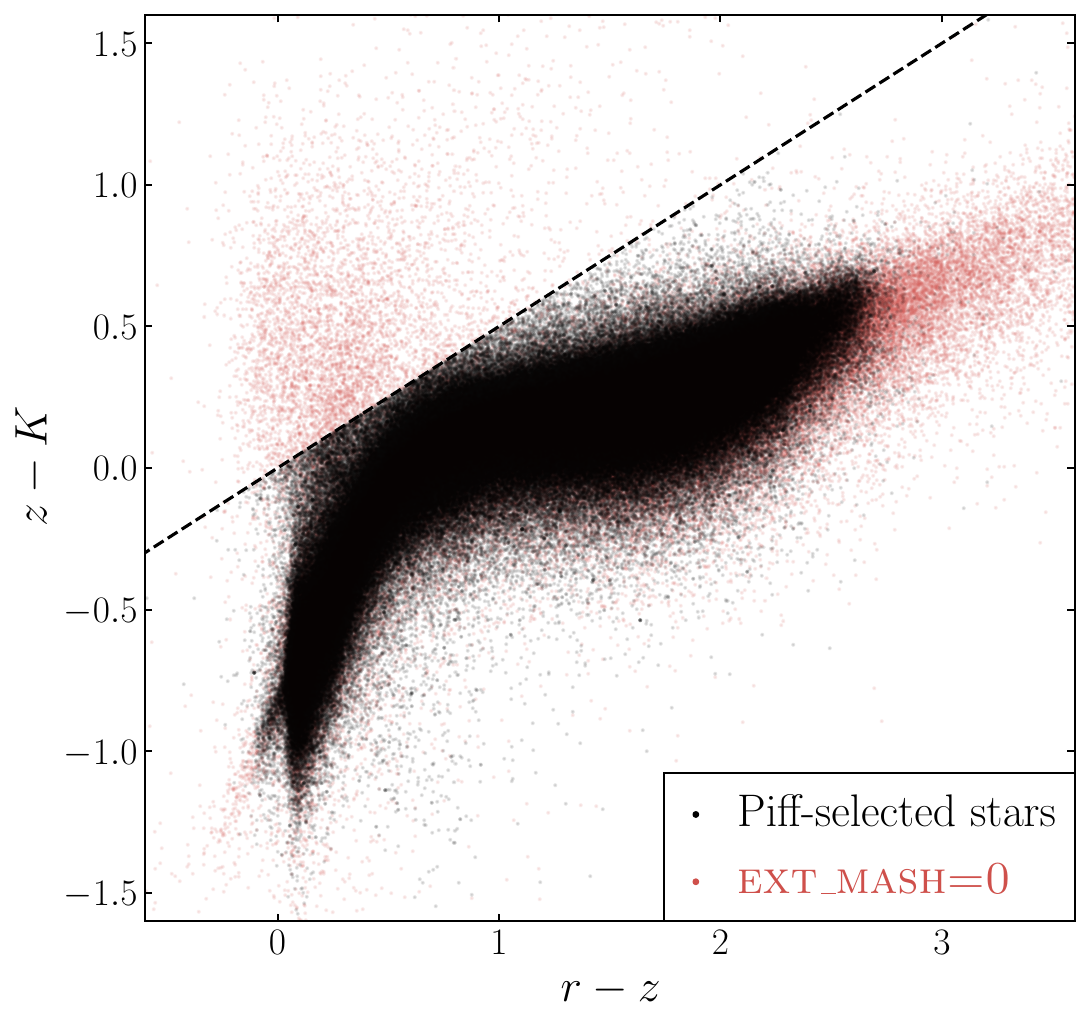}
\caption{
Color-color plot comparing Piff-selected reserve PSF stars (black) and Y6 Gold catalog objects classified as high-confidence stars (\textsc{ext\_mash}=0, red).
For PSF stars with a match to an IR source, we apply the cut given in \eqn{eq:ir_cut} (black dotted line), which removes likely quasars in the upper-left corner of the color-color space.
\label{fig:zk_rz_locus}
}
\end{figure}

However, the extended selection criteria do show some marked improvements over the simpler Y3 method in other respects.
The interloper non-stellar sources identified in \citetalias{y3-piff} using IR information are now largely cut from the PSF star sample, as shown in \fig{fig:zk_rz_locus}.
This color-color plot compares the distribution of objects in the Y6 Gold catalog classified as high-confidence stars via the \extmash\ algorithm (\extmash\texttt{=0}, red points) versus that of the PSF stars (black).
For both populations, the points shown are the subset with a match to a VHS/VIKING source (and thus have a measured $K$ magnitude).
The cut made in step \ref{step:ir_cut} cuts a distinct population of likely non-stellar sources in the upper-left corner of the $(r-z, z-K)$ space.
These interlopers must still be present among the PSF stars without a match in the IR catalogs.
However, they were found to make up $<1\%$ of the stars in the Y3 analysis where no IR cut was implemented.
After the cut among the 73\% of PSF stars with an IR match, this population of galaxies is likely reduced by more than half.

%%%%%%%%%%%%%%%%%%%%%%%%%%%%%%%%%

\subsection{Model quality assurance criteria}
\label{sec:exclusions}

After completing the calculation of a \piff\ model, we perform several quality assurance (QA) checks at the CCD image level
to ensure the model for that sensor is valid and extrapolates to galaxy images reasonably well.
If a PSF model for a particular CCD image fails any of the following QA criteria,
that image is excluded from any further analysis.

\begin{itemize}

\item
No \pixmappy\ WCS solution: Some exposures ($\sim 10$ nights) were taken during periods with insufficient calibration information to produce a reliable \pixmappy\ WCS solution.
\item
Poor PSF model fit: we flag and remove any image with a \piff\ model solution with $\chi^2_{\nu}$ > 1.5.
\item
Too few stars: We flag images for which fewer than 20 stars survived the outlier rejection.
\item Outlier PSF images in color-position space: When evaluated at the median galaxy color, some PSF models have poorly interpolated PSF images in some locations on the CCD.
These artifacts are likely due to the model being poorly constrained in color-position space.
% To remove these models, we measure the size of the PSF using adaptive moments on a grid with 128 pixel spacing across the CCD.
% We flag and remove any CCD where either one of the adaptive moments fits fails or any star deviates in $T_{*}$ by more than 0.15 from the median over the grid. 
\edit{To remove these images, we evaluate the PSF model at the median galaxy color on a grid with 128 pixel spacing across the CCD, and we measure the size $T$ of these model images using adaptive moments.
We flag and remove any CCD where either (i) any adaptive moments fit fails or (ii) any grid model image has a size deviation $>0.15$~arcsec$^2$ from the median size measured over the grid.}

\end{itemize}

%%%%%%%%%%%%%%%%%%%%%%%%%%%%%%%%%%%%%%%%%%%%%%%%%%%%%%%%%%%%%%%%%%%

\section{PSF diagnostics}
\label{sec:tests}

In this section, we present a comprehensive set of diagnostic tests of the PSF modeling.
These tests are chiefly focused on verifying that the PSF models are of sufficient quality for weak lensing shear requirements.
As such, they primarily consist of checks on the size and shape residuals of the PSF model (defined in \sect{sec:def_moms}), which contribute directly to the multiplicative and additive bias of shear measurements, respectively.

To quantify the quality of the PSF, we evaluate the PSF model at the same epoch, position and color of each real star reserved for testing (20\% of the full sample) and compare this PSF model with the observed surface brightness profile of the star.
Statistics for this catalog of reserve stars are given in \tab{tab:cat_stats}.

The goal of these diagnostics is to evaluate the quality of the models themselves decoupled from how they may be used in later analysis steps, e.g., in image coaddition or deconvolution during shear estimation.
Validating such \textit{applications} of the models is beyond the scope of this paper.
In particular, we note that all of the diagnostics in this paper are performed using single-epoch star images, not coadded images.
Therefore, potential PSF-like errors incurred during image coaddition are not tested here.
For validation of the PSF coaddition process used for the Y6 \mdet\ shear measurements \citep*{y6-mdet}, see \citet{Armstrong24}.

%%%%%%%%%%%%%%%%%%%%%%%%%%%%%%%%%

\subsection{Definition of moments}
\label{sec:def_moms}

To analyze the quality of the PSF models, we measure a set of moments on the PSF and reserve star images to estimate their size and shape/ellipticity. Here we define the set of moments we use for this measurement.

Since cosmic shear is a spin-2 field, PSF residuals that are spin-2 or a product of a spin-0 and spin-2 quantity can contribute an additive bias to the shear measurement.
Further, a spin-0 PSF residual (i.e., size error) can also contribute a multiplicative bias.
Thus, we specifically study PSF moments in terms of their spin-0 and spin-2 decompositions.
We use the HSM adaptive moments algorithm \citep{hirata2003, hsm} to measure the second-order moments\footnote{
To understand the indexing convention, we can define the complex quantity $\bm{m}_{jk} = (du + idv)^j (du - idv)^k$. The (unnormalized) moments are 
$$
M_{jk} = 
\begin{cases}
    \sum W(u,v) I(u,v)\mathrm{Re}[\bm{m}_{jk}]  & j \geq k, \\
    -\sum W(u,v) I(u,v)\mathrm{Im}[\bm{m}_{jk}] & j < k.
\end{cases}
$$
}
\begin{linenomath*}
    \begin{align}
        M_{00} &= \sum W(u,v) I(u,v) \\
        M_{11} &= \frac{1}{M_{00}}\sum W(u,v) I(u,v) (du^2 + dv^2) \\ \label{eq:m11}
        M_{20} &= \frac{1}{M_{00}}\sum W(u,v) I(u,v) (du^2 - dv^2) \\
        M_{02} &= \frac{1}{M_{00}}\sum W(u,v) I(u,v) (2 du dv),
    \end{align}
\end{linenomath*}
where $W(u,v)$ is the weight from the HSM fit, $I(u,v)$ is the image or PSF surface brightness profile and $du, dv$ are the positions relative to the HSM-measured centroid.
We define a scalar ``size'' quantity $T^{(2)}$ and the spin-2 complex ellipticity $\bm{e}^{(2)} = e^{(2)}_1 + i e^{(2)}_2$ as
\begin{linenomath*}
    \begin{align}
        T^{(2)} & \equiv \edit{2} M_{11} \\
        \bm{e}^{(2)} & \equiv \frac{M_{20} + iM_{02}}{M_{11} + \sqrt{M_{11}^2 - (M_{20}^2 + M_{02}^2)}}.
    \end{align}
\end{linenomath*}
\edit{$T^{(2)}$ is the \textit{unweighted} ``trace'' moment, which is double the weighted moment $M_{11}$ (see equation 4 in \citet{hirata2003}).}

\cite{Zhang22, Zhang23a, Zhang23b} have studied the effect of fourth-order moments on shear measurement errors and found they can impact shear measurement at equal or even greater magnitude than the conventional second-order moments. Following this work, we also compute analogous fourth-order spin-0 and spin-2 quantities analogous to $T^{(2)}$ and $\bm{e}^{(2)}$ above.
\piff\ measures the following fourth-order moments:
\begin{linenomath*}
    \begin{align}
        M_{22} &= \frac{1}{M_{00}}\sum W(u,v) I(u,v) (du^2 + dv^2)^2 \\
        M_{31} &= \frac{1}{M_{00}}\sum W(u,v) I(u,v) (du^2 + dv^2) (du^2 - dv^2) \\
        M_{13} &= \frac{1}{M_{00}}\sum W(u,v) I(u,v) (du^2 + dv^2) (2 du dv).
    \end{align}
\end{linenomath*}
Like the second-order moments, we can define a na\"ive fourth-order spin-0 size $\Tilde{T}^{(4)}$ and spin-2 complex ellipticity $\Tilde{\bm{e}}^{(4)}$ as
\begin{linenomath*}
    \begin{align}
        \Tilde{T}^{(4)} & \equiv \frac{M_{22}}{M_{11}} \\
        \Tilde{\bm{e}}^{(4)} & \equiv \frac{M_{31} + iM_{13}}{(M_{11})^2}.
    \end{align}
\end{linenomath*}
However, we find $\Tilde{T}^{(4)}$ and $\Tilde{\bm{e}}^{(4)}$ are highly correlated with $T^{(2)}$ and $\bm{e}^{(2)}$, respectively.
To create quantities that primarily probe higher-order spatial scales, we define the fourth-order moments used for diagnostics $T^{(4)}$ and $\bm{e}^{(4)} = e^{(4)}_1 + i e^{(4)}_2$ by subtracting the second-order terms
\begin{linenomath*}
    \begin{align}
        T^{(4)} & \equiv \frac{M_{22}}{M_{11}} - \edit{T^{(2)}} \\
        \bm{e}^{(4)} & \equiv \frac{M_{31} + iM_{13}}{(M_{11})^2} - 3\left(\frac{M_{20}+iM_{02}}{M_{11}}\right).
    \end{align}
\end{linenomath*}
Note that because these fourth-order moments are defined as ``distortion-like'' quantities, the term subtracted from $\Tilde{\bm{e}}^{(4)}$ is the second-order ``distortion,'' not the \edit{ellipticity} $\bm{e}^{(2)}$.
In the case of an elliptical Gaussian, $T^{(4)}$ and $\bm{e}^{(4)}$ are nearly zero.

To simplify notation throughout this work, when a statement applies to moments of both second- and fourth-order, the $(i)$ superscript is omitted. Further, for a given measured quantity of order $i$, $X^{(i)}$ (e.g. $T^{(2)}$), we define the residual $\delta X^{(i)}$ as the difference in this quantity measured on a star image and on the PSF model evaluated at the same epoch, position and color:

\begin{linenomath*}
    \begin{align}
        \delta X^{(i)} & \equiv X^{(i)}_{\rm data} - X^{(i)}_{\rm model}\\
        \frac{\delta X^{(i)}}{X^{(i)}} & \equiv \frac{X^{(i)}_{\rm data} - X^{(i)}_{\rm model}}{X^{(i)}_{\rm model}}.
    \end{align}
\label{eq:def_residual}
\end{linenomath*}

%%%%%%%%%%%%%%%%%%%%%%%%%%%%%%%%%

\subsection{Testing for overfitting}

\begin{figure}
\includegraphics[width=\columnwidth]{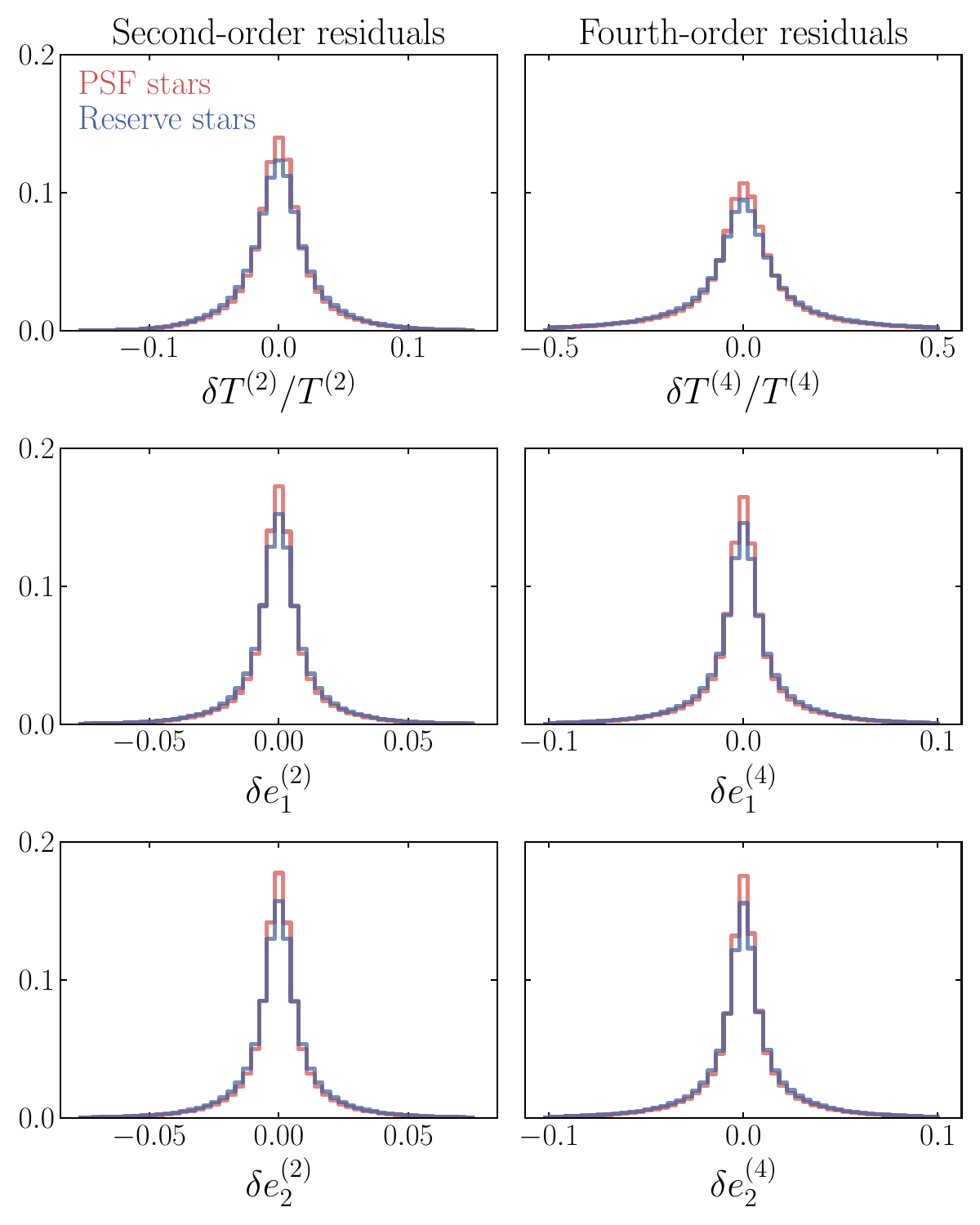}
\caption{
Comparison of modeling residuals, defined in \sect{sec:def_moms}, for stars used for PSF modeling (red) and the reserve stars (blue).
We do not find evidence of significant overfitting to the data.
}
\label{fig:res_nonres_2+4order}
\end{figure}

All PSF model diagnostics in this work use a randomly selected ``reserve'' sample of 20\% of the stars suitable for PSF modeling.
These reserved stars are not used in model fitting and thus provide information on the interpolation quality of the PSF models.
Aside from one caveat explained below, the stars reserved for testing undergo the same selection and QA process as the stars used for model fitting (hereafter referred to as ``reserve stars'' and ``PSF stars,'' respectively).

We perform a basic test for overfitting by comparing the values of the three main diagnostic PSF residuals, $\delta T/T$, $\delta e_1$ and $\delta e_2$ (second- and fourth-order) for the PSF and reserve stars.
The histograms of these values are shown in \fig{fig:res_nonres_2+4order}.

As described in \sect{sec:data:star_select}, one of the main methods we use to mitigate overfitting is to downweight bright stars so that they do not dominate the fit.
Since the distributions for the PSF and reserve stars are very similar for all residual types, we deem this strategy successful.
We do not find evidence that the PSF models are significantly overfitted to the data.
The main discrepancy between the two samples is that for every distribution except $\delta T^{(4)}/T^{(4)}$ the standard deviation is $\sim 7\%$ larger for the reserve stars than the PSF stars;
the standard deviation is $\sim 30\%$ larger for $\delta T^{(4)}/T^{(4)}$.
Presumably, this is in part because the reserve stars were not subject to outlier rejection (step \ref{step:outlier_reject} in \sect{sec:data:star_select}), so reserve stars that would have been identified as outliers are still present, leading to somewhat larger residuals than the PSF stars.\footnote{We note a more recent release of \piff\ (version 1.4) gives the option to run outlier rejection on the reserve stars after the iterative fitting has converged.
This enables complete apples-to-apples comparisons for future analyses.}
As such, a small fraction of the reserve stars are likely outlier objects with systematically larger PSF residuals and do not reflect an issue with overfitting.
Thus, using the reserve stars provides a conservative estimate of the impact of PSF errors on shear measurements.

%%%%%%%%%%%%%%%%%%%%%%%%%%%%%%%%%

\subsection{Residuals in the field of view}
\label{sec:fov}

The PSF is not constant across the DECam field of view (FOV).
Optical aberrations and sensor effects contribute to this variation.
Noteably, some of these effects are highly wavelength dependent.
In this section we present an analysis of the PSF behavior as a function of focal plane position, which isolates the impact caused by these optical and sensor effects.
All focal plane figures in this work are shown in sky coordinates (i.e. North up, West right).

\subsubsection{Full focal plane maps}
\label{sec:fov:focal_plane}

\begin{figure}
    \centering
    \includegraphics[width=\columnwidth]{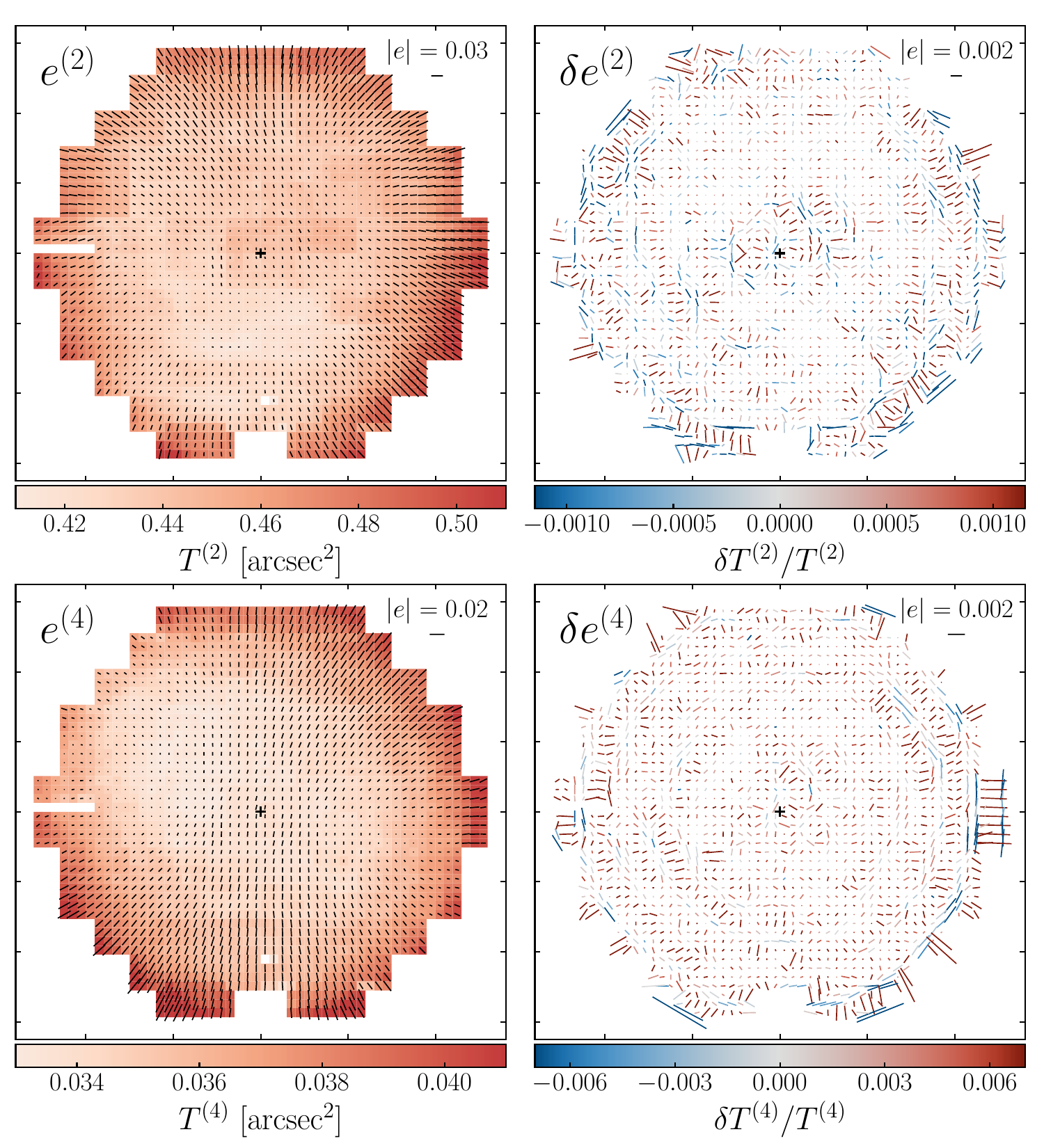}
    \caption{
    Whisker plots of the full focal plane showing the direction of the average $e^{(i)}$ (left panels) or $\delta e^{(i)}$ residual (right panels) in $\sim$ 500x500 superpixels for the $riz$ bands combined.
    The color of the map (left) or whiskers (right) denotes the average size or fractional size residual in the same pixel binning.
    The field of view is $\sim 2^{\circ}$.
    }
    \label{fig:fov_whisker}
\end{figure}

\begin{figure*}
\centering
\includegraphics[height=1.5in]{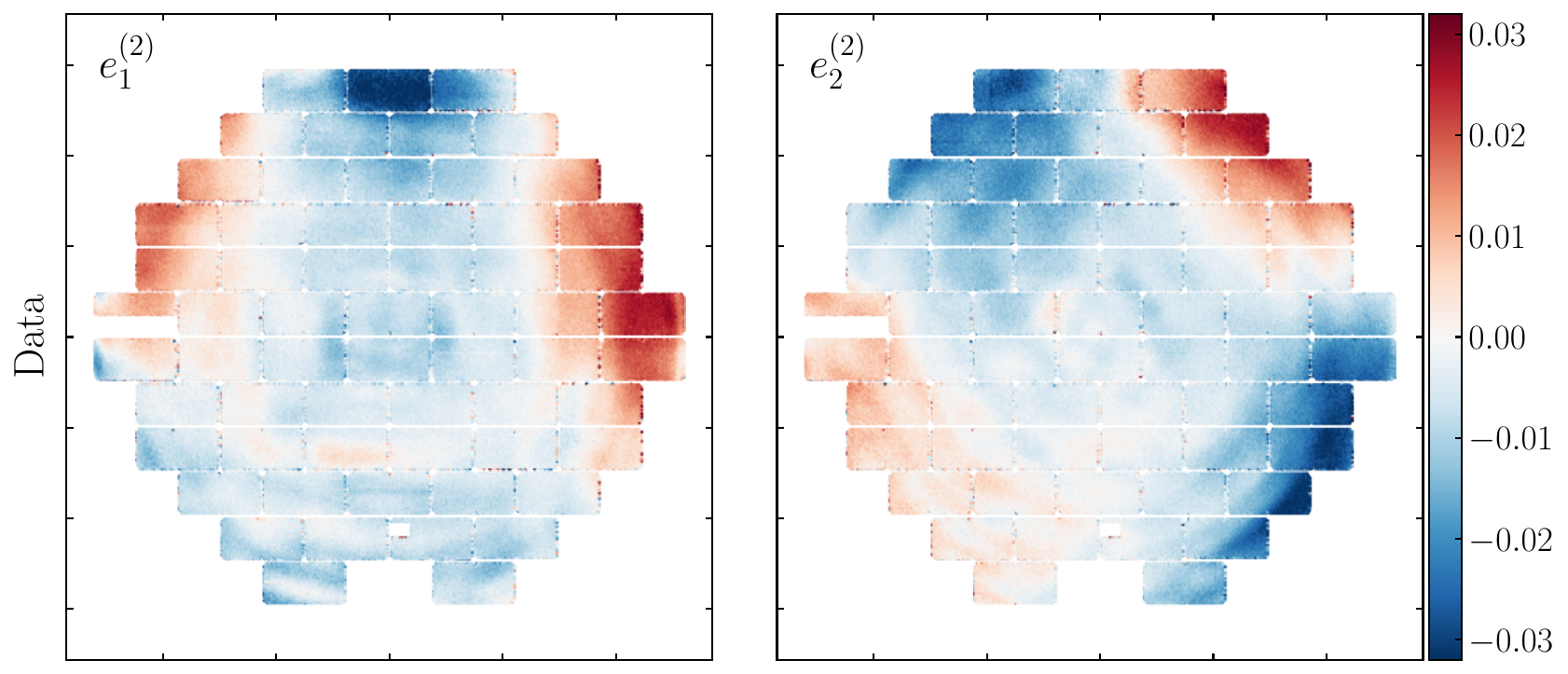}
\includegraphics[height=1.5in]{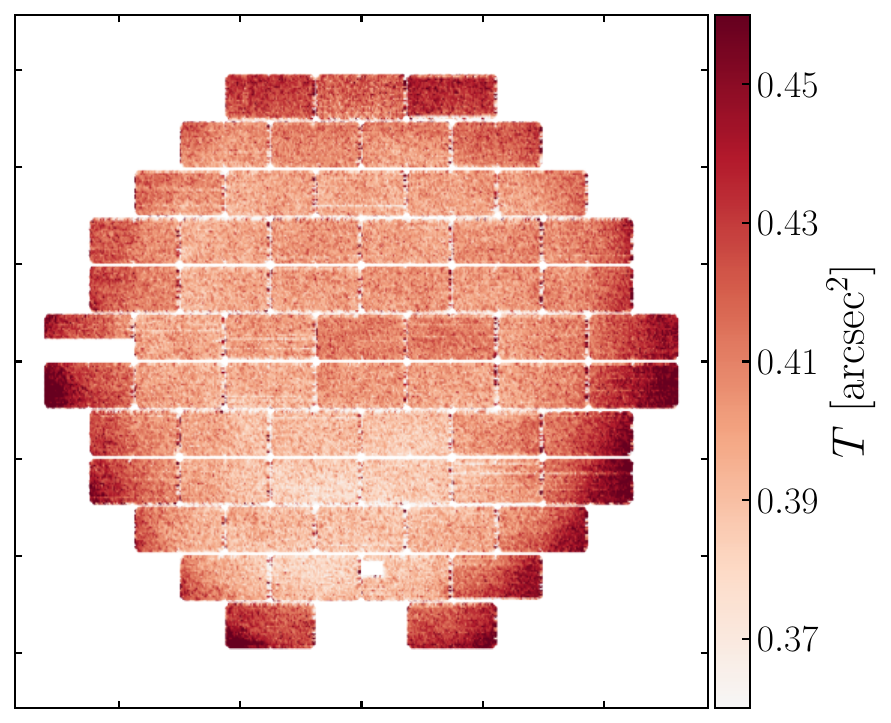}\\
\hspace{0.pt}
\includegraphics[height=1.51in]{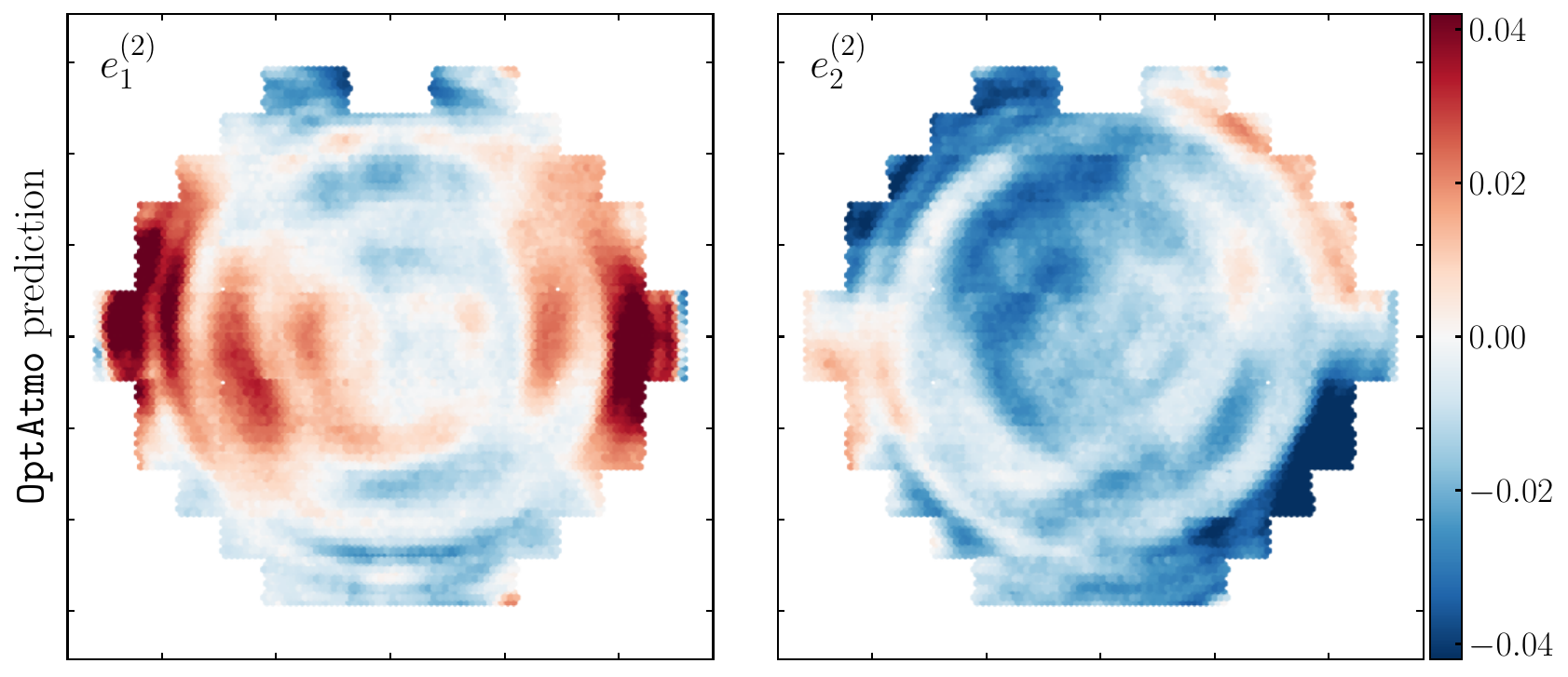}
\includegraphics[height=1.5in]{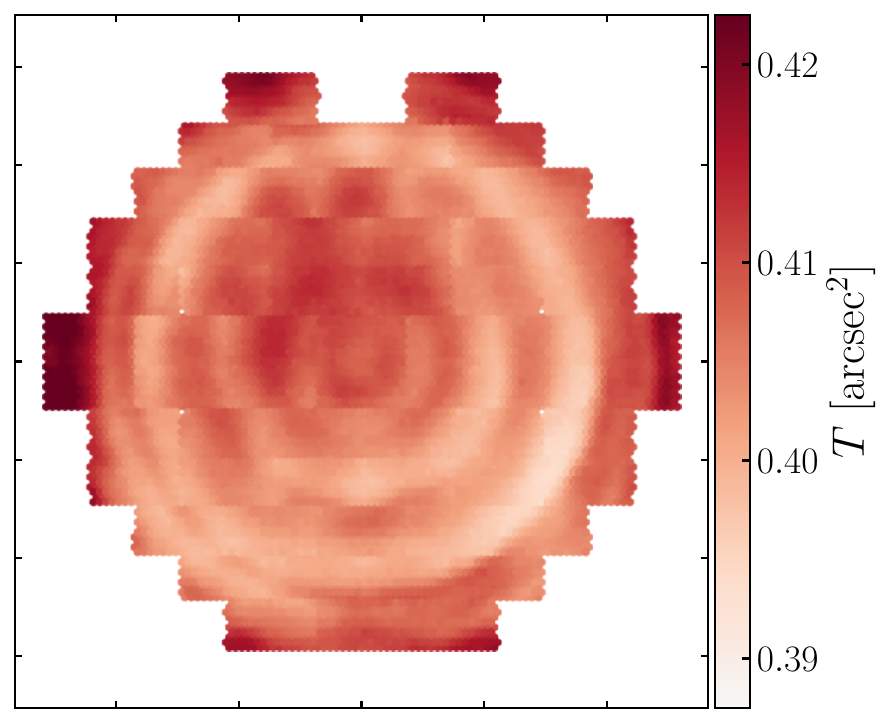}
\caption{
\textbf{Top}:
Maps of the two second-order components of the PSF shape $e^{(2)}_1$ and $e^{(2)}_2$ (left and middle) and the second-order size $T^{(2)}$ (right) as a function of position in the focal plane for $riz$ bands combined, binned in $\sim$ 50x50 superpixels.
\textbf{Bottom}:
Predicted second-order $i$-band PSF size and shape using the \optatmo\ wavefront-based physical PSF model, assuming perfect focus and alignment and a constant atmospheric kernel.
Qualitatively, large-scale higher-order aberration features agree well between the data and model prediction.
Features in the data may be smoother than the model due to variations in focus and atmospheric conditions over the survey.
Note the focal plane orientation is the same between the data and model; different CCDs were functional at the time the wavefront data were taken (Feb. 2014).
}
\label{fig:optatmo}
\end{figure*}

\begin{figure*}
\centering
\includegraphics[height=1.7in]{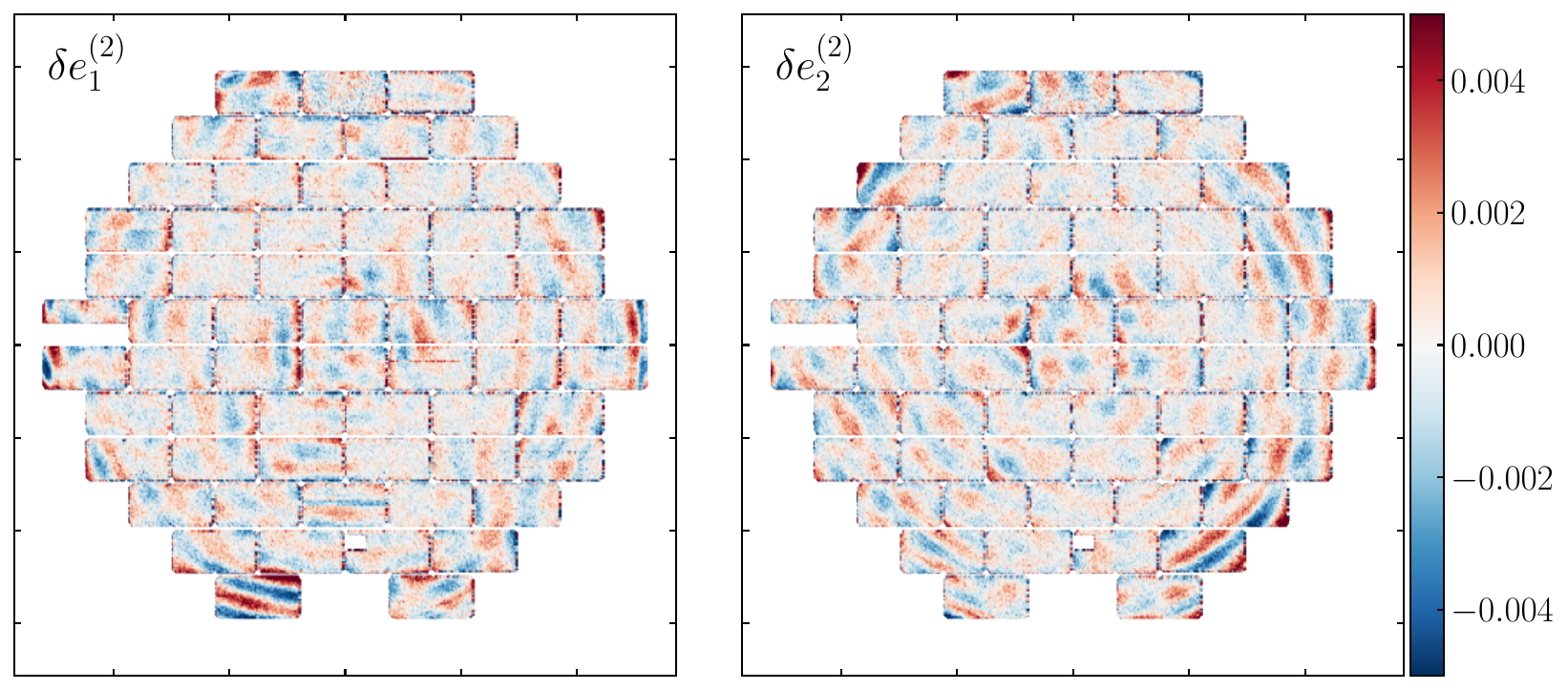}
\includegraphics[height=1.7in]{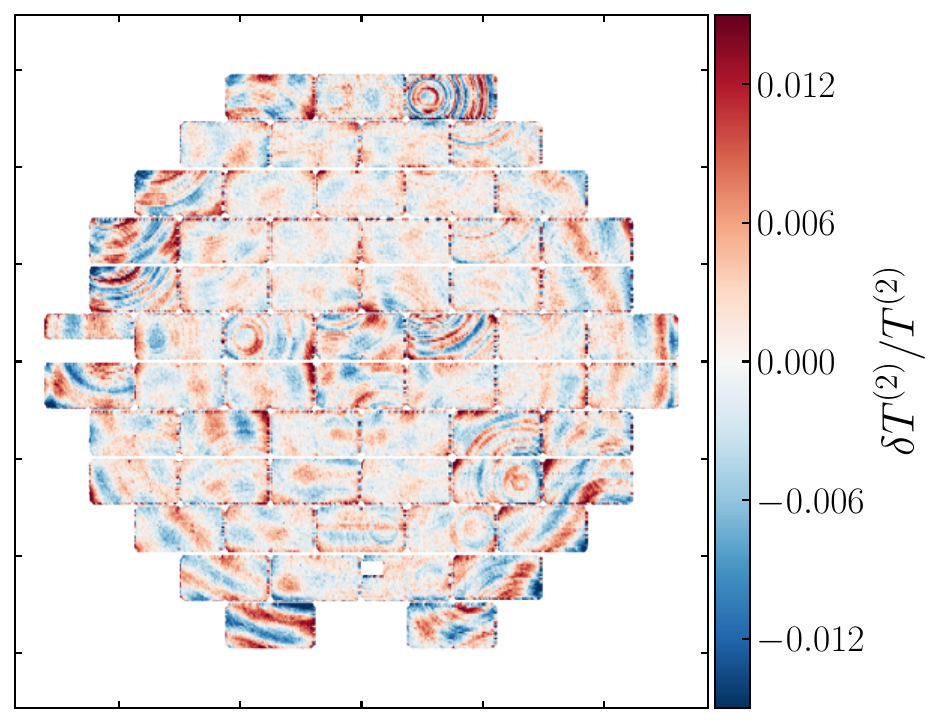}
\caption{
Maps of the two second-order components of the residual shape, $\delta e^{(2)}_1$ and $\delta e^{(2)}_2$ (left and middle), and the second-order fractional residual size $\delta T^{(2)}/T^{(2)}$ (right) as a function of position in the focal plane for $riz$ bands combined.
All three residual maps have an oscillating radial pattern corresponding to high spatial frequency modes largely caused by higher-order optical aberrations.
The magnitude of this pattern is slightly larger than in Y3 for $e^{(2)}_1$ and $e^{(2)}_2$ due to using second-order spatial interpolation instead of third-order as in Y3.
The size residuals show noticeable ``tree ring'' patterns in many CCDs.
These residual tree rings are likely due to diffusion within the CCD after photon-to-electron conversion; they are discussed in more detail in \sect{sec:fov:treerings}.
}
\label{fig:fov_res_2order}
\end{figure*}

\begin{figure*}
\centering
\includegraphics[height=1.7in]{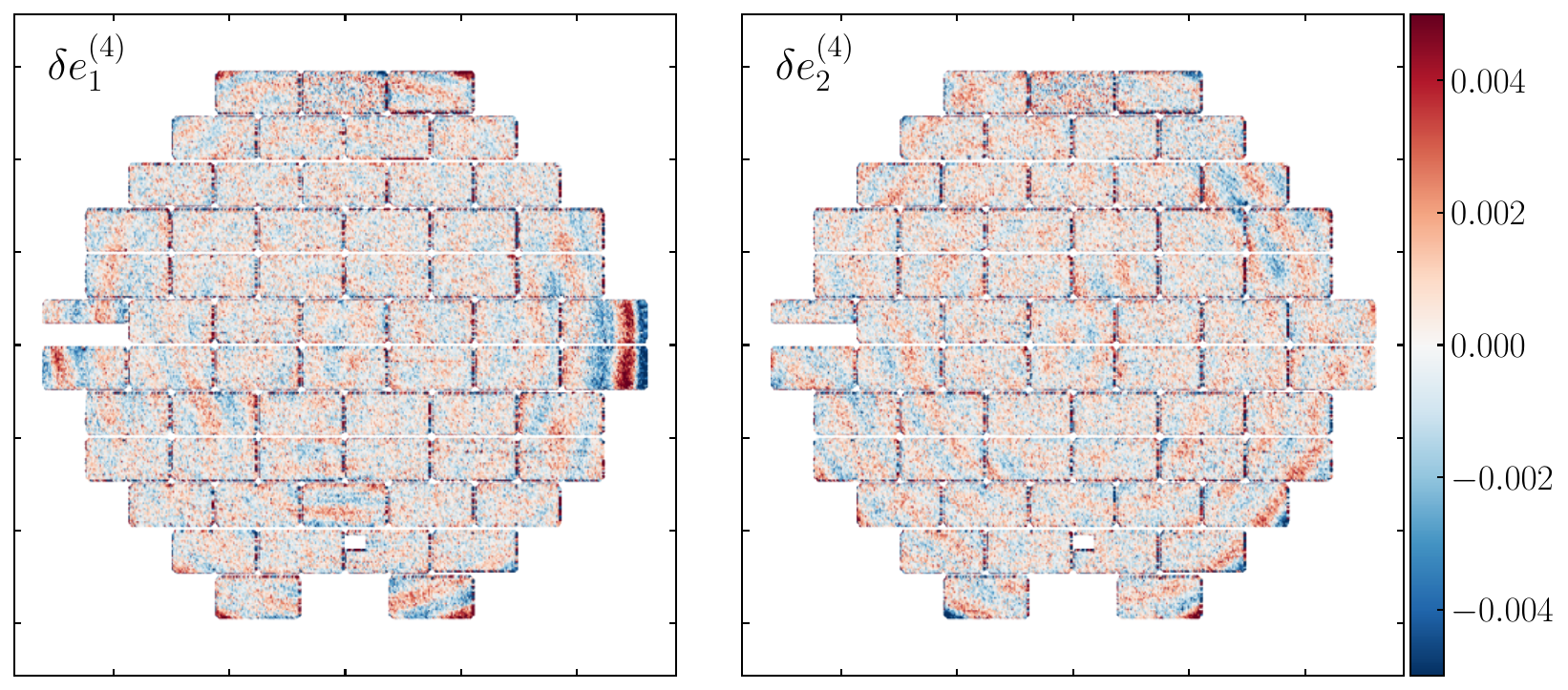}
\includegraphics[height=1.7in]{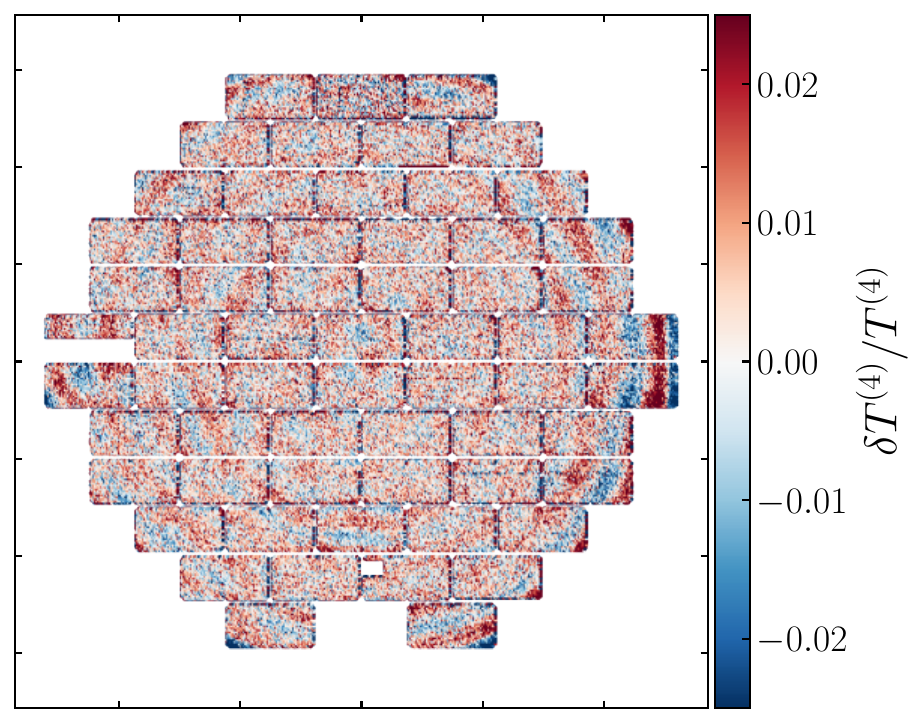}
\caption{
Same as \fig{fig:fov_res_2order} but for the fourth-order moment residuals.
All three moments' residuals exhibit the same oscillating radial pattern as the second-order residuals, as expected from higher-order optical aberrations.
Interestingly, the size residuals $\delta T^{(4)}/T^{(4)}$ do \textit{not} exhibit significant tree ring structures.
This provides evidence that the residual tree rings are largely due to diffusion, which should only cause Gaussian (i.e., second-order) size residuals.
}
\label{fig:fov_res_4order}
\end{figure*}

First, we characterize the PSF behavior before model subtraction.
The whisker plots in the left panels of \fig{fig:fov_whisker} are oriented in the direction of the reserve star ellipticity parameters, $e_1$ and $e_2$, averaged in $\sim$500x500 pixel bins.
The length of the whiskers correspond to the modulus of $\bm{e}$, and their color correspond to the average size $T$ in that pixel bin.
Second-order quantities are shown in the top row, fourth-order in the bottom row.
As expected from the optical design, the PSF grows in size with greater distance from the focal plane center.
The magnitude of ellipticity also grows, with the stars' shapes largely oriented in the radial direction.
Some azimuthal anisotropy is also present.

PSF residuals are shown in the right panels of \fig{fig:fov_whisker}, with the color of each whisker corresponding to the average fractional size residual $\delta T/T$ in the bin.
A radially-dependent oscillatory pattern is apparent in the size and shape residuals for both the second- and
fourth-order moments.
The residuals in both second- and fourth-order ellipticity are largely oriented in an oscillating tangential and radial pattern, i.e., in positive and negative tangential shear.
These whisker plots also make clear that the residuals in all moments are highly correlated; negative size residuals are correlated with shape residuals oriented tangent to the center of the focal plane and vice versa.

Such trends in the PSF size and shape and their residuals after model subtraction are consistent with expectations from \optatmo, a more complex optical and atmospheric PSF model using DECam wavefront measurements \citep{Davis16, roodman-inprep}.
The top row of \fig{fig:optatmo} shows the PSF $e^{(2)}_1$, $e^{(2)}_2$ and $T^{(2)}$ measurements, plotted separately as scalar quantities and at higher resolution ($\sim$ 50x50 superpixels) than \fig{fig:fov_whisker}.
The bottom row of \fig{fig:optatmo} shows the binned $e^{(2)}_1$, $e^{(2)}_2$ and $T^{(2)}$ values predicted from the \optatmo\ model, assuming perfect focus and alignment and a constant Kolmogorov atmospheric kernel. 

The model is sampled on a grid with $\sim$ 130x130 pixel spacing spanning the full focal plane.
Qualitatively, the large-scale aberration features agree between the modeled and measured moments.
They confirm the large concentric ring residuals seen in \figb{fig:fov_whisker}{fig:fov_res_2order} correspond to where the model's quadratic spatial interpolation (per CCD) cannot capture the higher-frequency spatial modes caused by optical aberrations.
There are some significant patterns in the model prediction that are not present in the data, such as the higher amplitude ringing in the size measurements and high amplitude ellipticity features at the edges of the focal plane.
These patterns may be reduced or smoothed out due to variations in focus and atmospheric conditions over the course of the survey.

To visualize the focal plane residuals in higher resolution, we show in \figb{fig:fov_res_2order}{fig:fov_res_4order} the second- and fourth-order $\delta e_1$, $\delta e_2$ and $\delta T / T$ residuals in the same $\sim$ 50x50 pixel bins as the top row of \fig{fig:optatmo}.
In the next two subsections, we study two main features more quantitatively: the variations dependent on radial distance from the focal plane center and the smaller concentric circle patterns, commonly called ``tree rings", which vary between CCDs.

%%%%%%%%%%%%%%%%%%%%%%%%%%%%%%%%%

\subsubsection{Radially-dependent residuals}
\label{sec:fov:rad_res}

\begin{figure*}
\centering
\includegraphics[width=\textwidth]{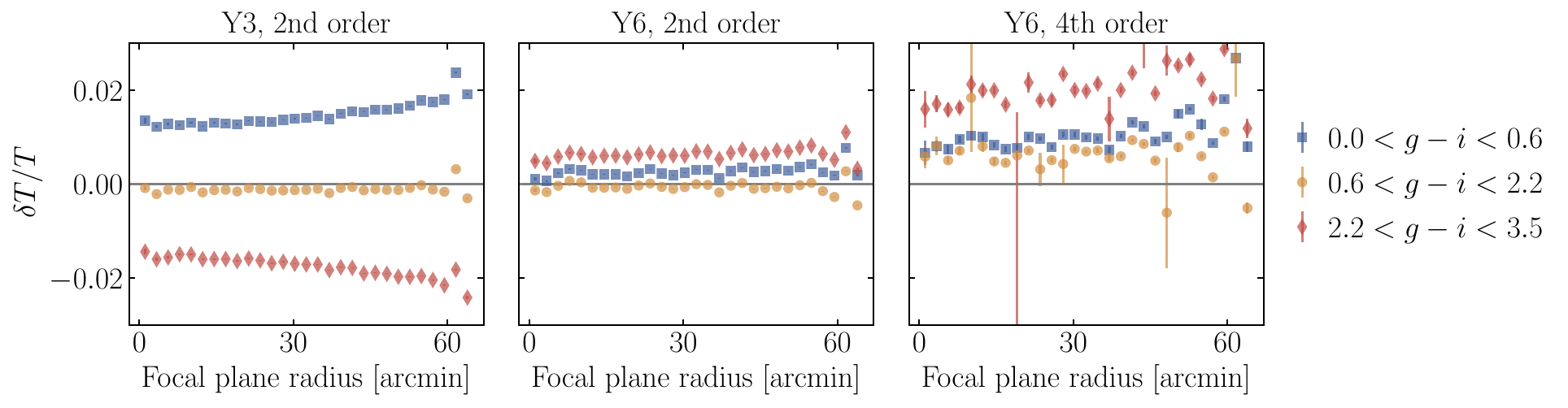}
\caption{
PSF fractional size residuals binned as a function of radial position in the focal plane in $g$ band, for which the effect is most significant.
Stars are split into three groups in $g-i$ color to study color dependence of the residuals: the bluest 25\% (blue squares), middle 50\% (yellow circles) and reddest 25\% (red diamonds) among stars detected in $g$ band.
The Y3 second-order size residuals (left) show a mean offset and large-scale radially-dependent trend, both of which are color-dependent.
The offset and global radial dependence are largely removed in the Y6 second-order residuals (middle), though nonlinear color-dependent residuals remain.
The remaining small-scale oscillatory residuals are likely due to high spatial frequency modes not captured by the second-order spatial interpolation (also seen in the 2D maps).
The Y6 fourth-order residuals (right) display similar trends in color-dependence and radially-dependent oscillations, but with larger magnitude and more noise.
}
\label{fig:fovrad}
\end{figure*}

We expect PSF size and shape to vary radially due to optical aberrations.
Further, chromatic aberrations make PSF characteristics wavelength dependent.
Without color-dependent modeling, the PSF model will only capture the PSF behavior at the mean color of stars over a given CCD image.
As discussed in \sect{sec:piff_color}, a major improvement in the PSF modeling between the DES Y3 and Y6 analyses is the addition of color-dependent interpolation.

\fig{fig:fovrad} shows the improvement made in addressing PSF size variation due to chromatic aberrations between Y3 and Y6.
In all panels, the fractional size residuals of reserve stars detected in $g$ band is plotted as a function of radial distance from the focal plane center.
Only stars detected in $g$ band are shown as this is where the residuals are most significant.
To study the impact of chromatic aberrations, the stars are split into quantiles in their $g-i$ color: the bluest 25\% ($0<g-i<0.6$, blue squares), middle 50\% ($0.6<g-i<2.2$, yellow circles) and reddest 25\% ($2.2<g-i<3.5$, red diamonds).

The left panel shows the second-order size residuals for the Y3 PSF models.
The mean shift of about $\pm1.5\%$ is likely due to chromatic seeing, but the radial trend is due to the optics.
The middle panel shows the second-order size residuals for the Y6 PSF models.
The color interpolation significantly improves residuals due to both atmospheric and optical chromatic effects.
The $\sim 3\%$ constant difference in color groups is decreased to the sub-percent level, and the large-scale radial trend is reduced by an order of magnitude.
What remains is the quadratic term of the chromatic seeing (studied further in \sect{sec:colorres}) and the achromatic small-scale oscillations explained above.
The right panel shows the Y6 fourth-order size residuals (fourth-order moments were not measured for the Y3 PSF models).
The difference in the mean size residual between the red and blue splits is $\sim 1\%$. 
For both splits, the mean size residual grows by about $0.5\%$ toward the edge of the focal plane.

By comparison, when binned by focal plane radius, the Y6 second-order and fourth-order size residuals for the $riz$ bands (not shown) are significantly smaller. 
For both second- and fourth-order residuals, the mean difference between the red and blue splits is $<0.1\%$, and the residuals do not show any color-dependent radial trend.
As seen in the $g$ band residuals, the oscillations due to higher-order optical aberrations grow in amplitude toward the edge of the focal plane, with a maximum peak-to-trough difference of $\lesssim 1\%$ for second-order size residuals and $\sim 1.5\%$ for fourth-order.

As \fig{fig:fov_whisker} shows, the ellipticity residuals $\delta \bm{e}$ also have significant radial dependence, particularly in the radial and tangential directions.
To study this, we project the stars' measured shapes and residuals into the tangential shear component
\begin{linenomath*}
\begin{align}
        e_{\rm t} &= -(e_1 \cos(2\alpha) + e_2 \sin(2\alpha)) \label{eq:tanshear1} \\
        \delta e_{\rm t} &= -(\delta e_1 \cos(2\alpha) + \delta e_2 \sin(2\alpha)), \label{eq:tanshear2}
\end{align}
\end{linenomath*}
where $\alpha$ is the azimuthal angle of the focal plane position, considered counterclockwise from due West.
In this projection, $(\delta)e_{\rm t}<0$ corresponds to a projected shape oriented in the radial direction and $(\delta)e_{\rm t}>0$ tangent to the radial direction.

\begin{figure}
\includegraphics[width=\columnwidth]{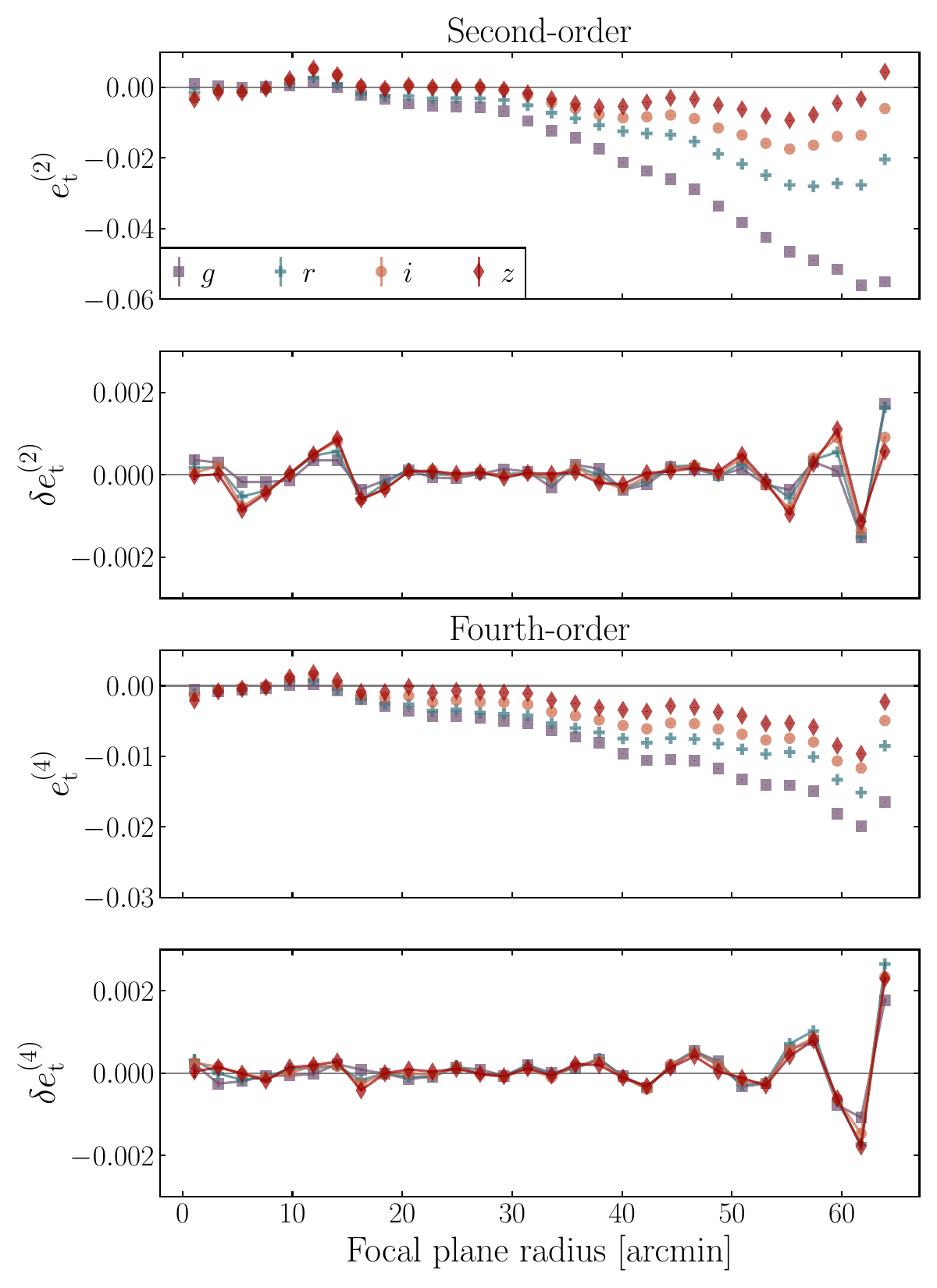}
\caption{
Projection of PSF shape measurement and residuals into the tangential shear component around the center of the focal plane, given in \eqnb{eq:tanshear1}{eq:tanshear2}.
The $griz$ bands are plotted in purple squares, blue pluses, orange circles and red diamonds, respectively.
The top two panels show the second-order tangential shear and its residuals; the bottom two panels show the fourth-order values.
Lines are drawn between points for the residuals purely as a visual aid.
}
\label{fig:fov_tanshear}
\end{figure}

\fig{fig:fov_tanshear} shows this projection of the reserve star shapes and residuals as a function of focal plane radius split in the separate $griz$ bands.
Similar to the left-hand panels in \fig{fig:fov_whisker}, the first and third panels show the measured second- and fourth-order $e_{\rm t}$ before subtracting the PSF model.
The shearing of the PSF in the radial direction is present in all bands and is especially prominent in $g$ band.
The maximum residuals after model subtraction, shown in the second and fourth panels,
are $\sim 10^{-3}$ for both second and fourth order, with large portions of the focal plane showing significantly smaller residuals.
Interestingly, $\delta e_{\rm t}$ is larger for redder bands.
Since the radial trend of $e_{\rm t}$ is smoother for bluer bands, the effect is better captured by the second-order spatial interpolation.

%%%%%%%%%%%%%%%%%%%%%%%%%%%%%%%%%

\subsubsection{Tree rings}
\label{sec:fov:treerings}

\begin{figure*}
\centering
\includegraphics[width=\textwidth]{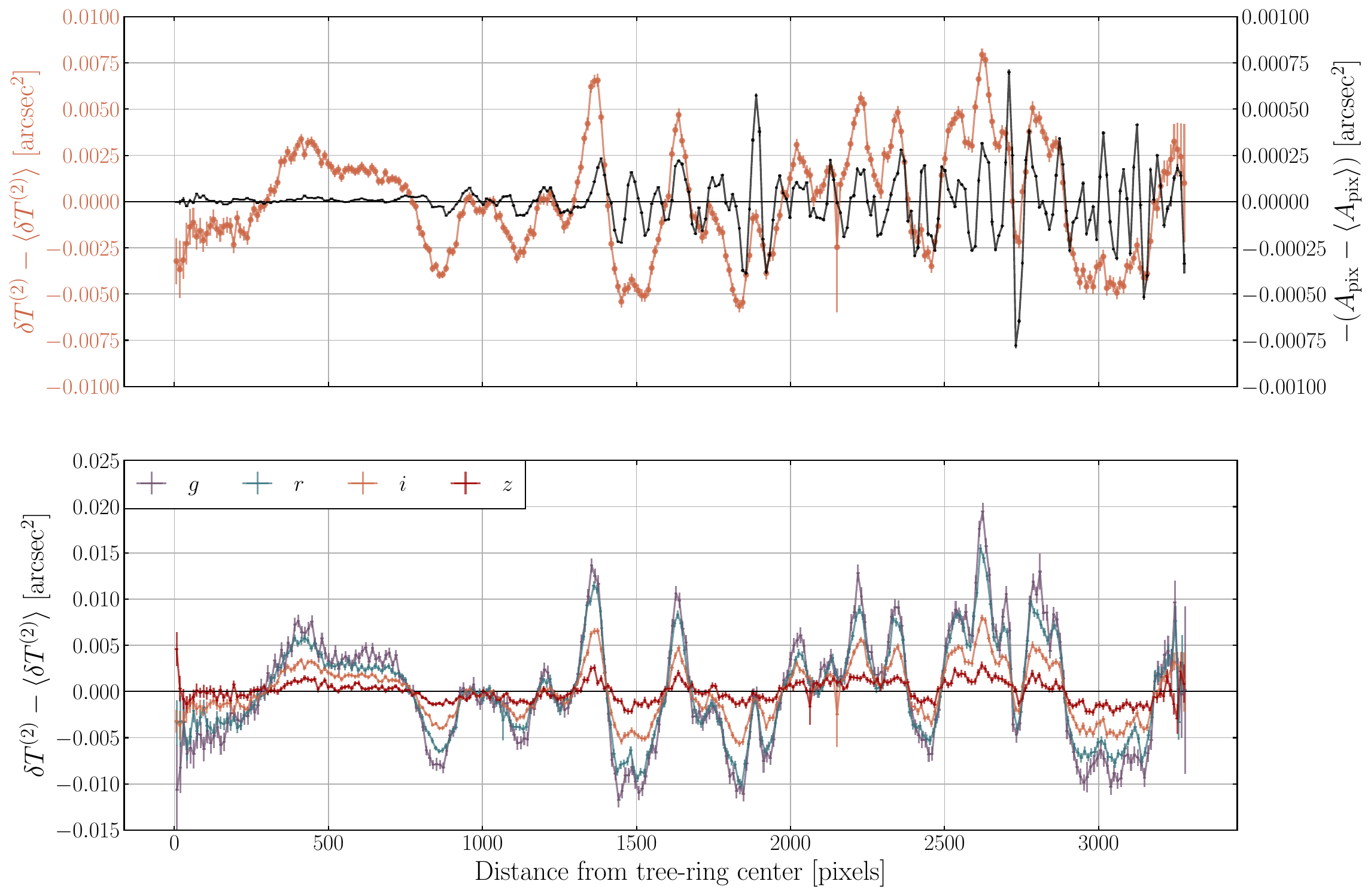}
\caption{
\textbf{Top:} PSF size residuals in $i$ band (orange) and fractional change in pixel area before WCS correction (black) as a function of distance from the center of the tree ring anomaly in CCD S29, the top right CCD in the detector plane orientation shown in \fig{fig:fov_res_2order}.
The two curves have highly correlated features but do not correspond one-to-one as expected from purely astrometric tree rings.
Note the scale of the right vertical axis is 10 times smaller than that of the left axis.
\textbf{Bottom:} Radial PSF size residuals in the $griz$ bands (same colors as \fig{fig:fov_tanshear}).
The tree ring size residuals are highly correlated between bands and grow in magnitude for bluer bands.
In both panels, $\delta T^{(2)}$, averaged over the entire CCD in the given band, is subtracted to remove the contributions to PSF residuals that are globally band-dependent.
}
\label{fig:treerings}
\end{figure*}

The next notable feature of the focal plane PSF residuals is the presence of high frequency concentric circular patterns on several of the CCDs called ``tree rings'' \citep{Kotov10}.
The principal form of the tree rings appear in the astrometric distortion patterns and are caused by variations in the electric field direction in the detectors' bulk silicon.
As each tree-ring center correspond to the center of the silicon boule the detector's wafer was diced from, the rings are likely due to fluctuations in the environment as the boule was grown, causing variations in the dopant concentration in the bulk.
The doping variations lead to distortions of the electric field in the silicon.  

Image defects caused by distortions of the lateral field (i.e., perpendicular to the surface) have been well studied for the DECam detectors \citep{plazas14a, plazas14b} and other thick, deep-depletion CCD detectors used for wide-field imaging surveys \citep{Kamata14, Magnier18, Park20, Esteves23}.
Nonuniform lateral electric fields distort the boundaries of pixels, leading to adjacent pixels having more or less area than their neighbors.
When left uncorrected, the nonuniform pixel areas lead to biased measurements of flux, astrometry and shape.

In our PSF modeling procedure, the pixel area effect, referred to as ``astrometric tree rings'' hereafter, is already accounted for via the \pixmappy\ local WCS solution, which has a model of the location of these tree rings in each CCD.
Our expectation is that our use of this WCS solution should completely account for the pixel-area effect of the tree rings. 
However, as \fig{fig:fov_res_2order} demonstrates, significant residuals remain in the second-order fractional size residuals, though notably not in the ellipticity residuals nor in any fourth-order moment residuals.
We refer to these remaining rings as ``anomalous'' tree rings.
The same size residuals were also seen in the Y3 PSF diagnostics.
They were attributed to two (non-mutually-exclusive) potential causes: (i) leakage of the WCS solution into the model via improper application of the pixel area effect to the diffusion component of the PSF, which may respond differently to the electric fields than the astrometry does, and (ii) variations in the charge diffusion itself, which may also depend on the same dopant non-uniformities that cause the astrometric tree rings.

To investigate the source of these PSF errors further, we study the tree-ring structure in sensor S29 (top right sensor in \fig{fig:fov_res_2order}, called CCDNUM 1 in the PSF catalog), which shows the largest residual effect.
We first compare the signature of the astrometric tree rings on pixel area before WCS correction and the residual size variation from the anomalous tree rings.
The top panel of \fig{fig:treerings} shows this comparison for the mean-subtracted $i$ band size residuals (orange) and the negative mean-subtracted pixel areas, calculated from the local Jacobian of the \pixmappy\ solution (black, see \eqn{eq:wcs_trans}).
Both quantities are binned by radial distance from the tree-ring center.
The two curves are clearly correlated, especially at radii $>1000$ pixels, with some shared zero-crossings and common high-frequency oscillations.
However, the amplitude of the size residuals is much greater, by a factor of a few to more than an order of magnitude.
Part of the larger amplitude seems to be caused by lower frequency oscillations also associated with the tree ring center, but the matching high frequency oscillations are still a factor of a few greater in the size residuals.
That the amplitude is so much greater for the size residuals, even the high-frequency modes, suggests that WCS leakage is not the primary cause of the anomalous tree rings.

The bottom panel shows the mean-subtracted anomalous tree rings split by band.
Both the low- and high-frequency oscillations are present in all bands, with the amplitude growing for shorter wavelengths.
In particular, the $g$ and $r$ bands are affected at a similar magnitude, and the effect quickly drops for the $i$ and $z$ bands.

Describing the Panoramic Survey Telescope and Rapid Response System (Pan-STARRS1) CCDs, \cite{Magnier18} suggest that variations in space-charge density within the silicon contribute to nonuniform levels of charge diffusion.
The isotropic Gaussian nature and wavelength dependence of DECam's size residuals are consistent with this interpretation.
When photoelectrons encounter areas of higher space-charge density, their drifting in the lateral electric field is slowed or interrupted, and they begin to diffuse \citep{Holland03}.
This diffusion results in purely Gaussian smearing of the measured surface brightness profile.
Further, shorter wavelength photons have a much shorter length scale for photoelectric conversion in silicon; photons in the $g$ and $r$ bands generally convert in the first 5-10 microns of the substrate while those in $i$ band convert at an intermediate depth of $\gtrsim 50$ microns \citep{Howell06}.
As such, short wavelength photons have more distance to diffuse and are particularly sensitive to undepleted regions close to the sensor's back side illumination surface.

We interpret the size residuals as being the consequence of a thin undepleted layer across the entirety of the CCD's back surface where there is little to no electric field.
Variations in the layer's thickness or space-charge density lead to more or less diffusion than average.
The $g$ and $r$ bands would be more significantly affected, but the $i$ and $z$ bands would still exhibit a nonzero effect due to the stochastic nature of photoelectric conversion.
This may also explain the low-frequency modes only present in the size residuals.
Pixel size variation requires a gradient in the space-charge density; thus, large regions of constant space-charge density could affect the PSF size without any corresponding change in pixel area.

In conclusion, the DECam sensors exhibit both astrometric and anomalous tree rings.
For most sensors, the astrometric tree rings are the dominant form and are well-corrected by the \pixmappy\ WCS solutions.
About 15\% of the science sensors exhibit anomalous tree rings with fractional size residuals $>\pm1.5\%$.

With the planned improvements to \piff\ discussed in \sect{sec:future}, variations in the diffusion strength could be explicitly included in the PSF model.  
Thus, if such an effect is seen in future experiments such as LSST, the \piff\ PSF model could account for it.

%%%%%%%%%%%%%%%%%%%%%%%%%%%%%%%%%

\subsection{Residuals by magnitude}
\label{sec:magres}

\begin{figure}
\includegraphics[width=\columnwidth]{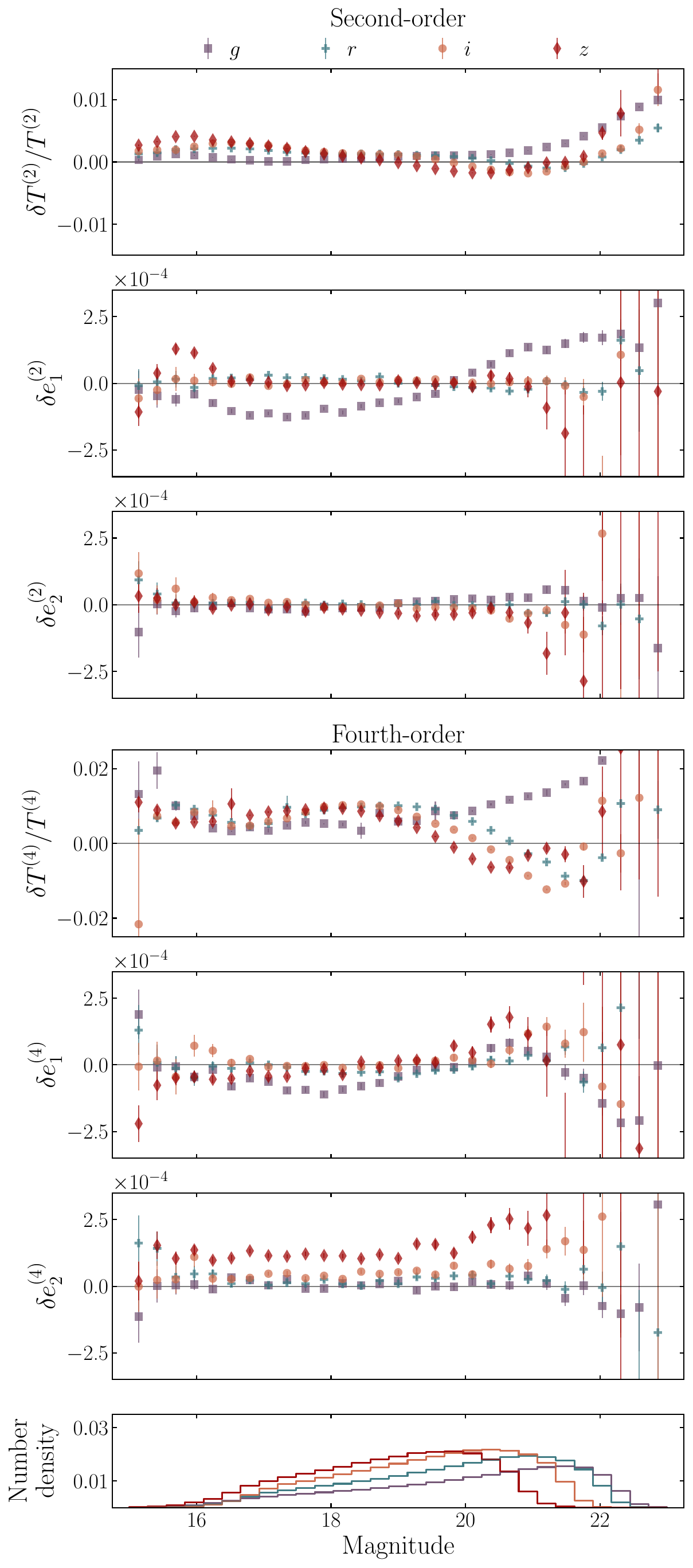}
\caption{
Second- and fourth-order PSF size and shape residuals as a function of their DES Y6 Gold catalog PSF magnitude, split by the $griz$ bands.
Markers for each band are the same as in \fig{fig:fov_tanshear}.
The bottom panel shows the histogram of stars in the same magnitude binning.
Unless otherwise noted, all histograms in this section are normalized by dividing the counts in each bin by the total number of stars in the $griz$ bands.
}
\label{fig:mag_bandsplit_2order}
\end{figure}

\begin{figure}
\includegraphics[width=\columnwidth]{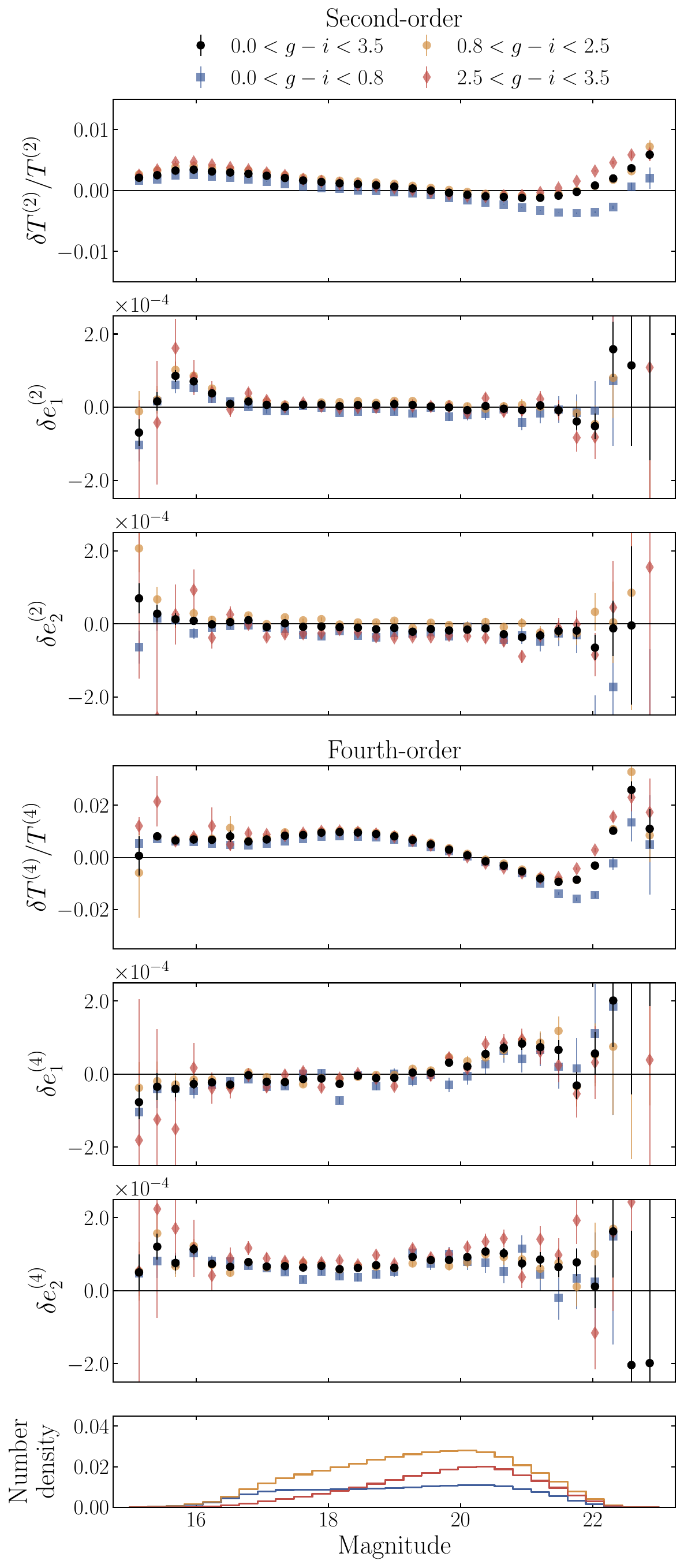}
\caption{
Second and fourth-order PSF moment residuals as a function of their magnitude for the $riz$ bands combined.
Stars are split into the same color quartiles as in \fig{fig:fovrad} to measure color dependence.
}
\label{fig:mag_gisplit_2order}
\end{figure}

In this section we study the PSF size and shape residuals as a function of apparent stellar magnitude.
The primary effect we expect to be modulated by star magnitude is the so-called ``brighter-fatter effect,'' which causes brighter sources to look larger due to charge deflection close to the CCD pixel gates.
As discussed in \sect{sec:data}, this effect is corrected to about 90\% during image processing \citep{imageproc}.
Further, the cut on bright stars described in \sect{sec:data} also mitigates this issue.
The following analysis probes the effectiveness of this correction and selection cut and uncovers further unknown effects that are magnitude dependent.

\fig{fig:mag_bandsplit_2order} shows for each band the average $\delta T^{(2)}/T^{(2)}$, $\delta e^{(2)}_1$ and $\delta e^{(2)}_2$ residuals as a function of magnitude.
The magnitude for each star is its Y6 Gold magnitude in that band.
The fractional size residuals at the bright end are small;
after brighter-fatter effect correction and the magnitude selection cut, we do not observe significant residuals due to the brighter-fatter effect.
However, there are size residuals at the faint end, up to $\sim 1\%$ in the $g$ and $z$ bands.
This is potentially a selection effect and not a real flux dependence of the PSF.
In the shape residuals, $g$ band is a clear outlier, with a distinct trend in $\delta e^{(2)}_1$.
We note, however, that the $g$ band residuals have approximately the same amplitude as the mean $\delta e^{(2)}_2$ trend averaged across the $riz$ bands found in \citetalias{y3-piff}.

In the fourth-order residuals shown in \fig{fig:mag_bandsplit_2order}, $g$ band continues to exhibit outlier behavior, with increasingly positive size residuals at faint magnitudes and similar trends between $\delta e^{(2)}_1$ and $\delta e^{(4)}_1$.
The causes for these trends remain unknown.
Across all bands, $\delta T^{(4)}/T^{(4)}$ remains relatively small at the bright end.
The $riz$ bands' size residuals show a common magnitude-dependent trend, dipping to negative residuals at intermediate magnitudes approximately where the samples in each band peak in number density.
Notably, the $\delta e^{(4)}_2$ residuals are mostly all positive;
here, $z$ band has significantly larger residuals than the other bands, though these are not strongly magnitude-dependent.
The positive $\delta e^{(4)}_2$ residuals correspond to the northwest/southeast direction; the cause is also unknown.
However, we again note that these residuals are very small.

\fig{fig:mag_gisplit_2order} shows the second- and fourth-order residuals for the $riz$ bands combined as a function of magnitude (black).
The sample is also split into three $g-i$ color bins, using the same 25th/75th percentile splitting as in \fig{fig:fovrad}.
The mean $\delta T^{(2)}/T^{(2)}$ and $\delta e^{(2)}_1$ are similar to that of the Y3 analysis (see Figure 10 of \citetalias{y3-piff}), but the mean $\delta e^{(2)}_2$ residuals are improved from Y3 by a factor of a few.
As expected, the improvements to the color-dependent residuals are much more pronounced, with an improvement of nearly an order of magnitude, particularly for $\delta T^{(2)}/T^{(2)}$ and $\delta e^{(2)}_2$.
The fractional size residuals at faint magnitudes are clearly color-dependent, though they follow the opposite trend from the naive expectation that blue stars would be larger than the model.

%%%%%%%%%%%%%%%%%%%%%%%%%%%%%%%%%

\subsection{Residuals by sky position}

\subsubsection{Differential Chromatic Refraction (DCR)}
\label{sec:sky:dcr}

\begin{figure*}
\centering
\includegraphics[width=\columnwidth]{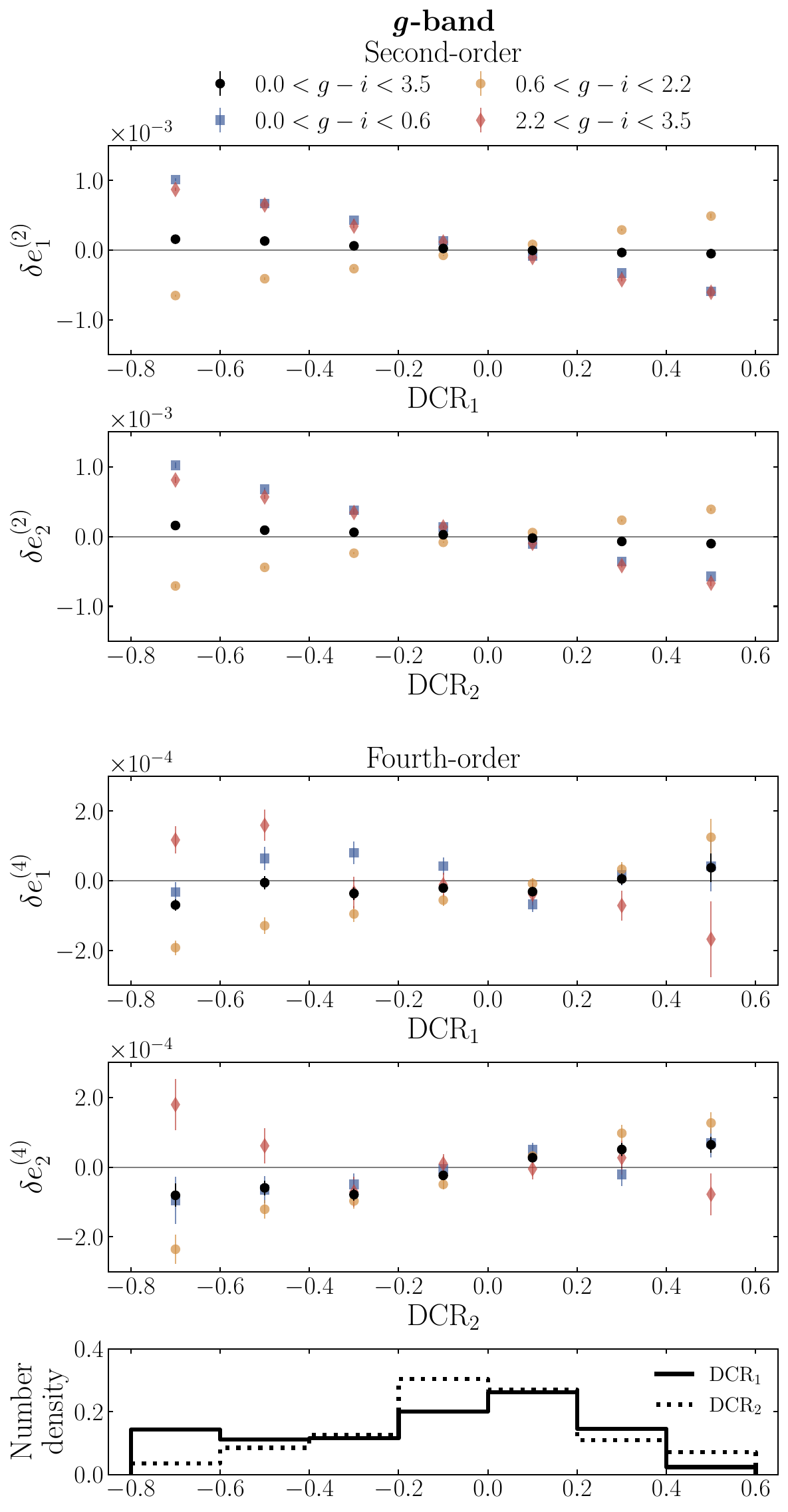}
\hspace{10pt}
\includegraphics[width=\columnwidth]{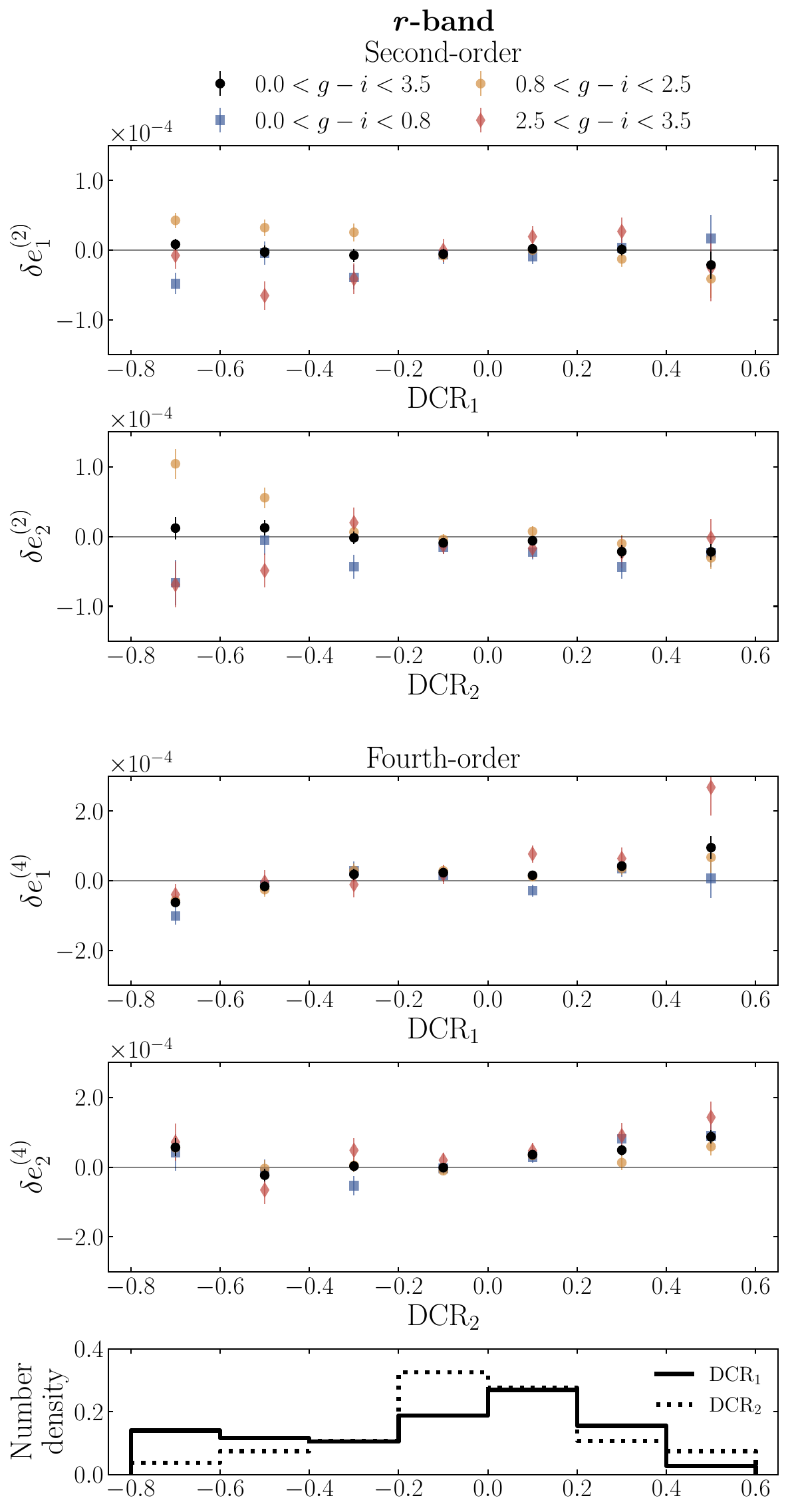}

\caption{
Second and fourth-order PSF shape residuals as a function of the mean direction and magnitude expected for DCR, defined in \eqnb{eq:dcr1}{eq:dcr2}.
As the magnitude of DCR increases for shorter wavelengths, the effect is particularly strong in $g$ band; we show residuals for $g$ band (left panels) and $r$ band (right panels) for comparison.
The four sets of points correspond to the same $g-i$ color quantile cuts as used in \fig{fig:mag_gisplit_2order}.
Note the $y$-scale for the $g$-band second-order residuals is $\sim$10 times larger than for the other panels.
The bottom panels show the histogram of stars in the same binning of DCR$_1$ and DCR$_2$, normalized by dividing the counts in each bin by the total number of stars in that band.
With the color-dependent PSF interpolation, $g$-band residuals are reduced to nearly the same magnitude as the Y3 $r$-band DCR residuals, and the Y6 $r$-band residuals are a further order of magnitude smaller.
}
\label{fig:dcr_2order}
\end{figure*}

Here we give a brief overview of the DCR effect, but refer readers to, e.g., \citet{Plazas12, Meyers15b} and references therein for a more thorough treatment.

As light propagates from vacuum into Earth's atmosphere, the change in refractive index induces refraction toward zenith.
The refractive index of air is also wavelength dependent; thus, light experiences dispersion at the vacuum/air boundary, with the amount of dispersion decreasing nonlinearly with wavelength.
Over a broadband filter with fixed bandpass width, e.g., 100~nm, the difference in refraction angle for wavelengths at the edge of the bandpass is greater for a blue filter than a red one (e.g., $g$ band versus $z$ band).
Thus, given two point sources of different color, there will be a greater difference in the angular dispersion (and thus induced ellipticity) of the two objects when observed in $g$ band than in $z$ band.

The dispersion happens in the direction toward zenith, i.e., in the observatory's local coordinates.
Thus, a transformation is necessary to measure the effect of DCR along the ellipticity directions used for shear measurement, $e_1$ and $e_2$, which are defined with respect to the equatorial coordinate system.
The transformation projecting the expected direction and magnitude of the DCR effect onto the $e_1$ and $e_2$ directions is described by two numbers
\begin{linenomath*}
\begin{align}
\mathrm{DCR}_1 &\equiv \tan^2(z) \cos(2 q) \label{eq:dcr1} \\
\mathrm{DCR}_2 &\equiv \tan^2(z) \sin(2 q), \label{eq:dcr2}
\end{align}
\end{linenomath*}
where $z$ is the zenith angle and $q$ is the parallactic angle.
The DCR magnitude scales as $\tan^2(z)$, and the $\cos(2q)$ and $\sin(2q)$ terms project the unit vector toward zenith onto the $e_1$ and $e_2$ directions, respectively.
To calculate DCR$_1$ and DCR$_2$ for each star, we calculate the local sidereal time for each exposure and then derive $z$ and $q$ for each star's right ascension and declination.

The PSF model residuals as a function of DCR$_1$ and DCR$_2$ are shown for the $g$ and $r$ bands in \fig{fig:dcr_2order}.
For $g$ band, the maximum amplitude of $\delta e_1$ and $\delta e_2$ is $\sim10^{-3}$.
Compared to the measured raw DCR effect (shown in \app{app:dcr}), the effect is corrected by $\sim80\%$.
The linear trend seen in the Y3 $r$ band PSF model residuals (see Figure 11 in \citetalias{y3-piff}) is removed.
The remaining residuals have a quadratic trend in color, with the reddest and bluest stars exhibiting similarly signed ellipticity residuals, opposite of stars in the central 50\% of the color distribution.
The Y6 $g$ band residuals have an absolute value similar to those of $r$ band in Y3, an improvement of about an order of magnitude.

The $r$ band residuals have a maximum amplitude of $\sim10^{-4}$ in $\delta e_1$ and $\delta e_2$.
Compared to the raw DCR signal, the DCR effect is corrected at the $>90\%$ level.
The $r$ band residuals are an order of magnitude smaller than in Y3, and the remaining residuals have a quadratic color dependence, similar to $g$ band (though curiously opposite in sign).
Note that the uncorrected DCR signal and PSF residuals are very similar in the $i$ and $z$ bands (not shown), suggesting the DCR$_1$ and DCR$_2$ quantities do not completely isolate the DCR signal, as we would expect it to be smaller in $i$ and $z$ bands.

For the fourth-order DCR residuals, the $g$ and $r$ bands have similar order of magnitude residuals.
Interestingly, unlike the second-order moments where the signal of all stars combined is consistent with zero, there is a trend in the mean signal in $\delta e_2$.
The $r$ band residuals display very little color dependence besides that red stars have more positive residuals compared to the mean, also seen in \fig{fig:mag_bandsplit_2order}.

We note the majority of stars have $|\mathrm{DCR}_i|<0.4$ where residuals across all bands are very small.

%%%%%%%%%%%%%%%%%%%%%%%%%%%%%%%%%

\subsubsection{Residuals in equatorial coordinates}
\label{sec:sky:radec}

\begin{figure*}
\centering
\includegraphics[width=0.90\columnwidth]{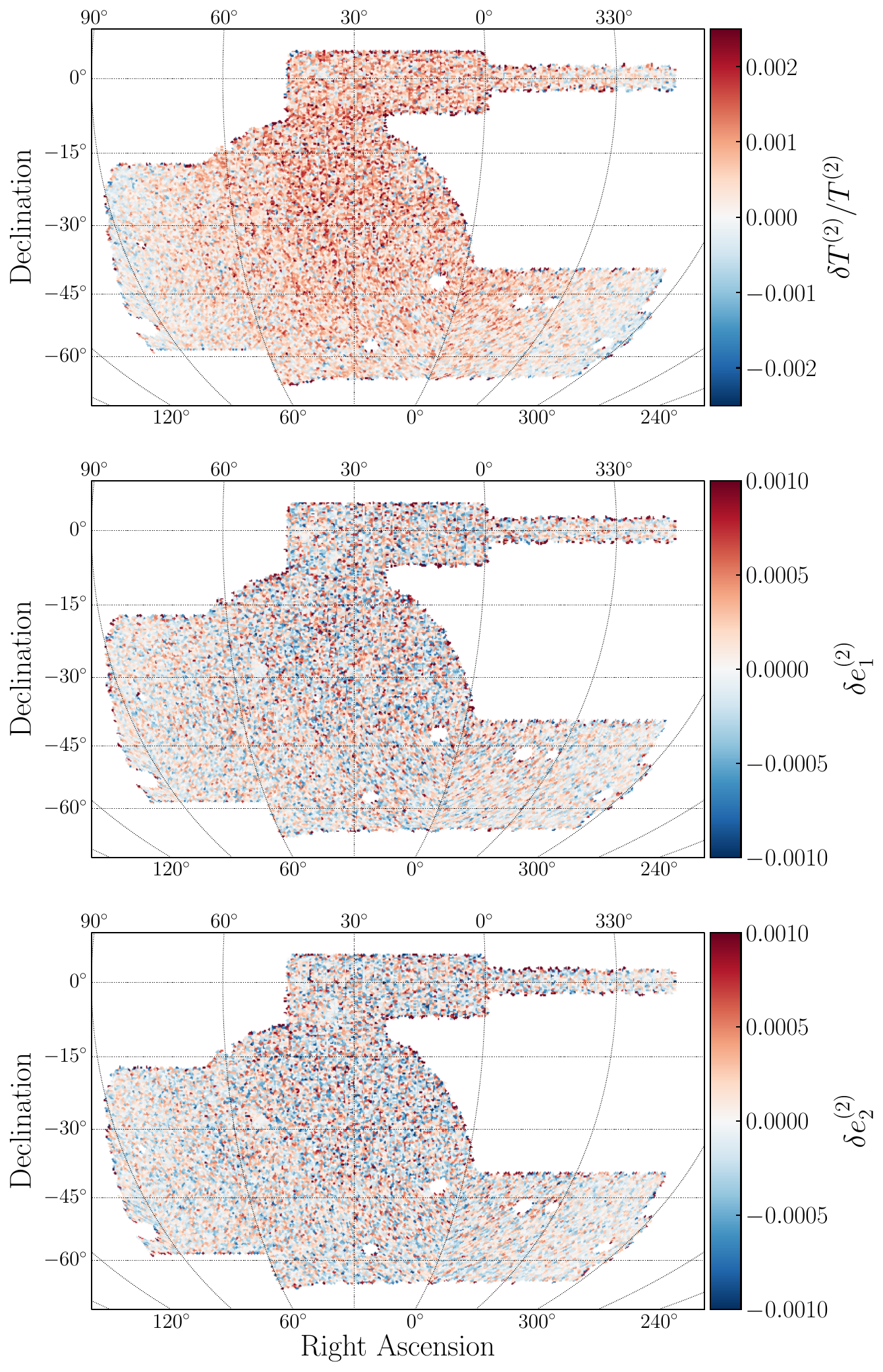}
\hspace{10pt}
\includegraphics[width=0.90\columnwidth]{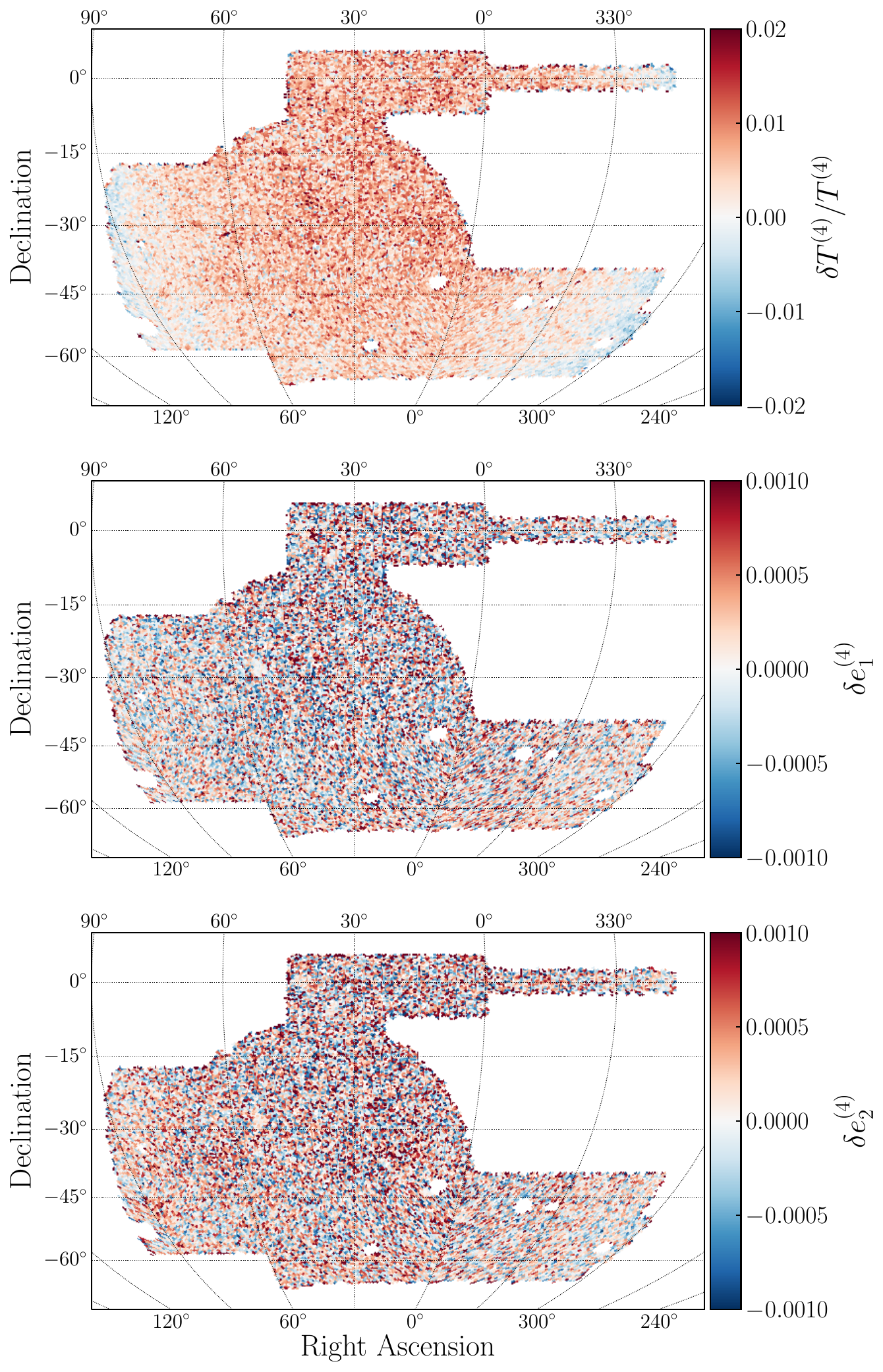}
\caption{
Maps of the second-order (left panels) and fourth-order (right panels) PSF moment residuals as a function of sky position for the $riz$ bands combined.
}
\label{fig:radec_2order}
\end{figure*}

Next, we investigate the variation of the PSF errors as a function of sky position in celestial coordinates.
Despite best efforts in observing, the survey exhibits nonuniformities across the survey footprint in, e.g., seeing, depth, etc., that might affect PSF quality.
Further, there are relevant astrophysical variations across the sky such as stellar density, which is higher near the edges of the footprint closer to the Galactic Plane to the East and West, and mean stellar color, which is also not randomly distributed.

The PSF size and shape residuals over the survey footprint are shown in \fig{fig:radec_2order}.
The residuals of $e_1$ and $e_2$ are consistent with random noise.
The variance of these residuals is lower in the areas of higher stellar density where the PSF model is more tightly constrained.

On the other hand, the second- and fourth-order size residuals have a small but significant trend correlated with stellar density.
The size residuals are biased positive across most of the footprint, particularly in areas of low stellar density, and become negative in regions of high stellar density.
By the sign convention of our residuals, this means the models are slightly too large where there are many stars and vice versa.
We find this is also present in the residuals of the stars used for PSF fitting (not shown), so this is not related to the lack of outlier rejection in the reserve stars.
We also find no significant correlation with the spatial distribution of mean stellar color.

The negative residuals in high stellar density regions may be caused by the model inferring a larger PSF due to a higher background level from faint, unresolved stars.
However, it is unclear how lower stellar density results in positive residuals.
In any case, the effect is quite small: a few tenths of a percent in $\dTt$ and $\sim 1\%$ in $\dTf$.
We leave further investigation to future work.

%%%%%%%%%%%%%%%%%%%%%%%%%%%%%%%%%

\subsection{Residuals by color}
\label{sec:colorres}

As discussed in \sect{sec:piff_color}, the PSF is known to be color dependent. The color dependence of the PSF is explored in several of the previous diagnostics presented in this work; in this section we inspect the PSF model residuals as a function of color explicitly.
The PSF models in Y6 include first-order color interpolation using $g-i$ color for $gri$ bands and $i-z$ color for $z$ band.

\begin{figure*}
\centering
\includegraphics[width=\columnwidth]{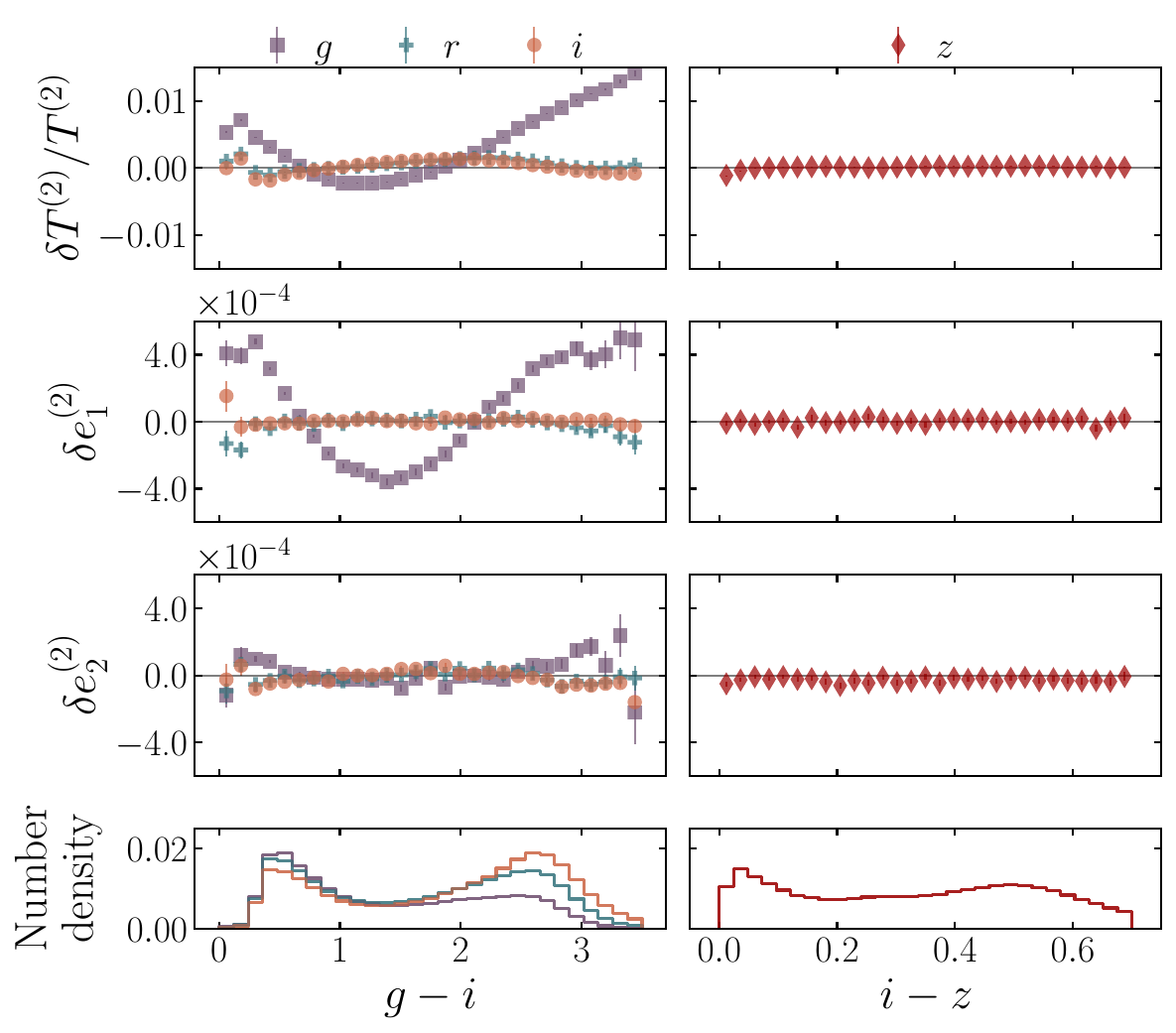}
\hspace{10pt}
\includegraphics[width=\columnwidth]{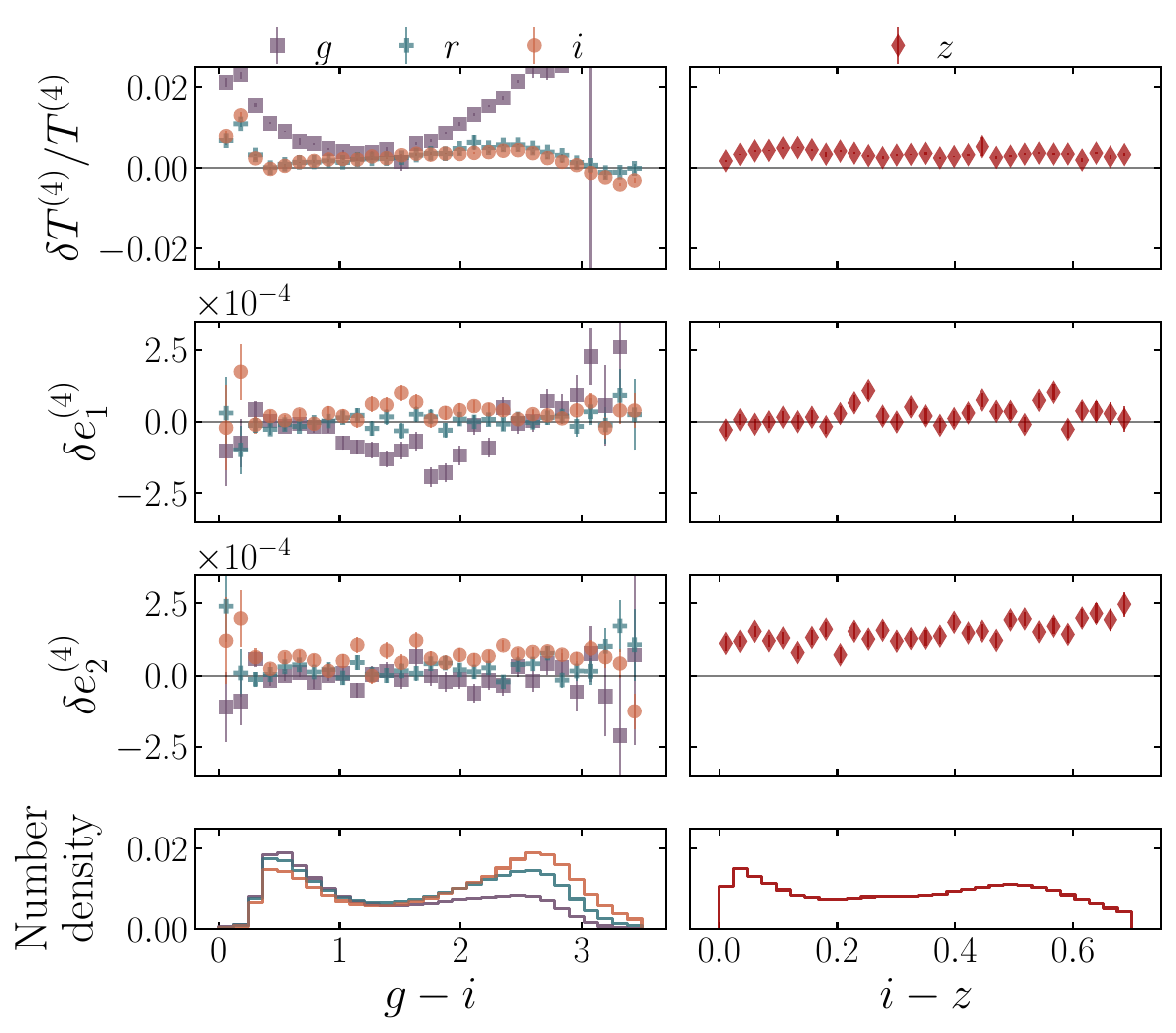}
\caption{
Second- and fourth-order PSF moment residuals as a function of $g-i$ (left column in set) or $i-z$ (right column in set) star color. The second-order residuals are shown in the left set of panels, and the fourth-order residuals are in the right set.
Markers for each band are the same as in \fig{fig:fov_tanshear}.
}
\label{fig:colorres_2order}
\end{figure*}

\fig{fig:colorres_2order} shows the PSF residuals for reserve stars observed in each band separately.  The left set of panels shows the second-order residuals, and the right set shows the fourth-order residuals.
They are plotted as a function of that band's interpolation color, i.e., $g-i$ for $gri$ (left column) and $i-z$ for $z$ (right column).
As expected, the color interpolation removes the linear component of the color dependence in each band.
The $riz$ bands' color dependence is well described by the linear model to the sub-percent level.
$g$ band exhibits residual higher-order color dependence at the $\sim 1\%$ level in \dTt\ and $\sim 0.0004$ for \deto, which follow an obvious quadratic pattern.
As is also seen in the magnitude-dependent residuals, $g$ band's color-dependent shape errors mainly manifest in \deto, not \dett.
The cause of this remains unclear.
Compared to the Y3 PSF models, \dTt, \deto and \dett\ are improved by a factor of $\gtrsim 5$ for all bands.

The corresponding fourth-order residuals in the right panels of \fig{fig:colorres_2order} also show $g$ band to be a clear outlier with \dTf$\sim 2-3\%$ and $\defo\sim 0.0002$, the latter particularly for intermediate $g-i$ values.
However, considering the size of these residuals and those shown in the previous subsections, we deem the $g$ band PSF modeling likely sufficient for photometric measurements, though further testing of this application will be required for analyses using the $g$ band models.

We note that these color-dependent diagnostics among the reserve stars do not test the effect of interpolating for a sample with a different color distribution from the stars -- namely,  galaxies. We explore this concern further with respect to measuring the galaxy-PSF two-point correlation functions in \app{app:weights}, where we define a novel weighting scheme for the PSF catalog to better reflect the PSF errors that galaxy images (as opposed to star images) will incur.

%%%%%%%%%%%%%%%%%%%%%%%%%%%%%%%%%

\subsection{Two-point correlation functions of PSF second- and fourth-order moments}
\label{sec:rho}

Considering that the most common statistic used to study cosmic shear is its two-point angular correlation functions $\xi_{\pm}(\theta)$, a very important diagnostic for the PSF modeling is to measure the two-point correlation functions of the PSF shapes and their residuals.
These spatial correlations will contribute an additive bias to $\xi_+$ and $\xi_-$.
In this section, we review the established formalism for defining two-point correlation functions of spin-2 PSF quantities, using both second- and fourth-order moments, and present the measurement of the resulting 36 PSF-PSF correlation functions, commonly known as $\rho$ statistics.
We note all references to $\rho$ statistics in the main text correspond to the $\xi_+$ component of the shear-shear correlation measurement.
We find the $\xi_-$ components to be negligible but present them in \app{app:alltherhos} for completeness.

%%%%%%%%%%%%%%%%%%%%%%%%%%%%%%%%%

\subsubsection{Formalism}
\label{sec:rho:formalism}

We refer the reader to \citet*{y6-mdet} and \cite*{y6-bfd} for full details on the estimation of PSF contamination on the Y6 galaxy shape catalogs and simply review the formalism here as it motivates measuring two-point correlation functions of certain PSF quantities detailed below.

As noted in \sect{sec:def_moms}, since cosmic shear is a spin-2 field, PSF quantities that are spin-2 or a product of spin-0 and spin-2 quantities can bias shear measurements.
The relevant quantities to study established in the literature \citep{paulinhenriksson08, Rowe10, sv-shearcat, Zhang22} are the second- and fourth-order PSF shape $\bm{e}$, shape residual after PSF model subtraction $\delta \bm{e}$ and the product of the PSF shape and the dimensionless spin-0 fractional size error $\bm{e}\frac{\delta T}{T}$.
To simplify notation we define these as
\begin{equation}
\left\{
    \begin{array}{rcl}
        \bm{p}_i &\equiv& \bm{e}^{(i)} \\
        \bm{q}_i &\equiv& \delta \bm{e}^{(i)} \\
        \bm{w}_{ij} &\equiv& \bm{e}^{(i)}\frac{\delta T^{(j)}}{T^{(j)}}
    \end{array}
  \right.,\, i,j=\{2,4\},
  \label{eq:psf_pqw}
\end{equation}
where $i, j$ denote the use of second- or fourth-order moments.
We model PSF contamination of an observed galaxy shape $\bm{g}_{\rm obs}$ as a sum of the true galaxy shape (including lensing-induced shear) and a net systematic residual $\delta\bm{g}_{\rm sys}$:
\begin{equation}
    \bm{g}_{\rm obs} = \bm{g}_{\rm true} + \delta\bm{g}_{\rm sys}.
    \label{eq:g_obs}
\end{equation}
We can expand $\delta\bm{g}_{\rm sys}$ as a sum of contributions by the second- and fourth-order spin-2 PSF quantities given in \eqn{eq:psf_pqw}
\begin{equation}
    \delta\bm{g}_{\rm sys} = \sum_{k=1}^8 c_k \bm{P}_k,
\label{eq:psfcontam}
\end{equation}
where $\bm{P}$ is the vector of PSF shape quantities $\{\bm{p}_2, \bm{q}_2, \bm{w}_{22}, \bm{p}_4, \bm{q}_4, \bm{w}_{24}, \bm{w}_{42}, \bm{w}_{44}\}$ and $\bm{c}$ is the vector of unknown model parameters $\{\alpha_2, \beta_2, \eta_{22}, \alpha_4, \beta_4, \eta_{24}, \eta_{42}, \eta_{44}\}$, which must be inferred from the data.
As the intrinsic galaxy shapes are uncorrelated with the PSF, we can measure the impact of the PSF on the measured galaxy shapes by correlating the shape catalog with each PSF quantity
\begin{equation}
    \langle \bm{g}_{\rm obs} \bm{P}_k \rangle = \underbrace{\langle \bm{g}_{\rm true} \bm{P}_k \rangle}_{\normalsize =0} +  \langle \delta\bm{g}_{\rm sys} \bm{P}_k \rangle.
\end{equation}

Substituting the r.h.s. of \eqn{eq:psfcontam} for $\delta\bm{g}_{\rm sys}$ results in a system of eight equations relating galaxy-PSF correlations with PSF-PSF correlations
\begin{equation}
    \langle \bm{g}_{\rm obs} \bm{P}_k \rangle = \sum_{l=1}^8  c_l \langle \bm{P}_k \bm{P}_l \rangle.
    \label{eq:tau_rho}
\end{equation}
\Eqn{eq:tau_rho} provides a way of estimating the values of $\bm{c}$, thus determining the impact of the PSF modeling on galaxy shear estimation.
As the $\bm{c}$ parameters are specific to a galaxy shape catalog, we refer to \citet*{y6-mdet} and \cite*{y6-bfd} for the PSF contamination modeling determining $\bm{c}$ for the \mdet\ and Bayesian Fourier Domain (BFD) catalogs, respectively.
Below we explore the features of the PSF-PSF correlations $\langle \bm{P}_k \bm{P}_l \rangle$.

%%%%%%%%%%%%%%%%%%%%%%%%%%%%%%%%%

\subsubsection{Measured $\rho$ statistics}

\begin{figure*}
\centering
\includegraphics[width=0.95\textwidth]{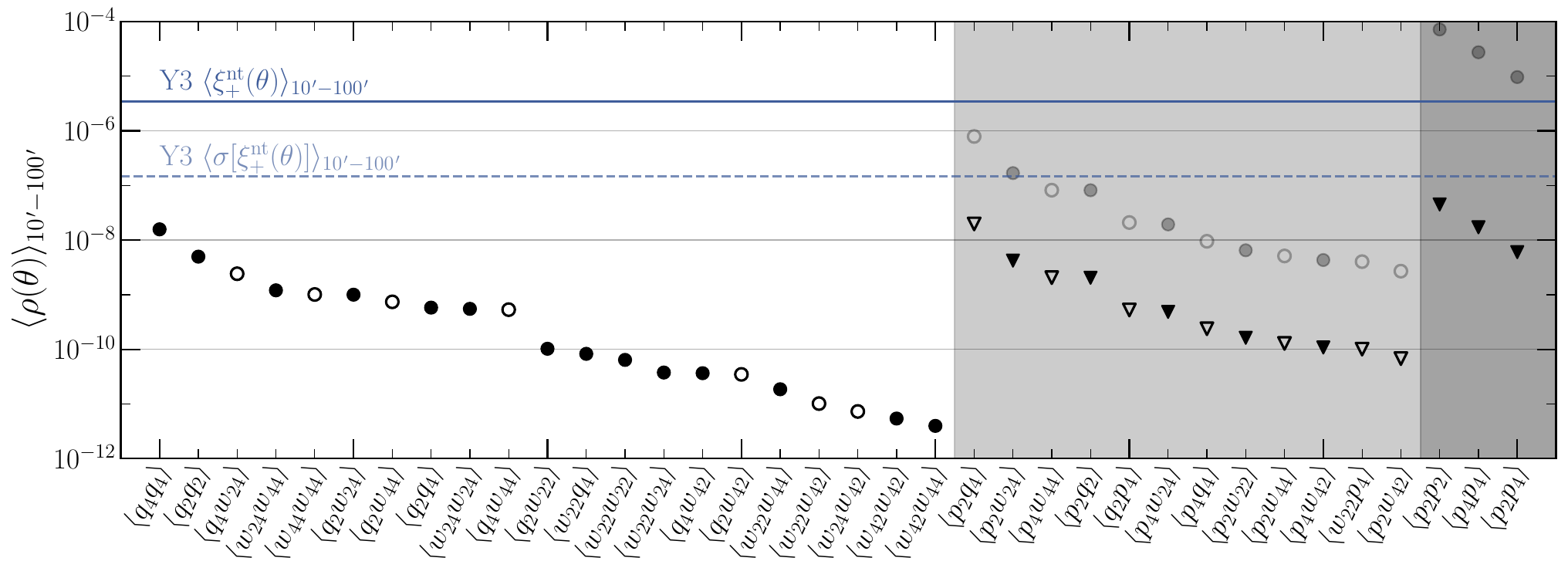}
\caption{
Comparison of all 36 PSF-PSF correlation functions, defined as $\langle \bm{P}_k \bm{P}_l \rangle$ in \eqn{eq:tau_rho}, for the $riz$ bands combined.
To form a point-estimate, the correlation functions are averaged over 10\arcmin-100\arcmin, approximately where the cosmic shear signal is most significant/informative.
Filled (open) markers correspond to positive (negative) values.
The Y3 non-tomographic $\xi_+^{\rm nt}$ (dark blue line) and standard deviation $\sigma[\xi_+^{\rm nt}]$ (light blue dashed line), both averaged over 10\arcmin-100\arcmin, are shown for an order-of-magnitude comparison.
The contamination of the shear measurement due to each function correlating one or two $p$ quantities (e.g., $\langle \bm{p}_2 \bm{q}_4 \rangle$ or $\langle \bm{p}_2 \bm{p}_4 \rangle$) is modulated by a factor of $\alpha$ or $\alpha^2$, respectively (see \eqn{eq:gsys_gsys}).
These subsets of the correlation functions are denoted by the light grey and dark grey shaded regions, respectively.
The grey circle markers show the ``raw'' measurement; the triangle markers show these correlations multiplied by the 1$\sigma$ upper limit of $\alpha = 0.025$ found in \citet*{y6-mdet}, e.g., $\alpha\langle \bm{p}_2 \bm{q}_4 \rangle$ or $\alpha^2\langle \bm{p}_2 \bm{p}_4 \rangle$.
}
\label{fig:rho_point_est}
\end{figure*}

\begin{figure*}
    \centering
    \includegraphics[width=\textwidth]{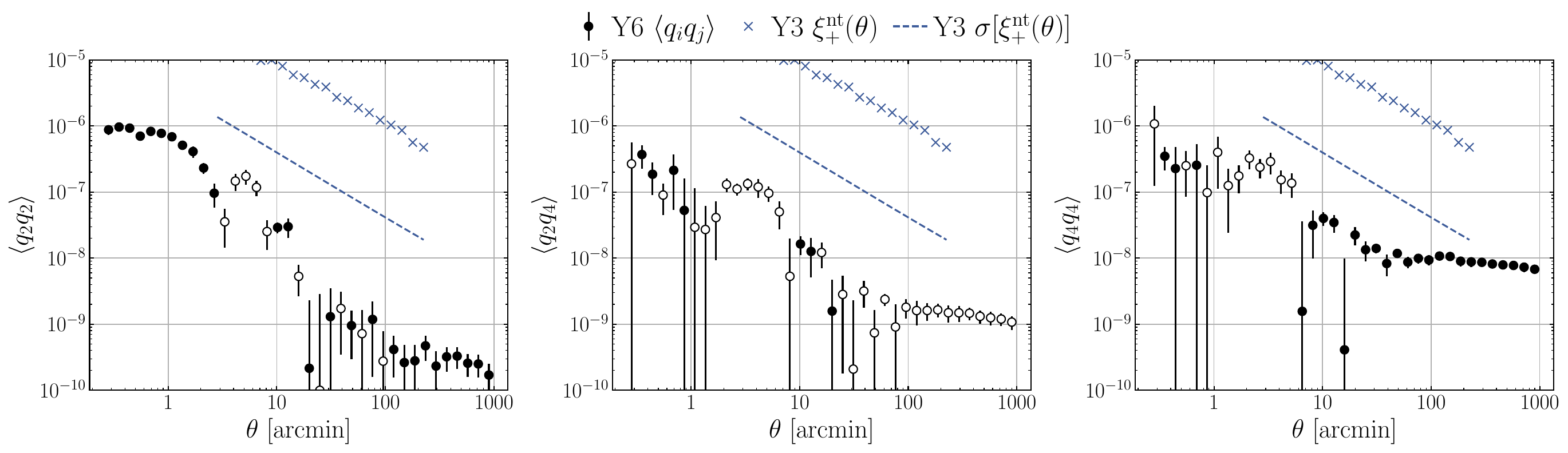}
    \caption{
        Shear-shear correlation functions of the $riz$ second- and fourth-order PSF shape residuals, $\bm{q}_2$ and $\bm{q}_4$ (black circles).
        Filled (open) markers correspond to positive (negative) values.
        The Y3 non-tomographic $\xi_+^{\rm nt}(\theta)$ signal (blue $\times$s) and standard deviation $\sigma[\xi_+^{\rm nt}(\theta)]$ (blue dashed line) are shown for an order-of-magnitude comparison.
        }
    \label{fig:qq_rhos}
\end{figure*}

The auto- and cross-correlations of the eight PSF quantities in $\bm{P}$ result in 36 two-point correlation functions.
In this work, we collectively call this set of functions ``$\rho$ statistics'' in keeping with the established literature but specify individual correlation functions by the two quantities being correlated (e.g., $\langle \bm{q}_2\bm{q}_2 \rangle$).
All correlation functions in this section and its associated appendices are measured using \textsc{treecorr}\footnote{\url{https://github.com/rmjarvis/TreeCorr}} \citep{Jarvis04, TreeCorr}.

To determine the relative impact of individual $\rho$ statistics, we define a one-point estimate of the magnitude of each correlation function: the mean of the function averaged over the angular separations between $\sim10'-100'$, approximately where the cosmic shear signal is most significant or informative.
These point estimates are shown in \fig{fig:rho_point_est} for all 36 $\rho$ statistics.
The functions are split into three groups: those with zero, one or two factors of $\bm{p}_2$ or $\bm{p}_4$ (white, light grey and dark grey backgrounds, respectively).
The motivation for this grouping is as follows.
Cosmic shear is the measurement of $\langle \bm{g}_{\rm obs}\bm{g}_{\rm obs} \rangle$ with the PSF bias contribution of $\langle \delta\bm{g}_{\rm sys}\delta\bm{g}_{\rm sys} \rangle$ by \eqn{eq:g_obs}.
Following \eqn{eq:psfcontam},
\begin{equation}
    \langle \delta\bm{g}_{\rm sys}\delta\bm{g}_{\rm sys} \rangle = \sum_k \sum_l c_k c_l \langle \bm{P}_k \bm{P}_l \rangle.
    \label{eq:gsys_gsys}
\end{equation}
Thus, the contribution of any PSF-PSF correlation function  $\langle \bm{P}_k \bm{P}_l \rangle$ to the cosmic shear signal is modulated by the coefficients $c_k c_l$.
The key point here is that, for shear estimation methods such as \mdet\ and BFD, the $\alpha_2$ and $\alpha_4$ parameters in $\bm{c}$, which describe the PSF ``leakage'' in the shear estimation, are of order $\sim 0.01$, about two orders of magnitude smaller than the $\beta$ and $\eta$ terms.
As such, when judging the relative impact of any given $\rho$ statistic on the shear measurement, these coefficients are important to take into account.
Hence, we separate $\rho$ statistics with one or two factors of $\bm{p}_i$ as they are modulated by a factor of $\alpha_i$ or $\alpha_i^2$, respectively, where $i$ denotes the second- or fourth-order version.

In \fig{fig:rho_point_est}, the ``raw'' point estimates of $\langle \bm{P}_k \bm{P}_l \rangle$ are plotted as circles.
For $\langle \bm{P}_k \bm{P}_l \rangle$ with one or two factors of $\bm{p}_i$, the raw measurement is plotted in grey; the measurement multiplied by the appropriate power of $\alpha$ is plotted as a black triangle.
Here, we use $\alpha=0.025$, the $1\sigma$ upper limit of $\alpha_2$ found in \cite*{y6-mdet}; thus, the triangle markers should be interpreted as approximate $1\sigma$ upper limits of the contribution to $\xi_+$.

For an order-of-magnitude comparison, the same point estimate is plotted for the Y3 non-tomographic $\xi_+^{\rm nt}$ as a solid blue line.
The average standard deviation of the $10\arcmin-100\arcmin$ bins $\sigma[\xi_+^{\rm nt}]$ is plotted as a light blue dashed line.
We use the Y3 non-tomographic $\xi_+$ signal as an estimate of the total statistical power of the full source galaxy sample; as such, the uncertainty is a conservative estimate and is lower than the expected Y6 $\xi_+$ error bars for any single redshift bin pairing.

$\langle \bm{q}_2\bm{q}_2 \rangle$ and $\langle \bm{q}_4\bm{q}_4 \rangle$ are the leading contributions of those correlations with no $\bm{p}_i$ factor.
This is expected as the $\bm{q}_i$ quantities, i.e., PSF model errors, directly add a bias to the shape measurement.
We study these in more detail below.
Interestingly, there are also significant contributions from correlations with the $\bm{w}_{ij}$ quantities, which were found to be negligible for the HSC Y3 analysis \citep{Zhang23b}.
Among those correlations with one or two $\bm{p}_i$ quantities, their contributions can be as or more significant than that of the PSF modeling errors, even with a very small leakage parameter of $\alpha=0.025$.
% As such, we note that even with perfect PSF modeling, it is important to optimize observing for conditions where the PSF ellipticity is minimized.
\edit{In the case of $\langle \bm{p}_i\bm{p}_j \rangle$, we note that observing conditions that impart large PSF ellipticity correlations can still negatively impact shape measurement and uncertainties even with perfect PSF modeling.}

Of course, \edit{averaging over 10\arcmin-100\arcmin\ largely} neglects the scale dependence of the correlation functions.
In \fig{fig:qq_rhos}, we show the full angular auto- and cross-correlation functions for the PSF residuals, $\bm{q}_2$ and $\bm{q}_4$ (black circles), measured over angular separations of $0.25'-1000'$.
For completeness, \app{app:alltherhos} shows the $\xi_+$ and $\xi_-$ components of all 36 $\rho$ correlation functions.

Overall, the $\rho$ statistics are all sufficiently small such that we do not expect them to significantly impact the Y6 cosmic shear measurements.
Compared to Y3, the $\langle \bm{q}_2\bm{q}_2 \rangle$ correlation (called $\rho_1$ in \citetalias{y3-piff}) is reduced by a factor of $\sim5-10$ at the angular separations of interest ($\sim10'-200'$) and presumably more at very large scales ($>250'$), which were not measured in the Y3 analysis.
This is likely due to reducing the mean residual PSF ellipticity caused by color-dependent bias.
The improvements at very small scales ($< 2'$) are more modest, with a reduction of $\sim10-20\%$.
On the other hand, $\langle \bm{q}_4\bm{q}_4 \rangle$ has significant power at large scales, with an amplitude of $\sim 10^{-8}$.
The cross-correlation $\langle \bm{q}_2\bm{q}_4 \rangle$ is an order of magnitude smaller and, interestingly, is negative.
Further, the anti-correlated feature between $1'-10'$ is present in all three $\langle \bm{q}_i\bm{q}_j \rangle$ correlations and is slightly enhanced compared to the Y3 $\langle \bm{q}_2\bm{q}_2 \rangle$.
This is likely due to the ringing seen in the focal plane shape residuals (\figb{fig:fov_res_2order}{fig:fov_res_4order}) caused by interpolating over the higher-order optical aberrations with second-order polynomials (third-order spatial polynomials were used in Y3).
We also plot the Y3 $\xi_+^{\rm nt}(\theta)$ signal (blue exes) and standard deviation (blue dashed line) for an order-of-magnitude comparison.
At worst, $\langle \bm{q}_4\bm{q}_4 \rangle$ at large scales approaches about 50\% of the $\xi_+^{\rm nt}$ uncertainty.
Otherwise, the $\rho$ statistics remain mostly at least an order of magnitude smaller than $\sigma[\xi_+^{\rm nt}]$.

We present further investigations into the $\rho$ statistics in several appendices.
\app{app:weights} studies how the color-dependent weighting scheme mentioned in \sect{sec:colorres} affects the $\rho$ statistics.
As mentioned previously, the full set of PSF-PSF correlations functions is shown in \app{app:alltherhos}.
Finally, the $\langle\bm{q}_i\bm{q}_j\rangle$ behavior in the separate photometric bands is presented in \app{app:rho_griz}.

%%%%%%%%%%%%%%%%%%%%%%%%%%%%%%%%%%%%%%%%%%%%%%%%%%%%%%%%%%%%%%%%%%%

\section{Planned Future Improvements}
\label{sec:future}

While the addition of color interpolation made a significant improvement in the quality of the PSF models for DES data from the Y3 to Y6 analyses,
there are still a number of upgrades planned for \piff, which we think will further improve the modeling for future data sets,
such as the Legacy Survey of Space and Time (LSST) with the Vera C. Rubin Observatory.

There has already been substantial effort in implementing a composite model into \piff.  This structure allows for multiple components to each
have a different functional form for the profile and a different interpolation scheme from the other components.  The multiple components can
be either convolved or summed (or a mix of both) as desired.  

The most obvious use case, which we are still working towards, is to model the PSF with three components: one describing the optical effects,
one modeling the atmosphere, and one modeling the diffusion in the sensors.
The optical component would use a double Zernike interpolation function to model the variation of the aberrations across the whole focal plane.
Furthermore, each CCD can have individual parameters setting its height and tilt relative to the average of the whole camera. 
The atmospheric component can use a simple von K\'arm\'an model using Gaussian process interpolation.  And finally, the diffusion can be modeled
with a simple Gaussian profile whose size varies spatially, capturing the tree ring patterns evident in \fig{fig:fov_res_2order}, but would be fixed for all exposures.

The optical and atmospheric components have already been shown to be effective for simulated data where they are the only relevant effects for the PSF.
We are still working on the system to robustly fit both components (or all three including the diffusion) on real data.

One significant advantage of this kind of composite model is that it will likely allow for more accurate modeling of the chromatic effects.
The physics of the wavelength-dependence in each component is fairly well understood, so we expect it to be more accurate to incorporate such
dependence directly in each component, rather than using a relatively crude linear interpolation in color as we have done for the Y6 analysis.
The requirements of the chromatic modeling for LSST will be much more stringent than for DES \citep{Meyers15b}, so we expect that the more
physically realistic modeling will be required to achieve sufficiently accurate PSF models.

There are a number of diagnostic tests that show small but nonzero systematic residuals in this analysis, which would need to be reduced
for LSST. We are hopeful that the above composite PSF model will be effective at reducing these residuals.

\begin{itemize}
\item There is a clear quadratic dependence with color in the $g$ band size residuals (\fig{fig:colorres_2order}).  We expect this could improve with more physically realistic chromatic modeling.
\item There are anomalous tree ring patterns in the size residuals, which seem to be constant in time and are believed to arise from variations in the diffusion due to doping variations in the sensors (\fig{fig:treerings}).  We plan to directly include this variation in diffusion strength in that component of the model.
\item The PSF error autocorrelations (both $\langle \bm{q}_2\bm{q}_2 \rangle$ and $\langle \bm{q}_4\bm{q}_4 \rangle$) show significant amplitude at scales below about 10\arcmin\ (\fig{fig:rho_point_est}).  This is largely due to the turbulent atmosphere, which we expect to be well modeled by the Gaussian process interpolation for that component.
\item The fourth-order size residual has a non-zero mean value.  This may be improved by including components whose functional forms extend beyond the current size of the \texttt{PixelGrid} model.  Both the optical and atmospheric components have profiles with wings that do not abruptly drop to zero, as the \texttt{PixelGrid} model currently does.  This may reduce the errors in the fourth order moments, since they are sensitive to the model at larger radii than the second order moments.
\item There are modestly larger size residuals in regions of higher stellar density (\fig{fig:radec_2order}).  We are less sure about the cause of this effect, but it may also be reduced by the improved interpolation of the atmospheric component.
\end{itemize}

%%%%%%%%%%%%%%%%%%%%%%%%%%%%%%%%%%%%%%%%%%%%%%%%%%%%%%%%%%%%%%%%%%%

\section{Conclusion}
\label{sec:conclusion}

In this work, we present the PSF models used for the DES Y6 WL analysis.
The models are constructed using the \piff\ software package, which has been updated since the version presented in \citetalias{y3-piff}.
The most notable improvement is the inclusion of interpolation with respect to color, which we find leads to significant improvements in the modeling.
This work also improves upon our selection of PSF stars by using \gaia, VHS and VIKING data to produce a more robust selection of isolated point sources.

Our improved diagnostic test suite now includes quantities based on fourth-order moments.
We find the fourth-order size and shape residuals often have similar amplitudes to the second-order residuals, and the two-point correlation function of fourth-order PSF residuals $\langle \bm{q}_4\bm{q}_4 \rangle$ is the highest amplitude $\rho$ statistic between $10\arcmin-100\arcmin$.
This is in agreement with other recent PSF modeling efforts for WL \citep{Zhang23b}, and we also conclude that fourth-order moments are very important to account for in validating the PSF and in any potential PSF contamination modeling.

We also now include more explicit tests of PSF quality with respect to color.
The size residuals from chromatic seeing and optical chromatic aberrations are improved by about an order of magnitude in $g$ band.
The shape residuals caused by DCR are corrected by $\sim80\%$ for $g$ band and $>90\%$ for the $riz$ bands; the $g$-band residuals from DCR now have an amplitude similar to the Y3 $r$-band residuals.
Compared to Y3, the two-point correlation function of second-order PSF residuals, $\langle \bm{q}_2\bm{q}_2 \rangle$ (called $\rho_1$ in \citetalias{y3-piff}), is reduced by a factor of $\sim5-10$ at angular separations of $\sim 10\arcmin-200\arcmin$ for the $riz$ bands combined.
This is likely due to reducing the mean residual PSF ellipticity caused by color-dependent bias.

The comprehensive set of diagnostics herein broadly demonstrates that the PSF models are sufficiently accurate for use in galaxy shear estimation.
However, it remains the case that the $g$-band PSF is the least accurate.
There is an evident quadratic dependence of the size residual on color.
It also shows somewhat larger $\rho$ statistics, especially at large scales.  
Thus, we recommend caution when using these PSF models for precise measurements of shapes in the $g$ band.
We believe, however, that the models are accurate enough to be used for $g$-band photometry, where these issues are not as problematic.

In this work, we also present an overview of ongoing \piff\ development.
While we are encouraged by the improvements made between the Y3 and Y6 modeling, we recognize that the model accuracy needs to be further improved for future surveys such as LSST.

%%%%%%%%%%%%%%%%%%%%%%%%%%%%%%%%%%%%%%%%%%%%%%%%%%%%%%%%%%%%%%%%%%%

\section*{Data Availability}

The \piff\ software is publicly available\footnote{\url{https://github.com/rmjarvis/Piff}} under an
open source license.
The PSF models presented in is paper were made using version 1.2.4.
Fourth-order diagnostic information was generated using version 1.3.3.
Installation instructions are found on the website, and we welcome
feature requests and bug reports from users.
The code we used to run \piff\ on DES images,
produce the PSF catalogs, and create all of the plots herein is also publicly available\footnote{\url{https://github.com/DarkEnergySurvey/despyPIFF} and \url{https://github.com/des-science/y6-psf-validation}}.
Catalogs of the PSF measurements on the reserve stars will be made available as part of
the DES Y6 coordinated release\footnote{\url{https://des.ncsa.illinois.edu/releases}}.

%%%%%%%%%%%%%%%%%%%%%%%%%%%%%%%%%%%%%%%%%%%%%%%%%%%%%%%%%%%%%%%%%%%

\section*{Acknowledgments}

We thank Roohi Dalal, Claire-Alice H\'ebert, Xiangchong Li, Shuang Liang and Tianqing Zhang for useful discussions.

TS is supported by the National Science Foundation Graduate Research Fellowship under Grant No. DGE-2146755.
TS thanks the LSST-DA Data Science Fellowship Program, which is funded by LSST-DA, the Brinson Foundation, and the Moore Foundation; their participation in the program has benefited this work.
TS and AR are supported at SLAC National Accelerator Laboratory under Department of Energy Contract No. DE-AC02-76SF00515.
MJ is partially supported by the U.S. Department of Energy grant DE-SC0007901 and funds from the University of Pennsylvania. Argonne National Laboratory’s work was supported under the U.S. Department of Energy contract DE-AC02-06CH11357.

Author Contributions:
TS performed almost all analysis and manuscript preparation and contributed to development of the \piff\ software package.
MJ contributed to manuscript preparation and \piff\ development and served as a project adviser.
AR contributed to analysis and manuscript preparation and served as a project adviser.
AA served as a project adviser.
MRB contributed to analysis and manuscript preparation. 
RG ran \piff\ on the full survey dataset and provided technical support for the analysis.
MY contributed to analysis.
KB, GB and MT contributed to manuscript preparation as collaboration internal reviewers.
GB, MG, ER, ES and MT contributed to analysis interpretation as members of the DES Y6 shear analysis team.
The remaining authors have made contributions to this paper that include, but are not limited to, the construction of DECam and other aspects of collecting the data; data processing and calibration; developing broadly used methods, codes, and simulations; running the pipelines and validation tests; and promoting the science analysis.

This document was prepared by the DES collaboration using the resources of the Fermi National Accelerator Laboratory (Fermilab), a U.S. Department of Energy, Office of Science, Office of High Energy Physics HEP User Facility. Fermilab is managed by Fermi Forward Discovery Group, LLC, acting under Contract No. 89243024CSC000002.

This work was based in part on observations at Cerro Tololo Inter-American Observatory,
National Optical Astronomy Observatory, which is operated by the Association of
Universities for Research in Astronomy (AURA) under a cooperative agreement with the National
Science Foundation.
This work was also based in part on data obtained from the ESO Science Archive Facility with DOI: \url{https://doi.org/10.18727/archive/57}.

Funding for the DES Projects has been provided by the U.S. Department of Energy, the U.S. National Science Foundation, the Ministry of Science and Education of Spain,
the Science and Technology Facilities Council of the United Kingdom, the Higher Education Funding Council for England, the National Center for Supercomputing
Applications at the University of Illinois at Urbana-Champaign, the Kavli Institute of Cosmological Physics at the University of Chicago,
the Center for Cosmology and Astro-Particle Physics at the Ohio State University,
the Mitchell Institute for Fundamental Physics and Astronomy at Texas A\&M University, Financiadora de Estudos e Projetos,
Funda{\c c}{\~a}o Carlos Chagas Filho de Amparo {\`a} Pesquisa do Estado do Rio de Janeiro, Conselho Nacional de Desenvolvimento Cient{\'i}fico e Tecnol{\'o}gico and
the Minist{\'e}rio da Ci{\^e}ncia, Tecnologia e Inova{\c c}{\~a}o, the Deutsche Forschungsgemeinschaft and the Collaborating Institutions in the Dark Energy Survey.

The Collaborating Institutions are Argonne National Laboratory, the University of California at Santa Cruz, the University of Cambridge, Centro de Investigaciones Energ{\'e}ticas,
Medioambientales y Tecnol{\'o}gicas-Madrid, the University of Chicago, University College London, the DES-Brazil Consortium, the University of Edinburgh,
the Eidgen{\"o}ssische Technische Hochschule (ETH) Z{\"u}rich,
Fermi National Accelerator Laboratory, the University of Illinois at Urbana-Champaign, the Institut de Ci{\`e}ncies de l'Espai (IEEC/CSIC),
the Institut de F{\'i}sica d'Altes Energies, Lawrence Berkeley National Laboratory, the Ludwig-Maximilians Universit{\"a}t M{\"u}nchen and the associated Excellence Cluster Universe,
the University of Michigan, the National Optical Astronomy Observatory, the University of Nottingham, The Ohio State University, the University of Pennsylvania, the University of Portsmouth,
SLAC National Accelerator Laboratory, Stanford University, the University of Sussex, Texas A\&M University, and the OzDES Membership Consortium.

\bibliographystyle{mnras_2author}
\bibliography{literature,des,des_y1kp,des_y3kp,y6}

%%%%%%%%%%%%%%%%%%%%%%%%%%%%%%%%%%%%%%%%%%%%%%%%%%%%%%%%%%%%%%%%%%%

\appendix

\section{Configuration Used for DES Y6 PSF Model}
\label{app:config}

We used \piff\ version 1.2.4 for the DES Y6 PSF solutions.
The input configuration file is given below.
Values that needed to be set differently for each exposure and CCD were specified
on the command line when we ran the \code{piffify} executable for each image.
The other parameters shown here were the same for all exposures.

\begin{lstlisting}[style=yaml]
modules:
    - pixmappy

input:

    image_file_name: # set on command line
    image_hdu: 1
    badpix_hdu: 2
    weight_hdu: 3

    cat_file_name: # set on command line
    cat_hdu: 1
    x_col: XWIN_IMAGE
    y_col: YWIN_IMAGE
    sky_col: BACKGROUND
    ra: TELRA
    dec: TELDEC
    gain: 1.0
    property_cols: ['GAIA_STAR','EXT_MASH',
                    'R_MAG','Z_MAG','K_MAG',
                    'GI_COLOR','IZ_COLOR']

    stamp_size: 32

    wcs:
        type: Pixmappy
        dir: 'pixmappy'
        use_DESMaps: True
        guts_file: 'y6a1.guts.astro'
        exposure_file: 'y6a1.exposureinfo.fits'
        affine_file: 'y6a1.affine.fits'
        resids_file: 'y6a1.astroresids.fits'
        default_color: 1.3

        exp: # set on command line
        ccdnum: # set on command line

select:
    initial_select:
        type: Properties

        # for z-band images: replace
        # (GI_COLOR > 0) & (GI_COLOR < 3.5) with
        # (IZ_COLOR > 0) & (IZ_COLOR < 0.7)
        where: '(GAIA_STAR == 1) & (EXT_MASH == 0) & 
                (GI_COLOR > 0) & (GI_COLOR < 3.5)'

    type: SizeMag
    fit_order: 1
    purity: 0.05

    # for z-band images: replace
    # (GI_COLOR < 0) | (GI_COLOR > 3.5) with
    # (IZ_COLOR < 0) | (IZ_COLOR > 0.7)
    reject_where: '(EXT_MASH > 0) |
                   ((K_MAG > -99) &
                    (R_MAG > -99) &
                    (Z_MAG > -99) &
                    ((Z_MAG - K_MAG) >
                     0.5*(R_MAG - Z_MAG))) |
                   (GI_COLOR < 0) | (GI_COLOR > 3.5)'

    max_snr: 100
    min_snr: 20
    hsm_size_reject: 3
    reserve_frac: 0.2
    seed: # used expnum as seed, set on command line

psf:
    type: Simple

    model:
        type: PixelGrid
        scale: 0.30
        size: 17

    interp:
        type: BasisPolynomial
        # for z-band images: replace
        # 'GI_COLOR' with 'IZ_COLOR'
        keys: ['u', 'v', 'GI_COLOR']
        order: [2, 2, 1]

    outliers:
        type: Chisq
        nsigma: 4
        max_remove: 0.01

output:
    file_name: # set on command line
\end{lstlisting}

%%%%%%%%%%%%%%%%%%%%%%%%%%%%%%%%%%%%%%%%%%%%%%%%%%%%%%%%%%%%%%%%%%%

\section{DCR before and after correction}
\label{app:dcr}

\begin{figure}
\includegraphics[width=\columnwidth]{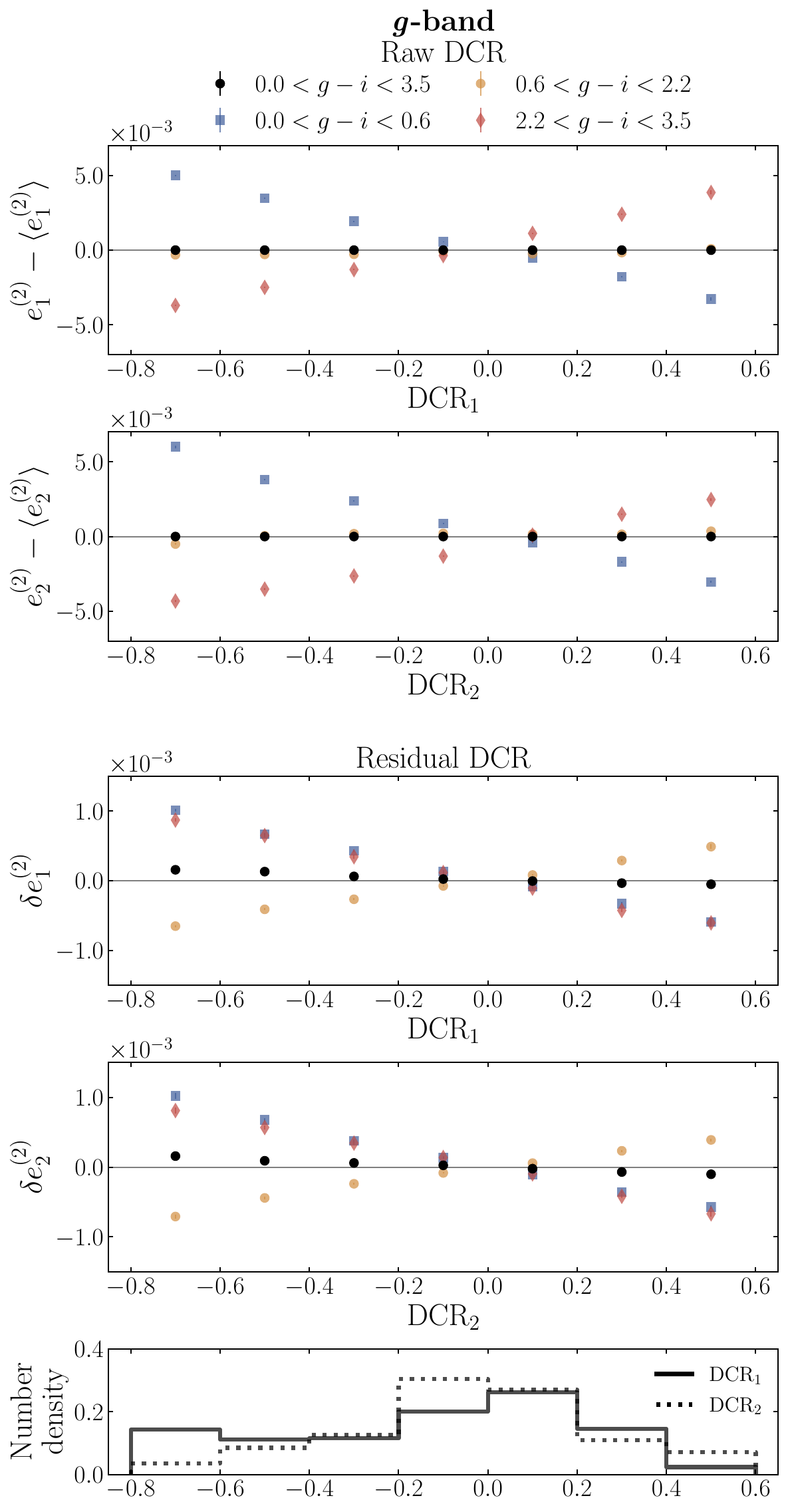}
\caption{
The mean-subtracted $e^{(2)}_1$ and $e^{(2)}_2$ in $g$ band (top two panels), binned by the DCR angle values described in \eqnb{eq:dcr1}{eq:dcr2}.
This isolates the uncorrected DCR effect in the survey.
The third and fourth panels reproduce the $g$-band $\delta e^{(2)}_1$ and $\delta e^{(2)}_2$ residuals shown in \fig{fig:dcr_2order}, i.e., the residuals after subtracting the color-dependent PSF model.
Comparing the raw and residual signals shows the DCR effect is corrected by $\sim 80 \%$ in $g$ band.
}
\label{fig:app:dcr}
\end{figure}

In this Appendix, we show the comparison between the raw DCR signal and the PSF shape residuals shown in \sect{sec:sky:dcr} to demonstrate how well the effect is corrected.
The top two panels of \fig{fig:app:dcr} show the mean-subtracted $e^{(2)}_1$ and $e^{(2)}_2$ in $g$ band, binned by the DCR angle values described in \eqnb{eq:dcr1}{eq:dcr2}.
This isolates the uncorrected DCR effect in the survey.
The bottom two panels reproduce the $g$-band $\delta e^{(2)}_1$ and $\delta e^{(2)}_2$ residuals shown in \fig{fig:dcr_2order}, i.e., the residuals after subtracting the color-dependent PSF model.
For $g$ band, the maximum amplitude of $\delta e^{(2)}_1$ and $\delta e^{(2)}_2$ is $\sim 10^{-3}$.
Compared to the measured raw DCR effect, the effect is corrected by $\sim80\%$.
For the $riz$ bands (not shown), the DCR effect is corrected by $> 90\%$.

%%%%%%%%%%%%%%%%%%%%%%%%%%%%%%%%%%%%%%%%%%%%%%%%%%%%%%%%%%%%%%%%%%%

\section{Weighting the PSF catalog to match galaxy distributions and shape measurement methodology}
\label{app:weights}

\begin{figure}
\includegraphics[width=\columnwidth]{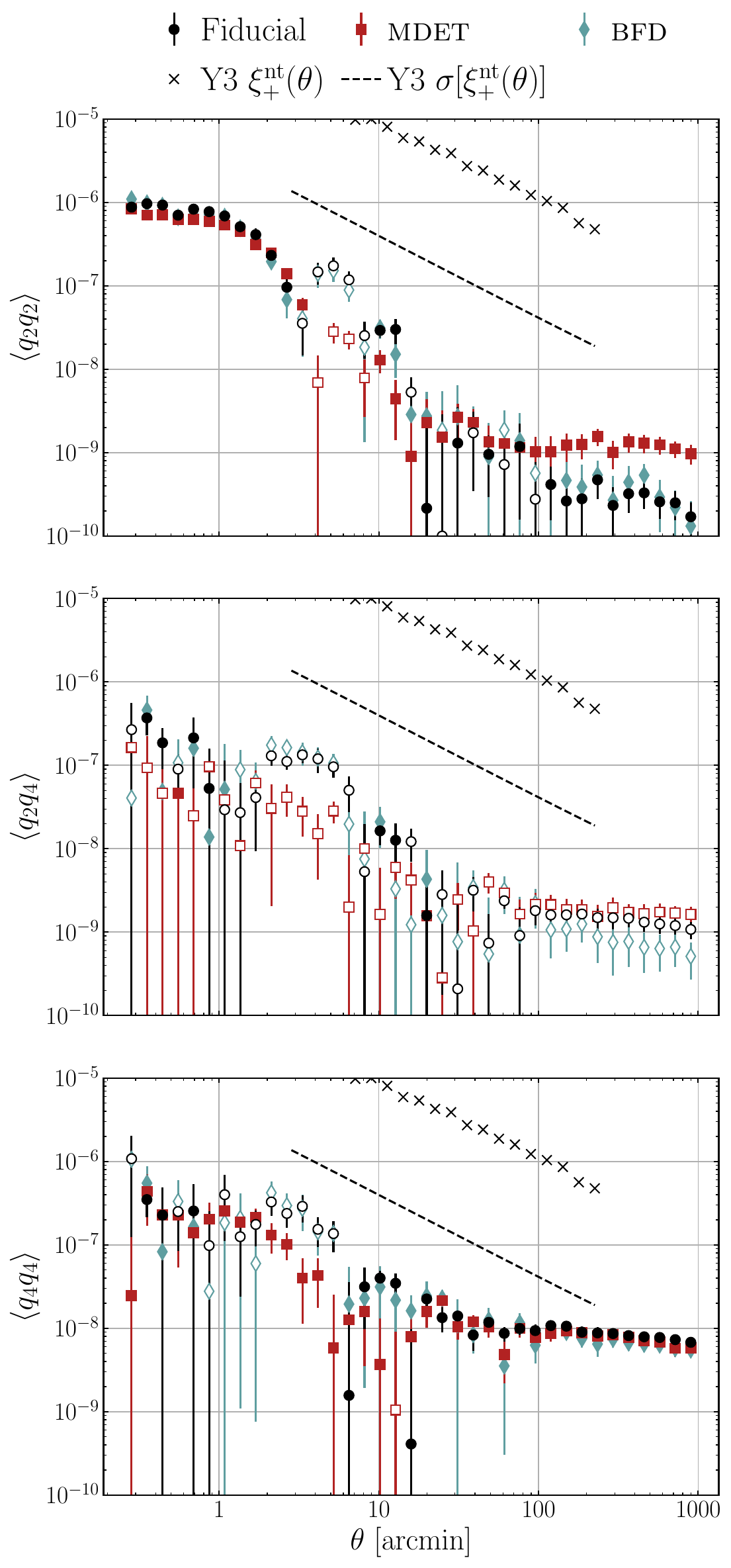},
\caption{
Shear-shear correlations for the second- and fourth-order PSF shape residuals.
Filled (open) markers correspond to positive (negative) values.
Shown are the correlations using three weighting schemes corresponding to the nominal PSF models (``fiducial'', black circles) and how the PSF models are used in the two Y6 shear estimation methods, \textsc{Metadetection} (\textsc{mdet}, red squares) and BFD (light blue diamonds).
The Y3 non-tomographic $\xi_+$ signal (black $\times$s) and standard deviation $\sigma[\xi_+^{\rm nt}(\theta)]$ (black dashed line) is shown for an order-of-magnitude comparison.
}
\label{fig:rho_shapecat_comp}
\end{figure}

A few difficulties arise when estimating PSF impact on shape measurement, particularly with color-dependent PSF models: (i) different shear estimation methods use the PSF in different ways, (ii) stars and galaxies do not have the same color distribution, and (iii) measurements from each band may have a different SNR contribution to any given galaxy shape measurement.

Using the PSF catalog with uniform weighting to measure $\rho$ statistics, as has been done for previous DES analyses, is a valid approach for characterizing the PSF modeling performance.
However, it does not capture the errors that may be incurred when using the PSF models on a set of objects the models are not ``trained on'' -- in our case, galaxies.
To calculate the PSF for each reserve star, the PSF is rendered at the exact epoch, location and color of the reserve star (besides the very small fraction of stars that are assigned the median star color).
Thus, the model residuals do not include any potential error from using an inaccurate color, which we will see below is the case for the \textsc{Metadetection} shear estimation algorithm.
Further, by construction these reserve stars have the exact same color distribution as the stars used to fit the PSF.
In practical use, galaxies have a very different color distribution from stars; indeed, its $g-i$ mode is in the intermediate values between the bimodal peaks of the stellar color distribution.
In this case, when using the PSF catalog without reweighting, the potential error from interpolating to areas in color space that are not well represented by the PSF stars is also not captured.

Here we develop a weighting scheme for the PSF catalog to more closely match how potential PSF biases would manifest in the \textsc{Metadetection} and BFD galaxy shape catalogs.
As such, we expect using these weights to result in more accurate parameters for the PSF contamination modeling (\eqn{eq:psfcontam}).

To mitigate shear-dependent detection bias, \textsc{Metadetection} measures shear on 200x200 pixel cells where the PSF is assumed to be constant over the cell.
A cell contains multiple galaxy images; thus, there is no single color to use for rendering the PSF.
Instead, the median galaxy color, determined over the entire galaxy sample, is used for every cell.
To capture the potential PSF bias from this, we create a second PSF catalog with the same set of reserve stars, where the PSF is rendered at the median galaxy colors ($g-i=1.1$ and $i-z=0.34$).

Even when using each galaxy's measured color in the PSF interpolation, as is the case in the BFD shear estimation method, the difference in the galaxy and stellar color distribution may still be a source of bias.
To estimate the impact of this, we calculate a set of weights for the PSF catalog such that their weighted color distribution matches that of the galaxy sample.
To do so, we compute the histograms of the galaxy and stellar samples in two colors ($g-i$ for $gri$ band stars and $i-z$ for $z$ band stars; both are computed for the galaxies) using the same binning for each.
For each histogram bin, we take the ratio of the bin value for the galaxies divided by that of the stars.
This ratio is then set as the weight for every star in that bin when computing the $\rho$ statistics (the weights are normalized during this calculation).
Thus, these reweighted $\rho$ statistics should better estimate potential problems due to interpolation from stellar colors to galaxy colors.

Finally, we also take into account that the relative importance of information from each band may be different between galaxies and stars.
With a uniformly weighted star catalog, how much each band contributes to the $\rho$ statistics is dictated by what fraction of stars were observed in that band.
For the galaxies, we make a simple estimate of the total S/N contributed by each band to the ensemble of galaxy shape measurements as
\begin{equation}
    \left(\frac{S}{N}\right)_{\rm gals,\, band} = \frac{\sum_{i} f_{i,\, \mathrm{band}}}{\sqrt{\sum_{i} \sigma^2_{i,\, \mathrm{band}}}},
\end{equation}
where $f_{i,\, \mathrm{band}}$ is the $i$th galaxy's flux measurement in a band (measured on a coadded image) and $\sigma^2_{i,\, \mathrm{band}}$ is the variance on that flux.
To form the final weights for the stars, we multiply the weights for stars measured in the $gri$ bands by the ratio of that band's galaxy S/N relative to that of $z$ band.

To demonstrate the effect of these adjustments, \fig{fig:rho_shapecat_comp} shows three leading PSF-PSF correlations, $\langle \bm{q}_2\bm{q}_2 \rangle$, $\langle \bm{q}_2\bm{q}_4 \rangle$, and $\langle \bm{q}_4\bm{q}_4 \rangle$ using three catalog/weighting schemes: uniform weighting where the true stellar color is used for interpolation (``fiducial'', black), weighted stars using the true stellar color (``BFD'', blue), and weighted stars where the median galaxy color is used for interpolation (``MDET'', red).
The nontomographic Y3 $\xi_+$ signal and its error is shown for an order-of-magnitude comparison.

The different catalogs and weights do not alter the $\rho$ statistics dramatically; however, some changes emerge.
The negative correlation feature at $\sim 2\arcmin-5\arcmin$ has a significantly different amplitude for the \mdet\ scheme compared to the fiducial and BFD schemes.
Further, the \mdet\ scheme better captures the constant mean correlation at large separation distance ($>100'$), likely due to the use of the median galaxy color causing a mean shear bias.
Though these are small effects, they may be more important for future surveys or in analyses using extended scales where PSF contamination must be modeled explicitly.

%%%%%%%%%%%%%%%%%%%%%%%%%%%%%%%%%%%%%%%%%%%%%%%%%%%%%%%%%%%%%%%%%%%

\section{Scale dependence of all second- and fourth-order PSF-PSF correlations on extended scales}
\label{app:alltherhos}

\begin{figure*}
\centering
\includegraphics[width=\textwidth]{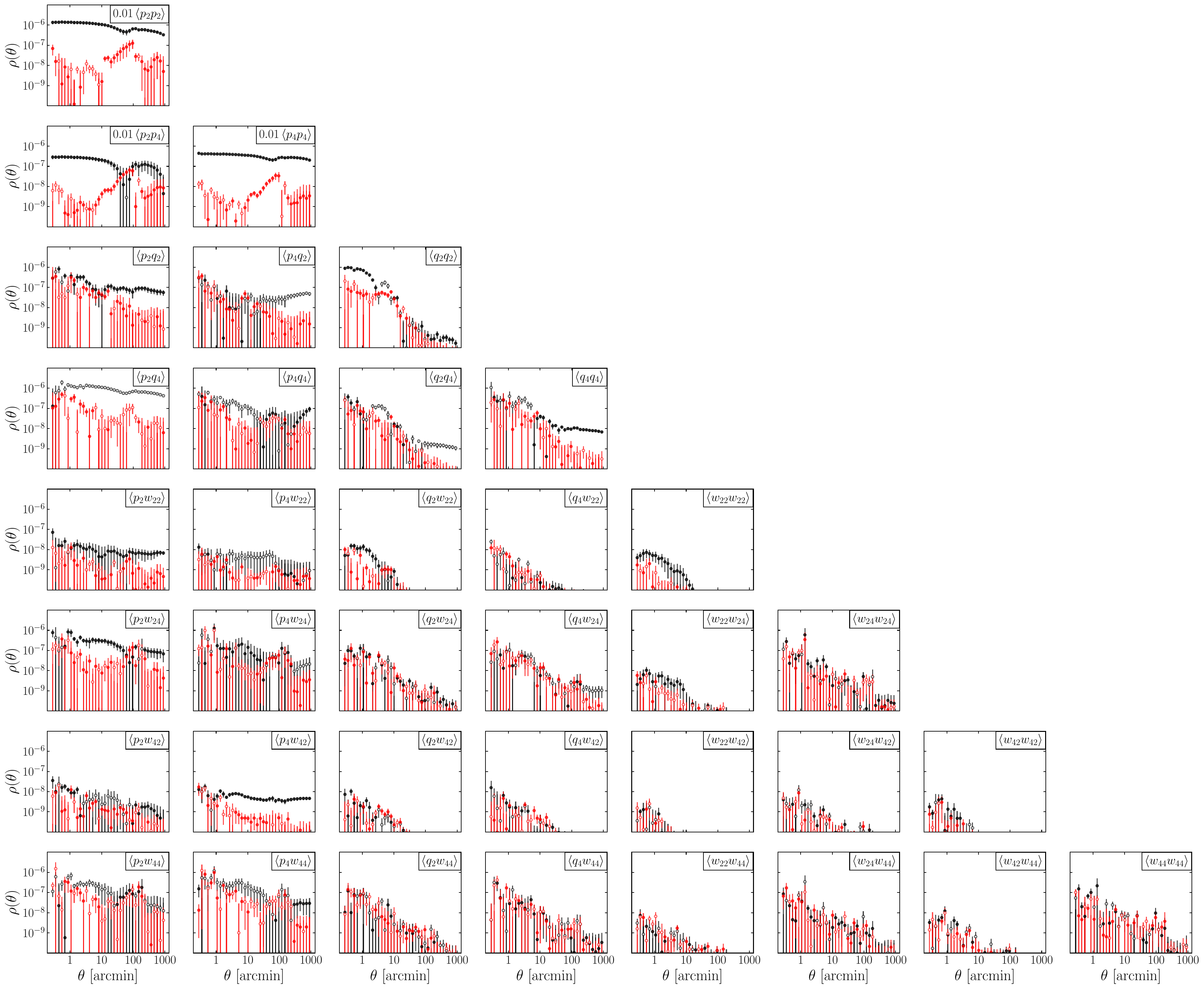}
\caption{
All 36 PSF-PSF correlation functions, defined as $\langle \bm{P}_k \bm{P}_l \rangle$ in \eqn{eq:tau_rho}, for the $riz$ bands combined.
The components corresponding to $\xi_+$ and $\xi_-$ are shown in black and red, respectively. Filled (open) markers denote positive (negative) values.
}
\label{fig:all_rhos}
\end{figure*}

In this Appendix, we present the 36 PSF-PSF correlation functions defined in \eqn{eq:tau_rho}. \fig{fig:all_rhos} shows these correlations measured over the extended $0.25'-1000'$ angular separation range.
Both the $\xi_+$ (black) and $\xi_-$ (red) components are shown.

%%%%%%%%%%%%%%%%%%%%%%%%%%%%%%%%%%%%%%%%%%%%%%%%%%%%%%%%%%%%%%%%%%%

\section{Color- and band-dependence of PSF-PSF correlations}
\label{app:rho_griz}

\begin{figure*}
\centering
\includegraphics[width=\textwidth]{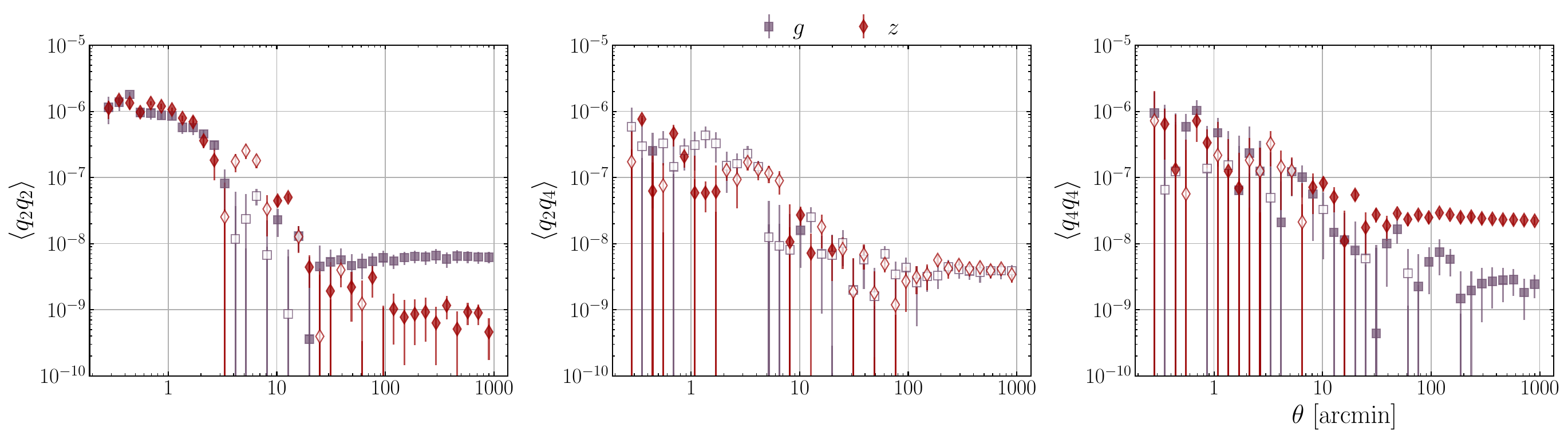}
\caption{
Shear-shear correlations for the second- and fourth-order PSF shape residuals, $\bm{q}_2$ and $\bm{q}_4$.
Filled (open) markers correspond to positive (negative) values.
$g$ and $z$ bands are shown to demonstrate the band dependence of the leading order $\rho$ statistics.
The $r$- and $i$-band correlations (not shown) fall within the ``envelope'' formed by the $g$- and $z$-band measurements.
}
\label{fig:rho_gz}
\end{figure*}

We expect the $\rho$ statistics to vary between different photometric bands as each is subject to somewhat different systematics.
In \fig{fig:rho_gz} we show the leading order $\rho$ statistics, $\langle \bm{q}_2\bm{q}_2 \rangle$ (top), $\langle \bm{q}_2\bm{q}_4 \rangle$ (middle) and $\langle \bm{q}_4\bm{q}_4 \rangle$ (bottom) for the $g$ and $z$ bands.
$r$ and $i$ bands are not shown as their correlation functions largely fall in the envelope formed by the $g$ and $z$ band correlation functions.
At small scales, the amplitude of the $\rho$ statistics is very similar, but they diverge at large scales.
For the second-order $\langle \bm{q}_2\bm{q}_2 \rangle$ correlation, the $g$-band values are significantly higher than those of $z$ band for $\theta >50'$.
On the other hand, for the fourth-order residuals, $z$ band's $\langle \bm{q}_4\bm{q}_4 \rangle$ dominates at $\theta > 100'$.
While the large-scale second-order correlation in $g$ band is likely due to residual chromatic effects causing a mean ellipticity, the cause of the high $\langle \bm{q}_4\bm{q}_4 \rangle$ in $z$ band remains unknown and requires future investigation.

%%%%%%%%%%%%%%%%%%%%%%%%%%%%%%%%%%%%%%%%%%%%%%%%%%%%%%%%%%%%%%%%%%%

\section{Author Affiliations}
\label{sec:affiliations}
\noindent
\input{affiliations}

\label{lastpage}

\end{document}

%% file: authors.tex
% Author list file generated with: mkauthlist 1.3.0+29.gb611c89 
% mkauthlist -sb --aux ../DES-2024-0874_author_order.csv -j emulateapj ../DES-2024-0874_author_list_v2.csv ../DES-2024-0874_author_list_v2.tex -f 

% \def\andname{}

\author{
T.~Schutt\altaffilmark{1,2,3},
M.~Jarvis\altaffilmark{4},
A.~Roodman\altaffilmark{1,2},
A.~Amon\altaffilmark{5},
M.~R.~Becker\altaffilmark{6},
R.~A.~Gruendl\altaffilmark{7,8},
M.~Yamamoto\altaffilmark{5,9},
K.~Bechtol\altaffilmark{10},
G.~M.~Bernstein\altaffilmark{4},
M.~Gatti\altaffilmark{4},
E.~S.~Rykoff\altaffilmark{1,2},
E.~Sheldon\altaffilmark{11},
M.~A.~Troxel\altaffilmark{9},
T.~M.~C.~Abbott\altaffilmark{12},
M.~Aguena\altaffilmark{13},
A.~Alarcon\altaffilmark{14},
F.~Andrade-Oliveira\altaffilmark{15},
D.~Brooks\altaffilmark{16},
A.~Carnero~Rosell\altaffilmark{17,13,18},
J.~Carretero\altaffilmark{19},
C.~Chang\altaffilmark{20,21},
A.~Choi\altaffilmark{22},
M.~Crocce\altaffilmark{23,14},
L.~N.~da Costa\altaffilmark{13},
T.~M.~Davis\altaffilmark{24},
J.~De~Vicente\altaffilmark{25},
S.~Desai\altaffilmark{26},
H.~T.~Diehl\altaffilmark{27},
S.~Dodelson\altaffilmark{20,27,21},
P.~Doel\altaffilmark{16},
C.~Doux\altaffilmark{4,28},
A.~Drlica-Wagner\altaffilmark{20,27,21},
A.~Fert\'e\altaffilmark{2},
J.~Frieman\altaffilmark{27,21},
J.~Garc\'ia-Bellido\altaffilmark{29},
E.~Gaztanaga\altaffilmark{23,30,14},
G.~Giannini\altaffilmark{19,21},
D.~Gruen\altaffilmark{31},
G.~Gutierrez\altaffilmark{27},
W.~G.~Hartley\altaffilmark{32},
K.~Herner\altaffilmark{27},
S.~R.~Hinton\altaffilmark{24},
D.~L.~Hollowood\altaffilmark{33},
K.~Honscheid\altaffilmark{34,35},
D.~Huterer\altaffilmark{36},
E.~Krause\altaffilmark{37},
K.~Kuehn\altaffilmark{38,39},
O.~Lahav\altaffilmark{16},
S.~Lee\altaffilmark{40},
M.~Lima\altaffilmark{41,13},
J.~L.~Marshall\altaffilmark{42},
J. Mena-Fern{\'a}ndez\altaffilmark{43},
R.~Miquel\altaffilmark{44,19},
J.~J.~Mohr\altaffilmark{45,31},
J.~Muir\altaffilmark{46},
J.~Myles\altaffilmark{5},
R.~L.~C.~Ogando\altaffilmark{47},
A.~Pieres\altaffilmark{13,47},
A.~A.~Plazas~Malag\'on\altaffilmark{1,2},
A.~Porredon\altaffilmark{25,48},
M.~Raveri\altaffilmark{49},
M.~Rodriguez-Monroy\altaffilmark{29},
S.~Samuroff\altaffilmark{50,19},
E.~Sanchez\altaffilmark{25},
D.~Sanchez Cid\altaffilmark{25},
I.~Sevilla-Noarbe\altaffilmark{25},
M.~Smith\altaffilmark{51},
E.~Suchyta\altaffilmark{52},
G.~Tarle\altaffilmark{36},
V.~Vikram\altaffilmark{53},
A.~R.~Walker\altaffilmark{12},
N.~Weaverdyck\altaffilmark{54,55},
and Y.~Zhang\altaffilmark{12}
\begin{center} (DES Collaboration) \end{center}
\begin{center} 
\small
\textit{The authors' affiliations are shown in Appendix \ref{sec:affiliations}.}
\end{center}
}

%% file: affiliations.tex
$^{1}$ Kavli Institute for Particle Astrophysics \& Cosmology, P. O. Box 2450, Stanford University, Stanford, CA 94305, USA\\
$^{2}$ SLAC National Accelerator Laboratory, Menlo Park, CA 94025, USA\\
$^{3}$ Department of Physics, Stanford University, 382 Via Pueblo Mall, Stanford, CA 94305, USA\\
$^{4}$ Department of Physics and Astronomy, University of Pennsylvania, Philadelphia, PA 19104, USA\\
$^{5}$ Department of Astrophysical Sciences, Princeton University, Peyton Hall, Princeton, NJ 08544, USA\\
$^{6}$ Argonne National Laboratory, 9700 South Cass Avenue, Lemont, IL 60439, USA\\
$^{7}$ Center for Astrophysical Surveys, National Center for Supercomputing Applications, 1205 West Clark St., Urbana, IL 61801, USA\\
$^{8}$ Department of Astronomy, University of Illinois at Urbana-Champaign, 1002 W. Green Street, Urbana, IL 61801, USA\\
$^{9}$ Department of Physics, Duke University Durham, NC 27708, USA\\
$^{10}$ Physics Department, 2320 Chamberlin Hall, University of Wisconsin-Madison, 1150 University Avenue Madison, WI  53706-1390\\
$^{11}$ Brookhaven National Laboratory, Bldg 510, Upton, NY 11973, USA\\
$^{12}$ Cerro Tololo Inter-American Observatory, NSF's National Optical-Infrared Astronomy Research Laboratory, Casilla 603, La Serena, Chile\\
$^{13}$ Laborat\'orio Interinstitucional de e-Astronomia - LIneA, Rua Gal. Jos\'e Cristino 77, Rio de Janeiro, RJ - 20921-400, Brazil\\
$^{14}$ Institute of Space Sciences (ICE, CSIC),  Campus UAB, Carrer de Can Magrans, s/n,  08193 Barcelona, Spain\\
$^{15}$ Physik-Institut — University of Zurich, Winterthurerstrasse 190, 8057 Zurich, Switzerland\\
$^{16}$ Department of Physics \& Astronomy, University College London, Gower Street, London, WC1E 6BT, UK\\
$^{17}$ Instituto de Astrofisica de Canarias, E-38205 La Laguna, Tenerife, Spain\\
$^{18}$ Universidad de La Laguna, Dpto. Astrofísica, E-38206 La Laguna, Tenerife, Spain\\
$^{19}$ Institut de F\'{\i}sica d'Altes Energies (IFAE), The Barcelona Institute of Science and Technology, Campus UAB, 08193 Bellaterra (Barcelona) Spain\\
$^{20}$ Department of Astronomy and Astrophysics, University of Chicago, Chicago, IL 60637, USA\\
$^{21}$ Kavli Institute for Cosmological Physics, University of Chicago, Chicago, IL 60637, USA\\
$^{22}$ NASA Goddard Space Flight Center, 8800 Greenbelt Rd, Greenbelt, MD 20771, USA\\
$^{23}$ Institut d'Estudis Espacials de Catalunya (IEEC), 08034 Barcelona, Spain\\
$^{24}$ School of Mathematics and Physics, University of Queensland,  Brisbane, QLD 4072, Australia\\
$^{25}$ Centro de Investigaciones Energ\'eticas, Medioambientales y Tecnol\'ogicas (CIEMAT), Madrid, Spain\\
$^{26}$ Department of Physics, IIT Hyderabad, Kandi, Telangana 502285, India\\
$^{27}$ Fermi National Accelerator Laboratory, P. O. Box 500, Batavia, IL 60510, USA\\
$^{28}$ Universit\'e Grenoble Alpes, CNRS, LPSC-IN2P3, 38000 Grenoble, France\\
$^{29}$ Instituto de Fisica Teorica UAM/CSIC, Universidad Autonoma de Madrid, 28049 Madrid, Spain\\
$^{30}$ Institute of Cosmology and Gravitation, University of Portsmouth, Portsmouth, PO1 3FX, UK\\
$^{31}$ University Observatory, Faculty of Physics, Ludwig-Maximilians-Universit\"at, Scheinerstr. 1, 81679 Munich, Germany\\
$^{32}$ Department of Astronomy, University of Geneva, ch. d'\'Ecogia 16, CH-1290 Versoix, Switzerland\\
$^{33}$ Santa Cruz Institute for Particle Physics, Santa Cruz, CA 95064, USA\\
$^{34}$ Center for Cosmology and Astro-Particle Physics, The Ohio State University, Columbus, OH 43210, USA\\
$^{35}$ Department of Physics, The Ohio State University, Columbus, OH 43210, USA\\
$^{36}$ Department of Physics, University of Michigan, Ann Arbor, MI 48109, USA\\
$^{37}$ Department of Astronomy/Steward Observatory, University of Arizona, 933 North Cherry Avenue, Tucson, AZ 85721-0065, USA\\
$^{38}$ Australian Astronomical Optics, Macquarie University, North Ryde, NSW 2113, Australia\\
$^{39}$ Lowell Observatory, 1400 Mars Hill Rd, Flagstaff, AZ 86001, USA\\
$^{40}$ Jet Propulsion Laboratory, California Institute of Technology, 4800 Oak Grove Dr., Pasadena, CA 91109, USA\\
$^{41}$ Departamento de F\'isica Matem\'atica, Instituto de F\'isica, Universidade de S\~ao Paulo, CP 66318, S\~ao Paulo, SP, 05314-970, Brazil\\
$^{42}$ George P. and Cynthia Woods Mitchell Institute for Fundamental Physics and Astronomy, and Department of Physics and Astronomy, Texas A\&M University, College Station, TX 77843,  USA\\
$^{43}$ LPSC Grenoble - 53, Avenue des Martyrs 38026 Grenoble, France\\
$^{44}$ Instituci\'o Catalana de Recerca i Estudis Avan\c{c}ats, E-08010 Barcelona, Spain\\
$^{45}$ Max Planck Institute for Extraterrestrial Physics, Giessenbachstrasse, 85748 Garching, Germany\\
$^{46}$ Perimeter Institute for Theoretical Physics, 31 Caroline St. North, Waterloo, ON N2L 2Y5, Canada\\
$^{47}$ Observat\'orio Nacional, Rua Gal. Jos\'e Cristino 77, Rio de Janeiro, RJ - 20921-400, Brazil\\
$^{48}$ Ruhr University Bochum, Faculty of Physics and Astronomy, Astronomical Institute, German Centre for Cosmological Lensing, 44780 Bochum, Germany\\
$^{49}$ Department of Physics, University of Genova and INFN, Via Dodecaneso 33, 16146, Genova, Italy\\
$^{50}$ Department of Physics, Northeastern University, Boston, MA 02115, USA\\
$^{51}$ Physics Department, Lancaster University, Lancaster, LA1 4YB, UK\\
$^{52}$ Computer Science and Mathematics Division, Oak Ridge National Laboratory, Oak Ridge, TN 37831\\
$^{53}$ Central University of Kerala, Kasaragod, Kerala, India\\
$^{54}$ Department of Astronomy, University of California, Berkeley,  501 Campbell Hall, Berkeley, CA 94720, USA\\
$^{55}$ Lawrence Berkeley National Laboratory, 1 Cyclotron Road, Berkeley, CA 94720, USA

%% file: main.bbl
\begin{thebibliography}{}
\makeatletter
\relax
\def\mn@urlcharsother{\let\do\@makeother \do\$\do\&\do\#\do\^\do\_\do\%\do\~}
\def\mn@doi{\begingroup\mn@urlcharsother \@ifnextchar [ {\mn@doi@}
  {\mn@doi@[]}}
\def\mn@doi@[#1]#2{\def\@tempa{#1}\ifx\@tempa\@empty \href
  {http://dx.doi.org/#2} {doi:#2}\else \href {http://dx.doi.org/#2} {#1}\fi
  \endgroup}
\def\mn@eprint#1#2{\mn@eprint@#1:#2::\@nil}
\def\mn@eprint@arXiv#1{\href {http://arxiv.org/abs/#1} {{\tt arXiv:#1}}}
\def\mn@eprint@dblp#1{\href {http://dblp.uni-trier.de/rec/bibtex/#1.xml}
  {dblp:#1}}
\def\mn@eprint@#1:#2:#3:#4\@nil{\def\@tempa {#1}\def\@tempb {#2}\def\@tempc
  {#3}\ifx \@tempc \@empty \let \@tempc \@tempb \let \@tempb \@tempa \fi \ifx
  \@tempb \@empty \def\@tempb {arXiv}\fi \@ifundefined
  {mn@eprint@\@tempb}{\@tempb:\@tempc}{\expandafter \expandafter \csname
  mn@eprint@\@tempb\endcsname \expandafter{\@tempc}}}

\bibitem[\protect\citeauthoryear{{Amon} \& {Gruen} et~al.,}{{Amon}
  et~al.}{2022}]{y3-cosmicshear1}
{Amon} A.,  et~al. 2022, \mn@doi [\prd] {10.1103/PhysRevD.105.023514}, \href
  {https://ui.adsabs.harvard.edu/abs/2022PhRvD.105b3514A} {105, 023514}

\bibitem[\protect\citeauthoryear{{Antilogus} \& {Astier} et~al.,}{{Antilogus}
  et~al.}{2014}]{Antilogus14}
{Antilogus} P.,  et~al. 2014, \mn@doi [Journal of Instrumentation]
  {10.1088/1748-0221/9/03/C03048}, \href
  {http://adsabs.harvard.edu/abs/2014JInst...9C3048A} {9, C3048}

\bibitem[\protect\citeauthoryear{{Armstrong} \& {Sheldon} et~al.,}{{Armstrong}
  et~al.}{2024}]{Armstrong24}
{Armstrong} R.,  et~al. 2024, \mn@doi [arXiv e-prints]
  {10.48550/arXiv.2407.01771}, \href
  {https://ui.adsabs.harvard.edu/abs/2024arXiv240701771A} {p. arXiv:2407.01771}

\bibitem[\protect\citeauthoryear{{Asgari} \& {Lin} et~al.,}{{Asgari}
  et~al.}{2021}]{Asgari21}
{Asgari} M.,  et~al. 2021, \mn@doi [\aap] {10.1051/0004-6361/202039070}, \href
  {https://ui.adsabs.harvard.edu/abs/2021A&A...645A.104A} {645, A104}

\bibitem[\protect\citeauthoryear{{Bechtol} et~al.}{{Bechtol}
  et~al.}{2025}]{y6-gold}
{Bechtol} K.,  et~al., 2025, in preparation.

\bibitem[\protect\citeauthoryear{{Bernstein} \& {Armstrong}
  et~al.,}{{Bernstein} et~al.}{2017}]{pixmappy}
{Bernstein} G.~M.,  et~al. 2017, \mn@doi [\pasp] {10.1088/1538-3873/aa6c55},
  \href {https://ui.adsabs.harvard.edu/abs/2017PASP..129g4503B} {129, 074503}

\bibitem[\protect\citeauthoryear{{Bertin} \& {Arnouts}}{{Bertin} \&
  {Arnouts}}{1996}]{BertinSExtractor1996}
{Bertin} E.,  {Arnouts} S.,  1996, \mn@doi [\aaps] {10.1051/aas:1996164}, \href
  {http://adsabs.harvard.edu/abs/1996A%26AS..117..393B} {117, 393}

\bibitem[\protect\citeauthoryear{{Carlsten} \& {Strauss} et~al.,}{{Carlsten}
  et~al.}{2018}]{Carlsten18}
{Carlsten} S.~G.,  et~al. 2018, \mn@doi [\mnras] {10.1093/mnras/sty1636}, \href
  {https://ui.adsabs.harvard.edu/abs/2018MNRAS.479.1491C} {479, 1491}

\bibitem[\protect\citeauthoryear{{DES Collaboration} \& {Abbott} et~al.,}{{DES
  Collaboration}}{2021}]{des-dr2}
{DES Collaboration} 2021, \mn@doi [\apjs] {10.3847/1538-4365/ac00b3}, \href
  {https://ui.adsabs.harvard.edu/abs/2021ApJS..255...20A} {255, 20}

\bibitem[\protect\citeauthoryear{{Dalal} \& {Li} et~al.,}{{Dalal}
  et~al.}{2023}]{Dalal23}
{Dalal} R.,  et~al. 2023, \mn@doi [\prd] {10.1103/PhysRevD.108.123519}, \href
  {https://ui.adsabs.harvard.edu/abs/2023PhRvD.108l3519D} {108, 123519}

\bibitem[\protect\citeauthoryear{{Davis}, {Rodriguez}  \& {Roodman}}{{Davis}
  et~al.}{2016}]{Davis16}
{Davis} C.~P.,  {Rodriguez} J.,   {Roodman} A.,  2016, in {Hall} H.~J.,
  {Gilmozzi} R.,   {Marshall} H.~K.,  eds,  Society of Photo-Optical
  Instrumentation Engineers (SPIE) Conference Series Vol. 9906, Ground-based
  and Airborne Telescopes VI. p. 990668, \mn@doi{10.1117/12.2233366}

\bibitem[\protect\citeauthoryear{{Diehl} \& {Angstadt} et~al.,}{{Diehl}
  et~al.}{2008}]{DECam-ccds}
{Diehl} H.~T.,  et~al. 2008, in {Dorn} D.~A.,  {Holland} A.~D.,  eds,  Society
  of Photo-Optical Instrumentation Engineers (SPIE) Conference Series Vol.
  7021, High Energy, Optical, and Infrared Detectors for Astronomy III. p.
  702107, \mn@doi{10.1117/12.790053}

\bibitem[\protect\citeauthoryear{{Edge} \& {Sutherland} et~al.,}{{Edge}
  et~al.}{2013}]{viking}
{Edge} A.,  et~al. 2013, The Messenger, \href
  {https://ui.adsabs.harvard.edu/abs/2013Msngr.154...32E} {154, 32}

\bibitem[\protect\citeauthoryear{{Esteves} \& {Utsumi} et~al.,}{{Esteves}
  et~al.}{2023}]{Esteves23}
{Esteves} J.~H.,  et~al. 2023, \mn@doi [\pasp] {10.1088/1538-3873/ad0a73},
  \href {https://ui.adsabs.harvard.edu/abs/2023PASP..135k5003E} {135, 115003}

\bibitem[\protect\citeauthoryear{{Flaugher} \& {Diehl} et~al.,}{{Flaugher}
  et~al.}{2015}]{DECam}
{Flaugher} B.,  et~al. 2015, \mn@doi [\aj] {10.1088/0004-6256/150/5/150}, \href
  {http://adsabs.harvard.edu/abs/2015AJ....150..150F} {150, 150}

\bibitem[\protect\citeauthoryear{{Gaia Collaboration} \& {Prusti}
  et~al.,}{{Gaia Collaboration}}{2016}]{gaia-instrument}
{Gaia Collaboration} 2016, \mn@doi [\aap] {10.1051/0004-6361/201629272}, \href
  {https://ui.adsabs.harvard.edu/abs/2016A&A...595A...1G} {595, A1}

\bibitem[\protect\citeauthoryear{{Gaia Collaboration} \& {Brown} et~al.,}{{Gaia
  Collaboration}}{2021}]{gaia-edr3}
{Gaia Collaboration} 2021, \mn@doi [\aap] {10.1051/0004-6361/202039657}, \href
  {https://ui.adsabs.harvard.edu/abs/2021A&A...649A...1G} {649, A1}

\bibitem[\protect\citeauthoryear{{Gatti}, {Wetzell}  et~al.}{{Gatti}
  et~al.}{2025}]{y6-bfd}
{Gatti} M.,  {Wetzell} V.,   et~al., 2025, in preparation.

\bibitem[\protect\citeauthoryear{{Giblin} \& {Heymans} et~al.,}{{Giblin}
  et~al.}{2021}]{Giblin21}
{Giblin} B.,  et~al. 2021, \mn@doi [\aap] {10.1051/0004-6361/202038850}, \href
  {https://ui.adsabs.harvard.edu/abs/2021A&A...645A.105G} {645, A105}

\bibitem[\protect\citeauthoryear{{Gruen} \& {Bernstein} et~al.,}{{Gruen}
  et~al.}{2015}]{Gruen15}
{Gruen} D.,  et~al. 2015, \mn@doi [Journal of Instrumentation]
  {10.1088/1748-0221/10/05/C05032}, \href
  {https://ui.adsabs.harvard.edu/abs/2015JInst..10C5032G} {10, C05032}

\bibitem[\protect\citeauthoryear{{Guyonnet} \& {Astier} et~al.,}{{Guyonnet}
  et~al.}{2015}]{Guyonnet15}
{Guyonnet} A.,  et~al. 2015, \mn@doi [\aap] {10.1051/0004-6361/201424897},
  \href {http://adsabs.harvard.edu/abs/2015A%26A...575A..41G} {575, A41}

\bibitem[\protect\citeauthoryear{{Hardy}}{{Hardy}}{1998}]{Hardy98}
{Hardy} J.~W.,  1998, {Adaptive Optics for Astronomical Telescopes}

\bibitem[\protect\citeauthoryear{{Hirata} \& {Seljak}}{{Hirata} \&
  {Seljak}}{2003}]{hirata2003}
{Hirata} C.,  {Seljak} U.,  2003, \mn@doi [\mnras]
  {10.1046/j.1365-8711.2003.06683.x}, \href
  {http://adsabs.harvard.edu/abs/2003MNRAS.343..459H} {343, 459}

\bibitem[\protect\citeauthoryear{Holland \& Groom et~al.,}{Holland
  et~al.}{2003}]{Holland03}
Holland S.,  et~al. 2003, \mn@doi [IEEE Transactions on Electron Devices]
  {10.1109/TED.2002.806476}, 50, 225

\bibitem[\protect\citeauthoryear{{Howell}}{{Howell}}{2006}]{Howell06}
{Howell} S.~B.,  2006, {Handbook of CCD Astronomy}.
 {Cambridge Observing Handbooks for Research Astronomers} Vol. 5

\bibitem[\protect\citeauthoryear{{Jarvis}}{{Jarvis}}{2015}]{TreeCorr}
{Jarvis} M.,  2015, {TreeCorr: Two-point correlation functions}, Astrophysics
  Source Code Library (\mn@eprint {ascl} {1508.007})

\bibitem[\protect\citeauthoryear{{Jarvis}, {Bernstein}  \& {Jain}}{{Jarvis}
  et~al.}{2004}]{Jarvis04}
{Jarvis} M.,  {Bernstein} G.,   {Jain} B.,  2004, \mn@doi [\mnras]
  {10.1111/j.1365-2966.2004.07926.x}, \href
  {https://ui.adsabs.harvard.edu/abs/2004MNRAS.352..338J} {352, 338}

\bibitem[\protect\citeauthoryear{{Jarvis} \& {Sheldon} et~al.,}{{Jarvis}
  et~al.}{2016}]{sv-shearcat}
{Jarvis} M.,  et~al. 2016, \mn@doi [\mnras] {10.1093/mnras/stw990}, \href
  {http://adsabs.harvard.edu/abs/2016MNRAS.460.2245J} {460, 2245}

\bibitem[\protect\citeauthoryear{{Jarvis} \& {Bernstein} et~al.,}{{Jarvis}
  et~al.}{2021}]{y3-piff}
{Jarvis} M.,  et~al. 2021, \mn@doi [\mnras] {10.1093/mnras/staa3679}, \href
  {https://ui.adsabs.harvard.edu/abs/2021MNRAS.501.1282J} {501, 1282}

\bibitem[\protect\citeauthoryear{{Kamata} \& {Nakaya} et~al.,}{{Kamata}
  et~al.}{2014}]{Kamata14}
{Kamata} Y.,  et~al. 2014, in {Holland} A.~D.,  {Beletic} J.,  eds,  Society of
  Photo-Optical Instrumentation Engineers (SPIE) Conference Series Vol. 9154,
  High Energy, Optical, and Infrared Detectors for Astronomy VI. p. 91541Z,
  \mn@doi{10.1117/12.2055763}

\bibitem[\protect\citeauthoryear{{Kotov} \& {Kotov} et~al.,}{{Kotov}
  et~al.}{2010}]{Kotov10}
{Kotov} I.~V.,  et~al. 2010, in Holland A.~D.,  Dorn D.~A.,  eds,  Society of
  Photo-Optical Instrumentation Engineers (SPIE) Conference Series Vol. 7742,
  High Energy, Optical, and Infrared Detectors for Astronomy IV. SPIE, pp 63 --
  70, \mn@doi{10.1117/12.856519}

\bibitem[\protect\citeauthoryear{{Lee} \& {Acevedo} et~al.,}{{Lee}
  et~al.}{2023}]{Lee23}
{Lee} J.,  et~al. 2023, \mn@doi [\aj] {10.3847/1538-3881/acca15}, \href
  {https://ui.adsabs.harvard.edu/abs/2023AJ....165..222L} {165, 222}

\bibitem[\protect\citeauthoryear{{Li} \& {Miyatake} et~al.,}{{Li}
  et~al.}{2022}]{Li22}
{Li} X.,  et~al. 2022, \mn@doi [\pasj] {10.1093/pasj/psac006}, \href
  {https://ui.adsabs.harvard.edu/abs/2022PASJ...74..421L} {74, 421}

\bibitem[\protect\citeauthoryear{{Li} \& {Zhang} et~al.,}{{Li}
  et~al.}{2023}]{Li23}
{Li} X.,  et~al. 2023, \mn@doi [\prd] {10.1103/PhysRevD.108.123518}, \href
  {https://ui.adsabs.harvard.edu/abs/2023PhRvD.108l3518L} {108, 123518}

\bibitem[\protect\citeauthoryear{{Liaudat}, {Starck}  \& {Kilbinger}}{{Liaudat}
  et~al.}{2023}]{Liaudat23}
{Liaudat} T.~I.,  {Starck} J.-L.,   {Kilbinger} M.,  2023, \mn@doi [Frontiers
  in Astronomy and Space Sciences] {10.3389/fspas.2023.1158213}, \href
  {https://ui.adsabs.harvard.edu/abs/2023FrASS..1058213L} {10, 1158213}

\bibitem[\protect\citeauthoryear{{Magnier} \& {Tonry} et~al.,}{{Magnier}
  et~al.}{2018}]{Magnier18}
{Magnier} E.~A.,  et~al. 2018, \mn@doi [\pasp] {10.1088/1538-3873/aaaad8},
  \href {https://ui.adsabs.harvard.edu/abs/2018PASP..130f5002M} {130, 065002}

\bibitem[\protect\citeauthoryear{Mandelbaum}{Mandelbaum}{2018}]{Mandelbaum18}
Mandelbaum R.,  2018, \mn@doi [Annual Review of Astronomy and Astrophysics]
  {10.1146/annurev-astro-081817-051928}, 56, 393

\bibitem[\protect\citeauthoryear{{Mandelbaum} \& {Hirata} et~al.,}{{Mandelbaum}
  et~al.}{2005}]{hsm}
{Mandelbaum} R.,  et~al. 2005, \mn@doi [\mnras]
  {10.1111/j.1365-2966.2005.09282.x}, \href
  {http://adsabs.harvard.edu/abs/2005MNRAS.361.1287M} {361, 1287}

\bibitem[\protect\citeauthoryear{{McMahon} \& {Banerji} et~al.,}{{McMahon}
  et~al.}{2013}]{vhs}
{McMahon} R.~G.,  et~al. 2013, The Messenger, \href
  {https://ui.adsabs.harvard.edu/abs/2013Msngr.154...35M} {154, 35}

\bibitem[\protect\citeauthoryear{{Melchior} \& {Suchyta} et~al.,}{{Melchior}
  et~al.}{2015}]{Melchior15}
{Melchior} P.,  et~al. 2015, \mn@doi [\mnras] {10.1093/mnras/stv398}, \href
  {https://ui.adsabs.harvard.edu/abs/2015MNRAS.449.2219M} {449, 2219}

\bibitem[\protect\citeauthoryear{{Meyers} \& {Burchat}}{{Meyers} \&
  {Burchat}}{2015a}]{Meyers15a}
{Meyers} J.~E.,  {Burchat} P.~R.,  2015a, \mn@doi [Journal of Instrumentation]
  {10.1088/1748-0221/10/06/C06004}, \href
  {https://ui.adsabs.harvard.edu/abs/2015JInst..10C6004M} {10, C06004}

\bibitem[\protect\citeauthoryear{{Meyers} \& {Burchat}}{{Meyers} \&
  {Burchat}}{2015b}]{Meyers15b}
{Meyers} J.~E.,  {Burchat} P.~R.,  2015b, \mn@doi [\apj]
  {10.1088/0004-637X/807/2/182}, \href
  {https://ui.adsabs.harvard.edu/abs/2015ApJ...807..182M} {807, 182}

\bibitem[\protect\citeauthoryear{{Morganson} \& {Gruendl} et~al.,}{{Morganson}
  et~al.}{2018}]{imageproc}
{Morganson} E.,  et~al. 2018, \mn@doi [\pasp] {10.1088/1538-3873/aab4ef}, \href
  {https://ui.adsabs.harvard.edu/abs/2018PASP..130g4501M} {130, 074501}

\bibitem[\protect\citeauthoryear{{Park} \& {Karpov} et~al.,}{{Park}
  et~al.}{2020}]{Park20}
{Park} H.,  et~al. 2020, \mn@doi [Journal of Astronomical Telescopes,
  Instruments, and Systems] {10.1117/1.JATIS.6.1.011005}, \href
  {https://ui.adsabs.harvard.edu/abs/2020JATIS...6a1005P} {6, 011005}

\bibitem[\protect\citeauthoryear{{Paulin-Henriksson} \& {Amara}
  et~al.,}{{Paulin-Henriksson} et~al.}{2008}]{paulinhenriksson08}
{Paulin-Henriksson} S.,  et~al. 2008, \mn@doi [\aap]
  {10.1051/0004-6361:20079150}, \href
  {http://adsabs.harvard.edu/abs/2008A%26A...484...67P} {484, 67}

\bibitem[\protect\citeauthoryear{{Plazas} \& {Bernstein}}{{Plazas} \&
  {Bernstein}}{2012}]{Plazas12}
{Plazas} A.,  {Bernstein} G.,  2012, \mn@doi [\pasp] {10.1086/668294}, \href
  {https://ui.adsabs.harvard.edu/abs/2012PASP..124.1113A} {124, 1113}

\bibitem[\protect\citeauthoryear{{Plazas}, {Bernstein}  \& {Sheldon}}{{Plazas}
  et~al.}{2014a}]{plazas14a}
{Plazas} A.~A.,  {Bernstein} G.~M.,   {Sheldon} E.~S.,  2014a, \mn@doi [Journal
  of Instrumentation] {10.1088/1748-0221/9/04/C04001}, \href
  {http://adsabs.harvard.edu/abs/2014JInst...9C4001P} {9, C04001}

\bibitem[\protect\citeauthoryear{{Plazas}, {Bernstein}  \& {Sheldon}}{{Plazas}
  et~al.}{2014b}]{plazas14b}
{Plazas} A.~A.,  {Bernstein} G.~M.,   {Sheldon} E.~S.,  2014b, \mn@doi [\pasp]
  {10.1086/677682}, \href {http://adsabs.harvard.edu/abs/2014PASP..126..750P}
  {126, 750}

\bibitem[\protect\citeauthoryear{{Roodman} et~al.}{{Roodman}
  et~al.}{2025}]{roodman-inprep}
{Roodman} A.,  et~al., 2025, in preparation.

\bibitem[\protect\citeauthoryear{{Rowe}}{{Rowe}}{2010}]{Rowe10}
{Rowe} B.,  2010, \mn@doi [\mnras] {10.1111/j.1365-2966.2010.16277.x}, \href
  {http://adsabs.harvard.edu/abs/2010MNRAS.404..350R} {404, 350}

\bibitem[\protect\citeauthoryear{Rykoff \& Tucker et~al.,}{Rykoff
  et~al.}{2023}]{Rykoff23}
Rykoff E.~S.,  et~al. 2023, \mn@doi [arXiv e-prints]
  {10.48550/arXiv.2305.01695}

\bibitem[\protect\citeauthoryear{{Secco} \& {Samuroff} et~al.,}{{Secco}
  et~al.}{2022}]{y3-cosmicshear2}
{Secco} L.~F.,  et~al. 2022, \mn@doi [\prd] {10.1103/PhysRevD.105.023515},
  \href {https://ui.adsabs.harvard.edu/abs/2022PhRvD.105b3515S} {105, 023515}

\bibitem[\protect\citeauthoryear{{Xin} \& {Ivezi{\'c}} et~al.,}{{Xin}
  et~al.}{2018}]{Xin18}
{Xin} B.,  et~al. 2018, \mn@doi [\aj] {10.3847/1538-3881/aae316}, \href
  {https://ui.adsabs.harvard.edu/abs/2018AJ....156..222X} {156, 222}

\bibitem[\protect\citeauthoryear{{Yamamoto}, {Becker}  et~al.}{{Yamamoto}
  et~al.}{2025}]{y6-mdet}
{Yamamoto} M.,  {Becker} M.~R.,   et~al., 2025, in preparation.

\bibitem[\protect\citeauthoryear{{Zhang}, {Mandelbaum}  \& {LSST Dark Energy
  Science Collaboration}}{{Zhang} et~al.}{2022}]{Zhang22}
{Zhang} T.,  {Mandelbaum} R.,   {LSST Dark Energy Science Collaboration} 2022,
  \mn@doi [\mnras] {10.1093/mnras/stab3584}, \href
  {https://ui.adsabs.harvard.edu/abs/2022MNRAS.510.1978Z} {510, 1978}

\bibitem[\protect\citeauthoryear{{Zhang} \& {Almoubayyed} et~al.,}{{Zhang}
  et~al.}{2023a}]{Zhang23a}
{Zhang} T.,  et~al. 2023a, \mn@doi [\mnras] {10.1093/mnras/stac3350}, \href
  {https://ui.adsabs.harvard.edu/abs/2023MNRAS.520.2328Z} {520, 2328}

\bibitem[\protect\citeauthoryear{{Zhang} \& {Li} et~al.,}{{Zhang}
  et~al.}{2023b}]{Zhang23b}
{Zhang} T.,  et~al. 2023b, \mn@doi [\mnras] {10.1093/mnras/stad1801}, \href
  {https://ui.adsabs.harvard.edu/abs/2023MNRAS.525.2441Z} {525, 2441}

\bibitem[\protect\citeauthoryear{{Zuntz} \& {Sheldon} et~al.,}{{Zuntz}
  et~al.}{2018}]{y1-shearcat}
{Zuntz} J.,  et~al. 2018, \mn@doi [\mnras] {10.1093/mnras/sty2219}, \href
  {https://ui.adsabs.harvard.edu/abs/2018MNRAS.481.1149Z} {481, 1149}

\makeatother
\end{thebibliography}
